\documentclass[a4paper,12pt]{article}
\usepackage[dvipdfmx]{graphicx}
\usepackage{amsfonts,amsthm,amsmath,amssymb,graphicx,float,mathtools,xcolor}
\usepackage[export]{adjustbox}
\numberwithin{equation}{section}
\usepackage{color}
\usepackage{bbm} 
\usepackage{tensor} 
\usepackage{verbatim} 
\usepackage[
	  pagebackref=false,
	  colorlinks=true,
      linkcolor=blue,
      urlcolor=blue,
      filecolor=black,
      citecolor=red,
      pdfstartview=FitV,
      pdftitle={},
        pdfauthor={},
        pdfsubject={},
        pdfkeywords={},
        pdfpagemode=None,
        bookmarksopen=true
      ]{hyperref}

\begin{document}

\newtheorem{definition}{Definition}[section]
\newcommand{\be}{\begin{equation}}
\newcommand{\ee}{\end{equation}}
\newcommand{\bea}{\begin{eqnarray}}
\newcommand{\eea}{\end{eqnarray}}
\newcommand{\LE}{\left[}
\newcommand{\R}{\right]}
\newcommand{\nn}{\nonumber}
\newcommand{\Tr}{\text{Tr}}
\newcommand{\N}{\mathcal{N}}
\newcommand{\G}{\Gamma}
\newcommand{\vf}{\varphi}
\newcommand{\LL}{\mathcal{L}}
\newcommand{\Op}{\mathcal{O}}
\newcommand{\HH}{\mathcal{H}}
\newcommand{\arctanh}{\text{arctanh}}
\newcommand{\up}{\uparrow}
\newcommand{\down}{\downarrow}
\newcommand{\ket}[1]{\left| #1 \right>}
\newcommand{\bra}[1]{\left< #1 \right|}
\newcommand{\ketbra}[1]{\left|#1\right>\left<#1\right|}
\newcommand{\rd}{\partial}
\newcommand{\de}{\partial}
\newcommand{\ba}{\begin{eqnarray}}
\newcommand{\ea}{\end{eqnarray}}
\newcommand{\db}{\bar{\partial}}
\newcommand{\we}{\wedge}
\newcommand{\ca}{\mathcal}
\newcommand{\lr}{\leftrightarrow}
\newcommand{\f}{\frac}
\newcommand{\s}{\sqrt}
\newcommand{\vp}{\varphi}
\newcommand{\hvp}{\hat{\varphi}}
\newcommand{\tvp}{\tilde{\varphi}}
\newcommand{\tp}{\tilde{\phi}}
\newcommand{\ti}{\tilde}
\newcommand{\ap}{\alpha}
\newcommand{\pr}{\propto}
\newcommand{\mb}{\mathbf}
\newcommand{\ddd}{\cdot\cdot\cdot}
\newcommand{\no}{\nonumber \\}
\newcommand{\la}{\langle}
\newcommand{\lb}{\rangle}
\newcommand{\ep}{\epsilon}
 \def\we{\wedge}
 \def\lr{\leftrightarrow}
 \def\f {\frac}
 \def\ti{\tilde}
 \def\ap{\alpha}
 \def\pr{\propto}
 \def\mb{\mathbf}
 \def\ddd{\cdot\cdot\cdot}
 \def\no{\nonumber \\}
 \def\la{\langle}
 \def\lb{\rangle}
 \def\ep{\epsilon}
 \def\be{\begin{equation}}
\def\ee{\end{equation}}
\def\ba{\begin{eqnarray}}
\def\ea{\end{eqnarray}}
\def\bal#1\eal{\begin{align}#1\end{align}}
\newcommand{\mcl}{\mathcal}
 \def\g{\gamma}
\def\Tr{\text{tr}}

\begin{titlepage}

\thispagestyle{empty}

\begin{flushright}
RIKEN-iTHEMS-Report-21,
YITP-21-***
\end{flushright}
\bigskip

\begin{center}
 \noindent{\large \textbf{
 Information Scrambling Versus Quantum Revival\\ \vspace{5mm} Through the Lens of Operator Entanglement}}\\

\vspace{1cm}
\renewcommand\thefootnote{\mbox{$\fnsymbol{footnote}$}}
Kanato Goto\footnote{kanato.goto@riken.jp}${}^{1}$, Ali Mollabashi\footnote{ali.mollabashi@yukawa.kyoto-u.ac.jp}${}^{2}$, Masahiro Nozaki\footnote{masahiro.nozaki@riken.jp}${}^{1,3}$,\\ Kotaro Tamaoka\footnote{tamaoka.kotaro@nihon-u.ac.jp}${}^{4}$ and Mao Tian Tan\footnote{mt4768@nyu.edu}${}^{5}$\\

\vspace{1cm}

${}^{1}${\small \sl RIKEN Interdisciplinary Theoretical and Mathematical Sciences (iTHEMS), \\Wako, Saitama 351-0198, Japan}\\
${}^{2}${\small \sl Center for Gravitational Physics, 
Yukawa Institute for Theoretical Physics (YITP), \\Kyoto University, Kitashirakawa Oiwakecho, Sakyo-ku, Kyoto 606-8502, Japan}\\
${}^{3}${\small \sl Kavli Institute for Theoretical Sciences and CAS Center for Excellence in Topological Quantum Computation, University of Chinese Academy of Sciences, Beijing, 100190, China}\\
${}^{4}${\small \sl Department of Physics, College of Humanities and Sciences, Nihon University, \\Sakura-josui, Tokyo 156-8550, Japan}\\
${}^{5}${\small \sl Center for Quantum Phenomena, Department of Physics, New York University, 726 Broadway, New York, New York 10003, USA}\\

\if(
\author{Kanato Goto}\email[]{kanato.goto@riken.jp}
\affiliation{\it RIKEN Interdisciplinary Theoretical and Mathematical Sciences (iTHEMS), Wako, Saitama 351-0198, Japan}
\author{Masahiro Nozaki}\email[]{masahiro.nozaki@riken.jp }
\affiliation{\it RIKEN Interdisciplinary Theoretical and Mathematical Sciences (iTHEMS), Wako, Saitama 351-0198, Japan}
\affiliation{\it Kavli Institute for Theoretical Sciences and CAS Center for Excellence in Topological Quantum Computation, University of Chinese Academy of Sciences, Beijing, 100190, China}
\author{Kotaro Tamaoka}\email[]{tamaoka.kotaro@nihon-u.ac.jp}

\affiliation{\it Department of Physics, College of Humanities and Sciences, Nihon University, Sakura-josui, Tokyo 156-8550, Japan}
)\fi

\setcounter{footnote}{0}
\renewcommand\thefootnote{\mbox{\arabic{footnote}}}
\vskip 2em
\end{center}
\begin{abstract}

In this paper, we look for signatures of quantum revivals in two-dimensional conformal field theories (2d CFTs) on a spatially compact manifold by using operator entanglement. It is believed that thermalization does not occur on spatially compact manifolds as the quantum state returns to its initial state which is a phenomenon known as quantum revival. We find that in CFTs such as the free fermion CFT, the operator mutual information exhibits quantum revival in accordance with the relativistic propagation of quasiparticles while in holographic CFTs, the operator mutual information does not
exhibit this revival and the quasiparticle picture breaks down. Furthermore, by computing the tripartite operator mutual information, we find that the information scrambling ability of holographic CFTs can be weakened by the finite size effect.
We propose a modification of an effective model known as the line tension picture to explain the entanglement dynamics due to the strong scrambling effect and find a close relationship between this model and the wormhole (Einstein-Rosen Bridge) in the holographic bulk dual.
\end{abstract}
\end{titlepage} 
\tableofcontents
\section{Introduction and Summary}
\section*{Introduction}
The study of observables defined on compact spaces is of paramount importance as the appearance or lack of periodic behavior that depends on the system size may yield new insights into the study of non-equilibrium phenomenon. 
Consider a density matrix that is expanded in the energy eigenbasis. After enough time has elapsed, physical quantities for certain systems are predominantly determined by the diagonal components in which case the system can be effectively described by a time-independent mixed state which is essentially a classical probability distribution. Furthermore, if this probability distribution is the Boltzmann distribution, the corresponding mixed state is the thermal Gibbs state. On the other hand, if the physical quantities also receive contributions from the off-diagonal components of the density matrix even at sufficiently late times, the system cannot be described by a thermal state. 
If some of the energy intervals are inversely proportional to the system size, the physical quantities can oscillate with a period that depends on the total system size. Thus, the periodic behavior of observables on a compact space indicates that 
the state cannot be approximated by a thermal one. This periodic behavior is known as quantum revival and it shows that the time-evolved state returns to the initial state. It has been shown that time-independent Hamiltonians with discrete spectra posesses energy gaps that can lead to quantum revival \cite{1957PhRv..107..337B,1961JMP.....2..235P,1978PhRvA..18.2379S}. In general, observables in integrable systems oscillate with a period that is determined by the system size \cite{PhysRevLett.44.1323,PhysRevA.42.6308,2014} while non-integrable systems thermalize \cite{Srednicki_1994,Rigol_2008,Ba_uls_2011}.
Exceptions to this simple dichotomy can occur in non-integrable systems that possess subsectors of the spectrum that are protected by symmetry in which case observables can oscillate and perfect thermalization is avoided \cite{Shiraishi_2017,Turner_2018,Pakrouski_2020}. 
This non-equilibrium phenomena is known as quantum many body scars \cite{Serbyn2021,moudgalya2021quantum}. 

A widely studied physical observable that can be used to quantify how different the time-evolved state is from the initial one is the entanglement entropy which measures the amount of bi-partite entanglement between a subsystem and its complement \cite{Calabrese_2005,Calabrese_2007,Coser_2014}. In two-dimensional  conformal field theories ($2$d CFTs) without gravity duals, the time evolution of the entanglement entropy after a global quench can be described by the relativistic propagation of quasiparticles\cite{Calabrese_2005,Calabrese_2007}. If these theories are placed on compact spaces, they can exhibit quantum revivals with periods that are determined by the system size since the quasi-particles that leave the system can re-enter it at a later time \cite{CardY_2014,St_phan_2011}. On the other hands, CFTs with gravity duals (holographic CFTs) defined on non-compact spaces strongly scramble information and so the time evolution of entanglement entropy cannot be described by the relativistic propagation of quasiparticles \cite{Asplund2015, 2014PhRvD..89f6015A}. Instead, the entanglement entropy in these theories are given by Ryu-Takayanagi (RT) surfaces \cite{Ryu_2006,Ryu_20062} (more generally, they are given by quantum extremal surfaces \cite{2007,Faulkner2013,Engelhardt2015}).


 
 
 In this paper, we seek to understand, through the lens of operator entanglement \cite{2016JHEP...02..004H,2017JPhA...50w4001D,Alba_2019,2017PhRvB..95i4206Z}, if the quantum revivals that are characteristic of quasi-particles can occur in 2d holographic CFTs defined on compact spaces
\footnote{The micro-states of black hole on compact spaces have been studied \cite{Maldacena_2003,Cotler_2017,Papadodimas_2015}.}. 
If these revivals are absent, we would like to know the degree to which information scrambling has been diminished by the finite extent of the system.
 
In the remainder of this introduction, key technical terminology that is used in the paper is introduced before the main results are summarized. Readers who are familiar with operator entanglement are welcome to skip straight to the summary of the results.
 

\subsection*{Operator entanglement}
\begin{figure}[htbp]
 \begin{center}
  \includegraphics[width=100mm]{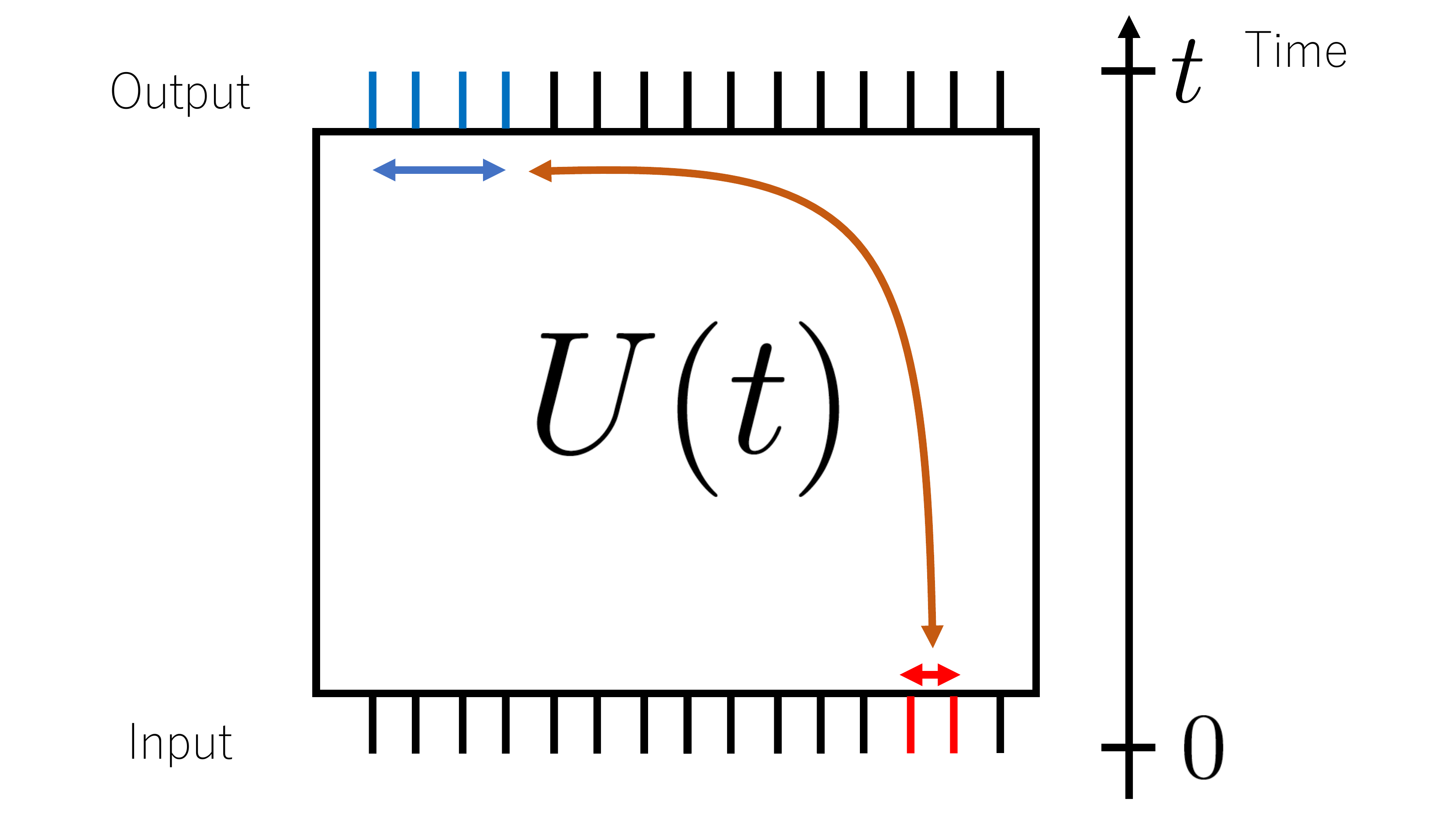}
 \end{center}
 \caption{A cartoon of the time evolution operator. \label{CBS}}
 \label{fig:one}
\end{figure}

The operator entanglement of a given operator is defined by the entanglement structure of the state dual to the operator. 
For simplicity, let us consider a time evolution operator $U(t)$ with a time-independent Hamiltonian $H$. 
By a channel-state map, the dual state to $U(t)$ is defined by
\cite{Hosur_2016, Zhou_2017,Nie_2019,Kudler_Flam_2020,PhysRevResearch.3.033182,maccormack2020operator}
\be\label{UnregulatedOperatorState}
U(t)= \sum_{a} e^{-it E_a} \ket{a}\bra{a}  \rightarrow \ket{U(t)} = \sum_{a}  e^{-it E_a} \ket{a}_{\text{out}} \ket{a^*}_{\text{in}},
\ee
where $\ket{a}$ is an eigenstate of the Hamiltonian, and $\ket{\cdot^*}$ is CPT conjugate to  $\ket{\cdot}$. 
Since the entanglement structure of $\ket{U(t)}$ depends on $e^{-it E_a}$, information about the time evolution operator is encoded in the dual state. 
This dual state is defined on a doubled Hilbert space $\mathcal{H}= \mathcal{H}_{\text{in}}\otimes \mathcal{H}_{\text{out}}$, and we refer to $\mathcal{H}_{\text{in}}$ and $\mathcal{H}_{\text{out}}$ as the input and output Hilbert space respectively.
We interpret the correlation between the input and output subsystems as the correlation between subsystems at different time slices of the time evolution operator as in Figure \ref{CBS}. 
In this paper, we compute the time evolution of the bipartite operator mutual information (BOMI) and the tripartite operator mutual information (TOMI) in order to study the correlations between the subsystems of the input and output Hilbert spaces.

BOMI and TOMI are defined by linear combinations of the operator entanglement entropy (OEE) which we define below. 
The symbol $\rho$ denotes the density operator of $\ket{U(t)}$.
We divide the Hilbert space $\mathcal{H}$ into a subsystem and its complement, $A$ and $\overline{A}$, and trace out the degrees of freedom in $\overline{A}$ of the density operator. 
Then, OEE is defined by the von Neumann entropy of this reduced density matrix $\rho_A$:
\be
S_A = -\Tr_A\left(\rho_A \log{\rho_A}\right).
\ee

Dividing the the Hilbert space $\mathcal{H}$ into $A$, $B$ and the  complement of their union $A\cup B$, the BOMI between $A$ and $B$, $I(A:B)$, is defined by
\be
I(A:B) =S_A+S_B-S_{A\cup B}.
\ee
If $A$ is taken to be a subsystem of $\mathcal{H}_{\text{in}}$, and $B$ is a subsystem of $\mathcal{H}_{\text{out}}$, we interpret $I(A:B)$ as the correlation between the subsystems before and after the time evolution. 
In this interpretation, $I(A:B)$ measures how much information is sent from $A$ to $B$.
The TOMI is defined by a linear combination of the BOMI. 
Let us divide the Hilbert space into $A$, $B_1$, $B_2$, and the space complement to $A\cup B_1\cup B_2$.
Then, the TOMI is defined by
\be \label{TOMI}
I(A:B_1:B_2)=I(A:B_1)+I(A:B_2)-I(A:B_1\cup B_2).
\ee
If $I(A:B_1\cup B_2)$, the BOMI which measures the global correlation, is larger than $I(A:B_1)+I(A:B_2)$, the sum of the BOMI which measures the local correlation, then the TOMI is negative. 
The negativity of the TOMI means that a piece of information about $A$ is locally hidden thanks to the scrambling effect of dynamics. 
In the holographic CFT, tripartite state-entanglement mutual information has to be non-positive \cite{Hayden_2013}\footnote{For non-positive tripartite information in infinite-range spin systems see \cite{Pappalardi:2018frz}.}. 
We will explain the details of operator entanglement in Section 2.
\subsection*{Quasi-particle picture of the operator entanglement}
\begin{figure}[htbp]
 \begin{center}
  \includegraphics[width=100mm]{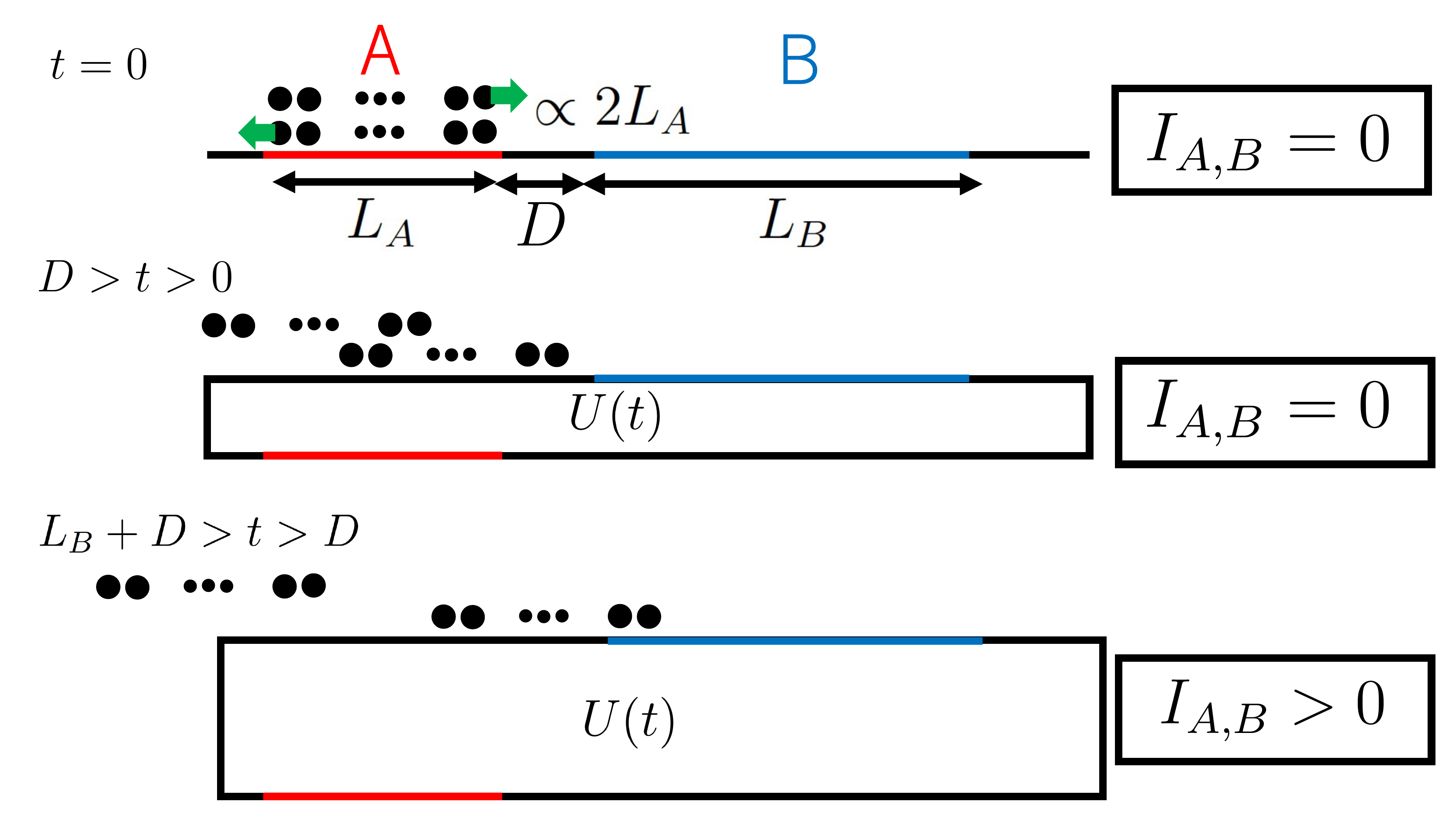}
 \end{center}
 \caption{
 A cartoon of quasiparticle picture. In this picture, the black dots are quasiparticles. \label{QP_Simple}}
 \label{fig:one}
\end{figure}

Here, let us explain the quasiparticle picture that describes the time evolution of BOMI and TOMI of the free fermion time evolution operator. 
This picture explain the time evolution of BOMI and TOMI in terms of the time evolution operator, not the dual state. 
In Figure \ref{QP_Simple}, $A$ denotes a subsystem before the time evolution and $B$ denotes a subsystem after the time evolution. 
At $t=0$, $A$ and $B$ are on the same time slice. 
In subsystem $A$, which has length $L_A$, the number of quasiparticles is proportional to $2L_A$\footnote{Since the quasiparticle picture does not predict the magnitude of the BOMI, here we assume that the quasiparticle density is one particle per unit length.
}.
For $t>0$, the time slice on which $B$ lives is different from the slice on which $A$ lives. The quasiparticles are on the time slice where $B$ lives and $L_A$ quasiparticles move to the right at the speed of light while the rest of quasiparticles moves to the left at the speed of light.
The value of $I(A:B)$ is proportional to the number of quasiparticles in $B$. The time evolution of BOMI, following this picture, exhibits quantum revival. 
We will explain the quasiparticle picture in greater detail in Section $3$.

\subsection*{Measure of information scrambling}
Let us define the amount of information scrambled to be the absolute value of TOMI. 
If we divide the output Hilbert space into $B_1$ and $B_2$, and the value of $I(A:B_{i=1,2})$ is zero, then the observers who are able to access only $B_{i=1,2}$ are not able to recover the information about $A$. The value of $I(A:B_1 \cup B_2)$ is the amount of information about $A$ which the observers, being able to access any regions of the output Hilbert space, are able to obtain.  
The absolute value of (\ref{TOMI}) is the amount of information locally hidden by the scrambling effect of dynamics. This is the definition of the amount of scrambled information.
We will explain the details of the scrambled information in Section $3$.

\subsection*{Line-tension picture of the operator entanglement}
\begin{figure}[h]
\begin{center}
\includegraphics[scale=0.3]{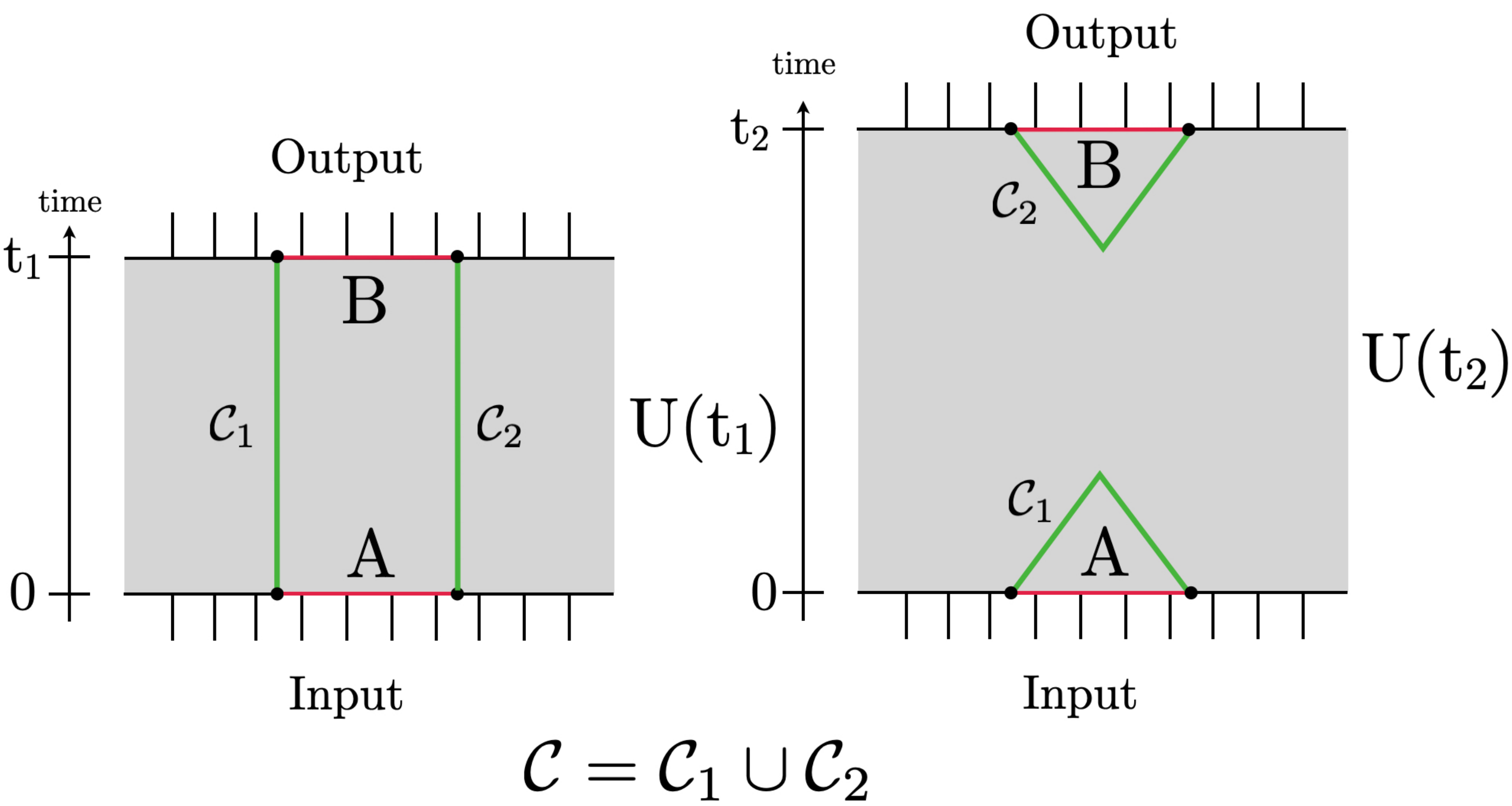}
\caption{A cartoon of line-tension picture. Integral of the line-tension ${\cal T}(v)$ over the minimal curve ${\cal C}$ gives the entanglement.}
\label{LT2_2}
\end{center}
\end{figure}
As we explained above,  in an  integrable system, the behavior of the BOMI and TOMI can be described by quasiparticles. On the other hand, in a chaotic system, the hydrodynamic behavior of the entanglement  is well described by the ``line-tension picture,''  introduced in \cite{2018arXiv180300089J,2018arXiv180409737Z,2018PhRvX...8b1014N,2017PhRvX...7c1016N}. The original line-tension picture is derived in the (two-dimensional) chaotic system defined on an infinite line, but in this paper we will generalize it to a compact space.  In the line-tension picture, the entanglement between the region $A$ at $t=0$ and $B$ at $t=t_1$  is given by the integral of the line-tension ${\cal T}(v)$ over the minimal curve ${\cal C}$ that connects the edges of $A$ and $B$ (see Figure \ref{LT2_2}). This reproduces the phase transition of the operator entanglement (or BOMI) between the connected phase (left figure in Figure \ref{LT2_2}) and the disconnected phase (right figure in Figure \ref{LT2_2}) in holographic CFTs, where the corresponding entanglement entropy (or the mutual information) is computed using the Ryu-Takayanagi surface in the double-sided black hole dual to the thermofield double state \cite{Maldacena_2003, HM}.  At late times, the disconnected phase is favored due to the minimality condition, and BOMI becomes zero.

When the holographic CFT is put on a compact space, as we will see in this paper, the Ryu-Takayanagi surface with non-trivial homology  gives an additional phase at late times which does not appear in non-compact spaces and it keeps BOMI at some positive value, contrary to the non-compact case where BOMI decays to zero.  As we will see Section \ref{LineTensionPictureSection}, this can be reproduced by the line-tension picture on the compact space.

\section*{Summary }
Here, we summarize the results of this paper.
\begin{itemize}
\item[1.] The time evolution of BOMI and TOMI of the time evolution operator in $2$d free fermion follows the relativistic propagation of quasiparticles: 
The time evolution of BOMI periodically behaves, and the period is given by the system size. 
Since the number of quasiparticles is conserved, the value of TOMI is zero. We confirm that quantum revival, following the relativistic propagation of quasi-particles, occurs in free fermion.
In holographic CFT, the late-time evolution of BOMI does not follow the relativistic propagation of quasiparticle, so that quantum revival does not occur.

\item[2.] 
Following the proposed quasi-particle picture for operator entanglement in \cite{Nie_2019}, we have adapted the method of \cite{Alba:2016, Alba:2017lvc} to find exact formulae for BOMI in integrable models with generic configurations. 

\item[3.] When we divide $\mathcal{H}_{\text{out}}$ into $B_1$ and $B_2$, the TOMI  which defined by (\ref{TOMI}), of the free fermionic time evolution operator vanishes since the number of quasiparticles is conserved. Thus, the finiteness of compact space does not affect the value of TOMI. In the holographic CFT, the absolute value of TOMI can become smaller than the non-compact case because the compactness of spacetime prevents dynamics from spreading and delocalizing the information about $A$. 
For example, in the picture, in terms of the time evolution operator, not the dual state, if $A$ is initially included in $B_1$, and the sum of sizes of $A$ and $B_1$ is smaller than the system size, then the late-time value of TOMI is independent of the system size. On the other hand, if the sum of sizes of $A$ and $B_1$ is larger than the system size, the late-time value of TOMI depends on the system size, so that the absolute value of TOMI in the compact spacetime is smaller than that of TOMI in the non-compact spacetime.
This is consistent with the time evolution of BOMI in the spin system with the strong scrambling ability  \cite{mascot2021local}.

\item[4.] When a chaotic system is defined on a compact space, we have a new phase of the entanglement entropy at late times, which does not appear when the system is defined on a non-compact space, and it prevents BOMI from decaying to zero. For a holographic CFT, this  phase is allowed by the the Ryu-Takayanagi surface with non-trivial homology, that holographically computes the entanglement entropy in a compact space. We will propose the line-tension picture for a chaotic system defined on a compact space that effectively describe the hydrodynamic behavior of the entanglement in the scaling limit. Our definition of the line-tension picture correctly captures the new phase of BOMI (or TOMI).

\item[5.]  We found that the line-tension picture, which captures the coarse-grained behavior of the operator entanglement, can be generalized to a CFT defined on a compact space. The new phase that appear in a chaotic system on a compact space can be captured by the homotopically non-trivial minimal curve that computes the line-tension, which wraps around a compact direction. In a holographic CFT, the operator entanglement can be holographically computed by the Ryu-Takayanagi surface on the wormhole geometry in the dual state given by the a channel-state map from the operator. We found that the homological equivalence between the minimal curve in the line-tension picture and the Ryu-Takayangi surface on the dual wormhole geometry.  
This suggests that the quantum circuit in the line tension picture and the wormhole in the gravity picture are closely related.
\end{itemize}

The articles \cite{Kuns:2014zka,Ugajin:2013xxa,Mandal:2016cdw} are preliminary studies of quantum revival in holographic CFT \footnote{For related studies see also \cite{Abajo-Arrastia:2014fma, daSilva:2014zva}.}. 
In these preliminary studies, the authors found that quantum revival occurs in holographic CFTs defined on strips by studying the time evolution of state entanglement instead of operator entanglement.
This can be taken to be an indication that the information about the initial state can be recovered by weakening the scrambling effect of the dynamics due to the finite size effect.
\section*{Organization of this paper}

So far, we have explained the background of this study and summarized the results obtained.
In Section 2, we explain how to compute operator mutual information in path-integral formalism.
In Section 3, we study the time evolution of operator mutual information in $2$-dimensional free field theory and holographic CFT.
In Section 4, we explain the line tension picture, which is an effective theory to explain the time evolution of operator mutual information in holographic CFT.
In Section 5, we discuss the results obtained in this paper, and then explain some of the future directions.

\section{Operator mutual information in path-integral}
Let us begin by recapitulating the computation of OEE in $2$d CFTs using the path integral formalism as described in \cite{Nie_2019}. The unitary operator state \eqref{UnregulatedOperatorState} can be regulated by introducing a UV cutoff $\epsilon$ as\footnote{It is understood that the second Hilbert space is CPT conjugated.}
\begin{equation}
    |U_\epsilon (t)\rangle = \mathcal{N} e^{-\left(\frac{i t+\epsilon}{2}\right)H_{\text{tot}}}\sum_a |a\rangle_{1} |a\rangle_{2}
\end{equation}
where $H=H_1\otimes 1+1\otimes H_2$ is the total Hamiltonian acting on the doubled Hilbert space $\mathcal{H}$ and $\mathcal{N}$ is the normalization factor
\begin{equation}
    \mathcal{N}^{-2} = \text{tr}_{\mathcal{H}_1}e^{-2\epsilon H_1}.
\end{equation}
The corresponding density matrix in Euclidean signature is
\begin{equation}
    \rho^E = \mathcal{N}^2 e^{\frac{\tau_1}{2}H_{\text{tot}}}\sum_{a,b}|a\rangle\langle b|_1 \otimes |a\rangle\langle b|_2 \, e^{-\frac{\tau_2}{2} H_{\text{tot}}  }
\end{equation}
where the analytic continuation 
\begin{equation}
    \tau_1 \rightarrow -\epsilon -it,\qquad
    \tau_2 \rightarrow \epsilon-it,
\end{equation}
must be performed at the end of the calculation. The matrix elements of the Euclidean density matrix can be written as
\begin{equation}
    {}_{1}\langle\psi_1|
    {}_{2}\langle\psi_2|
    \rho^E |\psi_2'\rangle_{2}|\psi_1'\rangle_{1}= \mathcal{N}^2 \langle\psi_2'^*|e^{-\tau_2 H} |\psi_1'\rangle \langle\psi_1|e^{\tau_1 H}|\psi_2^*\rangle
\end{equation}
where the conjugated fields are defined by
\begin{equation}
    {}_{2}\langle\psi_2|
    e^{\frac{\tau_1 H_2}{2}}|a\rangle_{2} =
    {}_{2}\langle a|
    e^{\frac{\tau_1 H_2}{2}}|\psi_2^*\rangle_{2},\qquad
    {}_{2}\langle b|e^{-\frac{\tau_2 H_2}{2}}
    |\psi_2'\rangle_2 = {}_{2}\langle \psi_2'^*|e^{-\frac{\tau_2 H_2}{2}}
    |b\rangle_2.
\end{equation}
Let $A$ and $B$ be subsystems of $\mathcal{H}_1$ and $\mathcal{H}_2$ respectively. The reduced density matrix is obtained by integrating over the degrees of freedom that lie on the complement of $A\cup B$ and has the matrix elements
\begin{align}
    &\langle \psi_{1,A\cup B}| \langle \psi_{2,A\cup B}| \rho_{A\cup B} |\psi_{2,A\cup B}'\rangle|\psi_{1,A\cup B}'\rangle \\ \nonumber
    =& \mathcal{N}^2 \int \mathcal{D} \psi_{1,\overline{A\cup B}}\mathcal{D} \psi_{2,\overline{A\cup B}}^* \langle \psi'^*_{2,A\cup B},\psi^*_{2,\overline{A\cup B}}| e^{-\tau_2 H}|\psi'_{1,A\cup B}, \psi_{1,\overline{A\cup B}}\rangle\langle\psi_{1,A\cup B},\psi_{1,\overline{A\cup B}}|e^{\tau_1 H}|\psi_{2,A\cup B}^*,\psi_{2,\overline{A\cup B}}^*\rangle
\end{align}

\begin{figure}
    \centering
    \includegraphics[trim={0 8cm 0 8cm},clip,width=\textwidth]{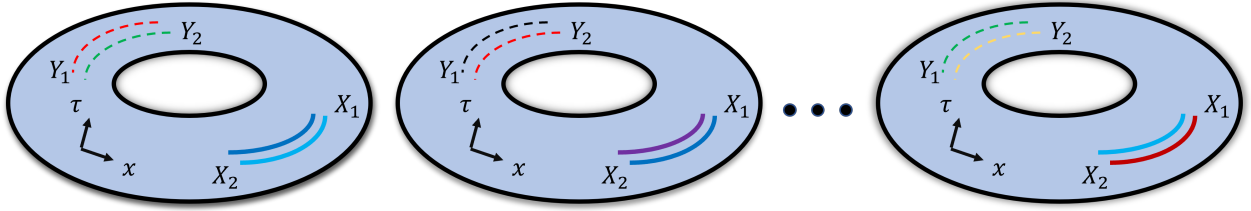}
    \caption{The n\textsuperscript{th} unitary operator R\'{e}nyi  entropy for a field theory defined on a circle $S^1$ is given by the n\textsuperscript{th} moment of the reduced density matrix which is equivalent to a path integral on n copies of a torus with two periods $2\pi R$ and $\beta = 2\epsilon$. Each torus has two cuts corresponding to subsystems $A$ and $B$ as shown in the diagram with the dashed curves corresponding to the latter and the full curves corresponding to the former. The cuts that are identified are shown in the same color.}
    \label{UnitaryOPEETorus}
\end{figure}

This is a path integral defined on a cylinder with inverse temperature $\beta = 2\epsilon$. The trace of the n\textsuperscript{th} moment is a path integral over a n-sheeted Riemann surface as shown in Figure \ref{UnitaryOPEETorus}. Such a path integral can in turn be written as a correlation function of twist operators in the cyclic orbifolded theory $\text{CFT}^n/\mathbb{Z}^n$ on the original unreplicated spacetime \cite{Calabrese_2004,Calabrese_2009} as 
\begin{equation}
    \text{tr}_{A\cup B} (\rho_{A\cup B})^n = C_0 \langle\sigma_n(X_1)\bar{\sigma}_n(X_2)\sigma_n(Y_2,\tau_1)\bar{\sigma}_n(Y_1,\tau_1) \rangle_{T^2}
\end{equation}
where $T^2$ is a torus with a spatial circumference $2\pi R$ and an inverse temperature $\beta=2\epsilon$. Here $\sigma_n$ and $\bar{\sigma}_n$ are twist and anti-twist operators whose scaling dimensions are given by $\Delta_n=\frac{c}{12}\left(n-\frac{1}{n}\right)$. The proportionality constant $C_0$ can be fixed by requiring the BOMI between two non-overlapping intervals to vanish at $t=0$ in the $\epsilon\rightarrow 0$ limit. Note also that the order of the twist and anti-twist operators in the second Hilbert space have been reversed relative to the first Hilbert space. The n\textsuperscript{th} operator R\'{e}nyi entropy is thus given by
\begin{align}
    S_{A\cup B}^{(n)} =& \frac{1}{1-n}\log\left[C_0\langle\sigma_n(X_1)\bar{\sigma}_n(X_2)\sigma_n(Y_2,\tau_1)\bar{\sigma}_n(Y_1,\tau_1) \rangle_{T^2} \right], \\ \nonumber
    S_{A\cup B} =& \lim_{n\rightarrow1}S_{A\cup B}^{(n)}.
\end{align}
Similarly, the single interval R\'{e}nyi entanglement entropies are 
\begin{align}
    S_A^{(n)} =& \frac{1}{1-n}\log\left[C_0'\langle\sigma_n(X_1)\bar{\sigma}_n(X_2) \rangle_{T^2} \right], \\ \nonumber
    S_B^{(n)} =&\frac{1}{1-n} \log\left[C_0'\langle\sigma_n(Y_2,\tau_1)\bar{\sigma}_n(Y_1,\tau_1) \rangle_{T^2} \right]
\end{align}
where the proportionality constant $C_0'$ can be fixed by requiring the entanglement entropies to match up with the usual thermal entropy when the total system size is taken to be infinite.

\section{The time evolution of BOMI and TOMI in various field theories}
In this section, we study the time evolution of BOMI and TOMI in two-dimensional free field theories and holographic CFT.
In particular, the time evolution of the BOMI and TOMI in free field theories is studied using analytical and numerical methods.
In addition, we propose a quasiparticle picture that describes the time evolution of BOMI and TOMI in free field theories.

\subsection{Quasiparticle description of operator entanglement dynamics}
While the entanglement entropy of a generic quantum system is difficult to compute, it has been shown that the evolution of entanglement entropy in integrable models are well-described by localized excitations known as quasiparticles \cite{Calabrese_2005,2007JSMTE..10....4C,Calabrese_2009,Coser_2014,2017PNAS..114.7947A,2018ScPP....4...17A,10.21468/SciPostPhysLectNotes.20}. This quasiparticle description of entanglement entropy can also be used to describe the evolution of BOMI in free theories like the free fermion and the compact boson \cite{Nie_2019}. 

\begin{figure}
    \centering
    \includegraphics[width=0.4\textwidth]{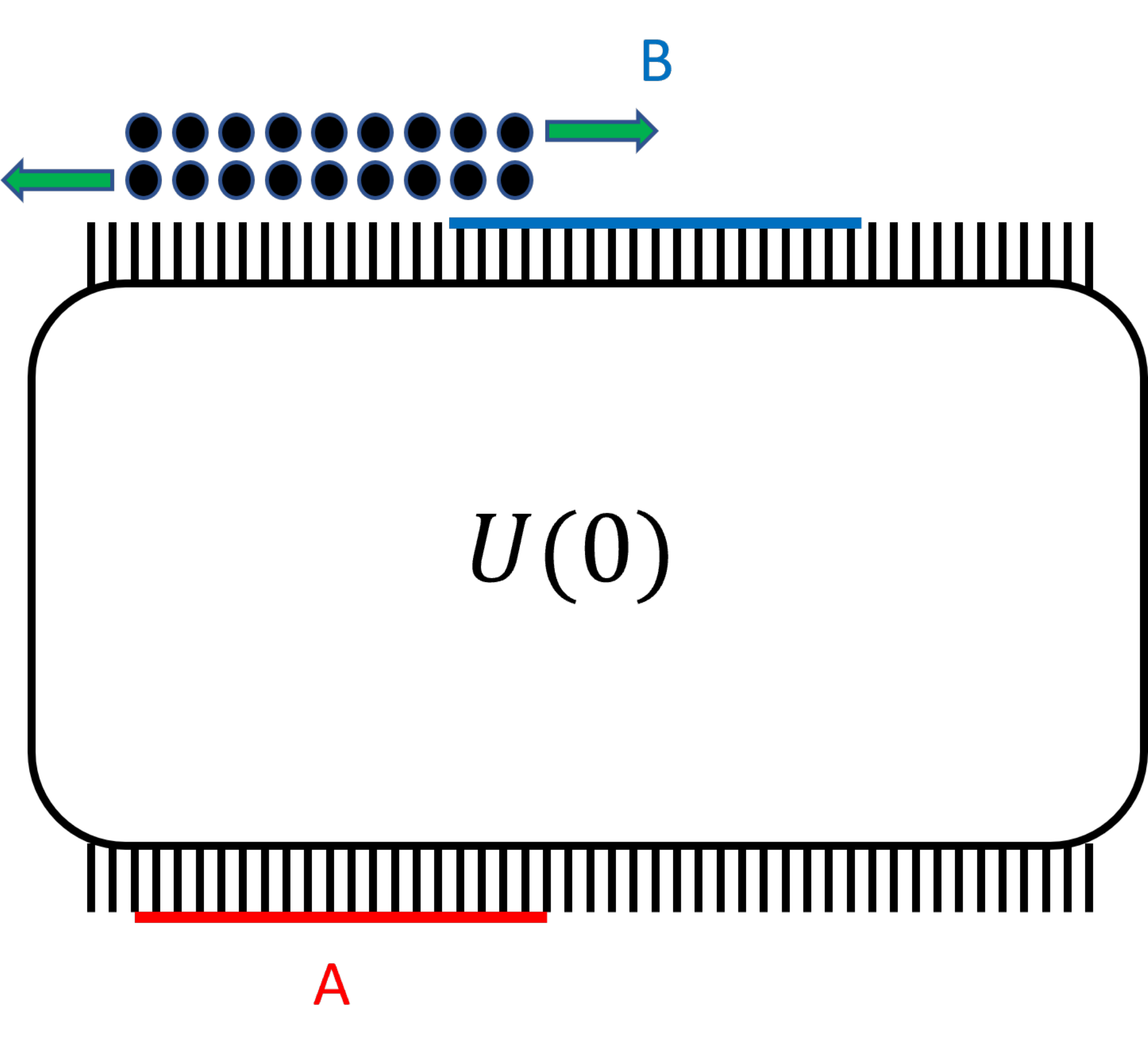}
    \hspace{1cm}
    \includegraphics[width=0.4\textwidth]{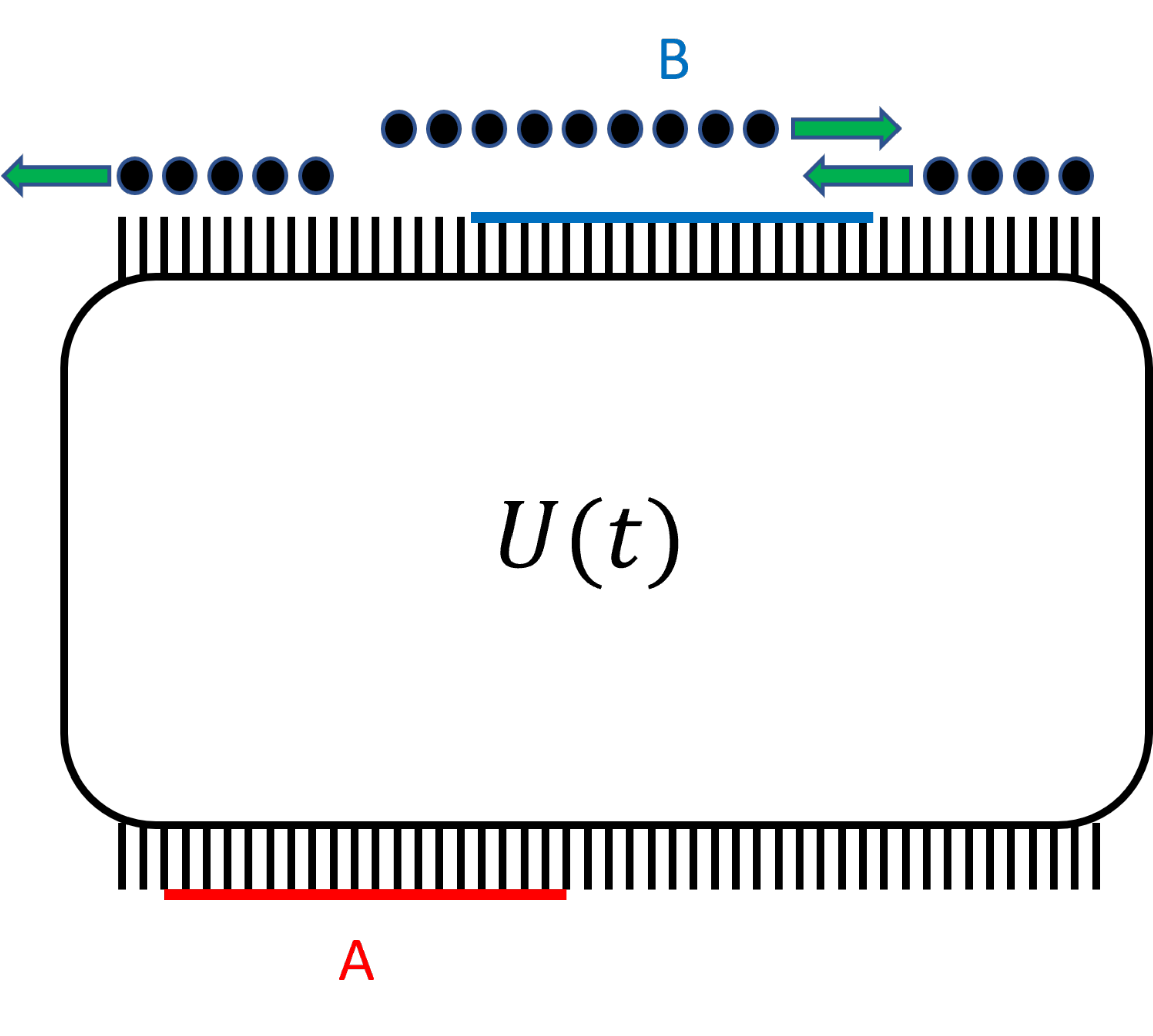}
    \caption{Illustration of the quasiparticle picture for the time evolution of OEE. The intervals $A$ and $B$ are subsystems of the input and output Hilbert spaces respectively. Since the spatial direction is periodic, the left edge of the unitary operator must be identified with the right. (\textbf{Left}): Initially, all quasiparticles, denoted by the black dots, are spatially located in subsystem $A$. The top row of quasiparticles are right moving while the bottom row consists of left moving quasiparticles. (\textbf{Right}): The two sets of quasiparticles move independently. At any given instant in time, the BOMI is proportional to the number of quasiparticles in subsystem $B$.}
    \label{OperatorEntanglementQuasiparticlePicture}
\end{figure}

Consider two subsystems $A$ and $B$ of the input and output Hilbert spaces respectively. Initially, all quasiparticles are uniformly distributed in subsystem $A$ as shown in Figure \ref{OperatorEntanglementQuasiparticlePicture}. Half of the quasiparticles are left moving while the other half are right moving. As time progresses, these two sets of quasiparticles will move independently at the speed of light. At any instant in time, the BOMI between the pair of subsystems $A$ and $B$ is proportional to the number of quasiparticles that are contained in subsystem $B$. These quasiparticles should be thought of as localized quanta of information. An immediate corrollary of this is that in theories where the OEE is well-described by the quasiparticle picture, there is little to no information scrambling since information is propagating in localized packets \cite{Nie_2019}.

The quasiparticle picture predicts two interesting properties of BOMI for integrable CFTs defined on one compact spatial dimension. Firstly, since the spatial manifold is a circle, the quasiparticles will simply go around the circle over and over again. Therefore, the BOMI will display periodic perfect revivals\footnote{In integrable systems that are not CFTs, the quasiparticles will have a range of velocities. As time progresses, these quasiparticles will disperse so the revivals will not be perfect.}. Secondly, if the total length of subsystems $A$ and $B$ is greater than the length of the whole system, there will always be some quasiparticles contained in subsystem $B$ and BOMI can never vanish.

\subsection{Free fermion: analytical results}\label{DiracFermionAnalyticalResultSection}
The OEE of a free massless Dirac fermion living on a torus can be computed by extending the bosonization approach in \cite{Herzog2013,Casini_2005} to the doubled Hilbert space. As explained in the previous section, the reduced density matrix of the operator state with regulator $\epsilon$ is defined on a manifold where the Euclidean time has a period of $\beta=2\epsilon$. If space is taken to be a circle $S^1$ with circumference $2\pi R$, then spacetime manifold is a torus with a holomorphic coordinate given by
\begin{equation}
    w = x+i\tau_E
\end{equation}
and $x \sim x+2\pi R$ and $\tau_E \sim \tau_E + 2\epsilon$. Before analytic continuation to Lorentzian signature, the anti-holomorphic coordinate is simply the complex conjugate of this. Rescaling the torus coordinates by $\frac{1}{2\pi R}$ results in the periodicities
\begin{equation}\label{RescaledCoordinates}
    \frac{w}{2\pi R} \sim \frac{w}{2\pi R}+1,\qquad
     \frac{w}{2\pi R} \sim \frac{w}{2\pi R} +\tau
\end{equation}
where the modular parameters are $(\tau,\bar{\tau})=\left(\frac{i\epsilon}{\pi R},-\frac{i\epsilon}{\pi R}\right)$. 

Along each of the two cycles of the torus, one can impose either periodic or anti-periodic boundary on the fermions. These boundary conditions are also known as the Ramond (R) and the Neveu-Schwarz (NS) boundary conditions respectively. There are a total of four possible combinations of boundary conditions as shown in table \ref{FermionSpinStructure}, although the partition function for the first spin structure vanishes due to the zero-mode \cite{DiFrancesco:639405}. 
\begin{table}[]
    \centering
    \begin{tabular}{|c|c|}
    \hline
         $\nu$& sector  \\
         \hline
         1&(R,R)  \\
    2&(R,NS)   \\
    3&(NS,NS)  \\
    4&(NS,R)  \\
    \hline
    \end{tabular}
    \caption{Spin structures of the fermion on a torus.}
    \label{FermionSpinStructure}
\end{table}

In our coordinate system \eqref{RescaledCoordinates}, the first cycle corresponds to the spatial direction while the second cycle corresponds to the Euclidean time direction. Therefore, the second and third spin-structures correspond to imposing the physical anti-periodic boundary conditions along the thermal circle while the fourth spin-structure corresponds to imposing periodic boundary conditions along the thermal circle.

As explained in the previous section, the computation of OEE boils down to the computation of correlation functions of twist-operators \cite{Calabrese_2004,Calabrese_2009} or equivalently, the partition function defined on the $n$-sheeted Riemann surface where $n$ corresponds to the R\'{e}nyi index. The entanglement cuts that are relevant to the computation of OEE consists of a single interval in the input Hilbert space and a single interval in the output Hilbert space, as well as their union. The coordinates of the boundaries of the intervals on the torus are given by $w_{X_1}=X_1,w_{X_2}=X_2,w_{Y_1}=Y_1+i\tau_1,$ and $w_{Y_2}=Y_2+i\tau_1$. After performing the analytic continuation $\tau_1\rightarrow -\epsilon-it$, the coordinates become
\begin{align}
    w_{X_1} &= X_1, \qquad w_{X_2} = X_2, \\ \nonumber
    w_{Y_1} &= Y_1 + t-i\epsilon, \qquad w_{Y_2} = Y_2 + t-i\epsilon, \\ \nonumber
    \bar{w}_{Y_1} &= Y_1 - t+i\epsilon, \qquad \bar{w}_{Y_2} = Y_2 - t+i\epsilon.
\end{align}
The OEE for a subsystem consisting of $p$ intervals $(u_a,v_a)$ for $a = 1\ldots p$ is obtained from the correlation function of $2p$ twist operators located at the corresponding coordinates $w_{u_a}$, $w_{v_a}$ and their anti-holomorphic counterparts
\begin{align}\label{TwistOperatorLogarithm}
    &\log \langle\sigma_n(w_{u_1},\bar{w}_{u_1}) \bar{\sigma}_n(w_{v_1},\bar{w}_{v_1})\ldots
    \sigma_n(w_{u_p},\bar{w}_{u_p}) \bar{\sigma}_n(w_{v_p},\bar{w}_{v_p}) \\ \nonumber =& \frac{p}{6}\frac{1-n^2}{n}\log (2\pi R)+\frac{n^2-1}{12n}\log \left|\frac{\prod\limits_{a<b}\theta_1\left(\frac{w_{u_a}-w_{u_b}}{2\pi R}\bigg|\tau\right)\theta_1\left(\frac{w_{v_a}-w_{v_b}}{2\pi R}\bigg|\tau\right)(\varepsilon\partial_z\theta_1(0|\tau))^p}{\prod\limits_{a,b}\left(\frac{w_{u_a}-w_{v_b}}{2\pi R}\bigg|\tau\right)}\right|^2 \\ \nonumber
    +& \sum_{k=-\frac{n-1}{2}}^{\frac{n-1}{2}}\log \left|\frac{\theta_\nu\left(\frac{k}{n}\sum_a\frac{w_{u_a}-w_{v_a}}{2\pi R}\bigg|\tau\right)}{\theta_\nu(0|\tau)}\right|^2
\end{align}
The first term comes from rescaling the coordinates as in \eqref{RescaledCoordinates} so as to have a torus with periods $1$ and $\tau$. The second and third term correspond to the partition function on the n-sheeted Riemann surface which was computed using bosonization \cite{Herzog2013,Casini_2005}. A UV-cutoff $\varepsilon$ was also introduced to regulate coincident points in the correlation function. In the decompactification limit $R\rightarrow \infty$, \eqref{TwistOperatorLogarithm} reproduces the OEE for conformal field theories defined on the real line \cite{Nie_2019} when the physical boundary conditions $\nu=2,3$ are imposed.

The OEE involving two intervals $A = [X_2,X_1]$ and $B = [Y_2,Y_1]$ residing in the first and second Hilbert space respectively can be obtained by setting 
\begin{align}
    w_{u_1} = w_{X_2}, \qquad w_{v_1} = w_{X_1}, \\ \nonumber
    w_{u_2} = w_{Y_1}, \qquad w_{v_2} = w_{Y_2},
\end{align}
with similar expressions for the anti-holomorphic components. Note that the position of the twist and anti-twist operators are exchanged for the subsystem of the second Hilbert space as explained in the previous section. These correlation functions can finally be combined to give the BOMI
\begin{align}\label{BOMI}
    I^{(n)}(A:B) =& \frac{n+1}{12n}\log \left|\frac{\theta_1\left(\frac{w_{X_2}-w_{Y_1}}{2\pi R}|\tau\right)\theta_1\left(\frac{w_{X_1}-w_{Y_2}}{2\pi R}|\tau\right)}{\theta_1\left(\frac{w_{X_2}-w_{Y_2}}{2\pi R}|\tau\right)\theta_1\left(\frac{w_{Y_1}-w_{X_1}}{2\pi R}|\tau\right)}\right|^2 \\ \nonumber +&\frac{1}{1-n}\sum_{k=-\frac{n-1}{2}}^{\frac{n-1}{2}}\log\left|\frac{\theta_\nu\left(\frac{k}{n}\frac{w_{X_2}-w_{X_1}}{2\pi R}|\tau\right)\theta_\nu\left(\frac{k}{n}\frac{w_{Y_1}-w_{Y_2}}{2\pi R}|\tau\right)}{\theta_\nu(0|\tau)\theta_\nu\left(\frac{k}{n}\frac{w_{X_2}-w_{X_1}+w_{Y_1}-w_{Y_2}}{2\pi R}|\tau\right)}\right|^2.
\end{align}
Taking the decompactification limit $R\rightarrow\infty$ for the boundary conditions $\nu=2,3$ yields the BOMI for the $c=1$ Dirac fermion defined on the real line \cite{Nie_2019}.

\subsubsection{Bipartite Operator Mutual Information}
In this subsection, plots of the BOMI \eqref{BOMI} for the $c=1$ free Dirac fermion defined on the circle $S^1$ are shown for various choices of the total system size as well as various choices of subsystems $A$ and $B$. Only the results for the physical boundary conditions $\nu=2,3$ with anti-periodic boundary conditions are shown here. The corresponding plots for the boundary condition $\nu=4$ with NS boundary conditions imposed along the thermal cycle are relegated to the appendix \ref{app:4NS}.

\subsubsection{Symmetric Intervals}
\begin{figure}
    \centering
    \textbf{Symmetric Intervals}\par\medskip
    \includegraphics[trim={4cm 0 28cm 0},clip,valign=t,width=0.32\textwidth]{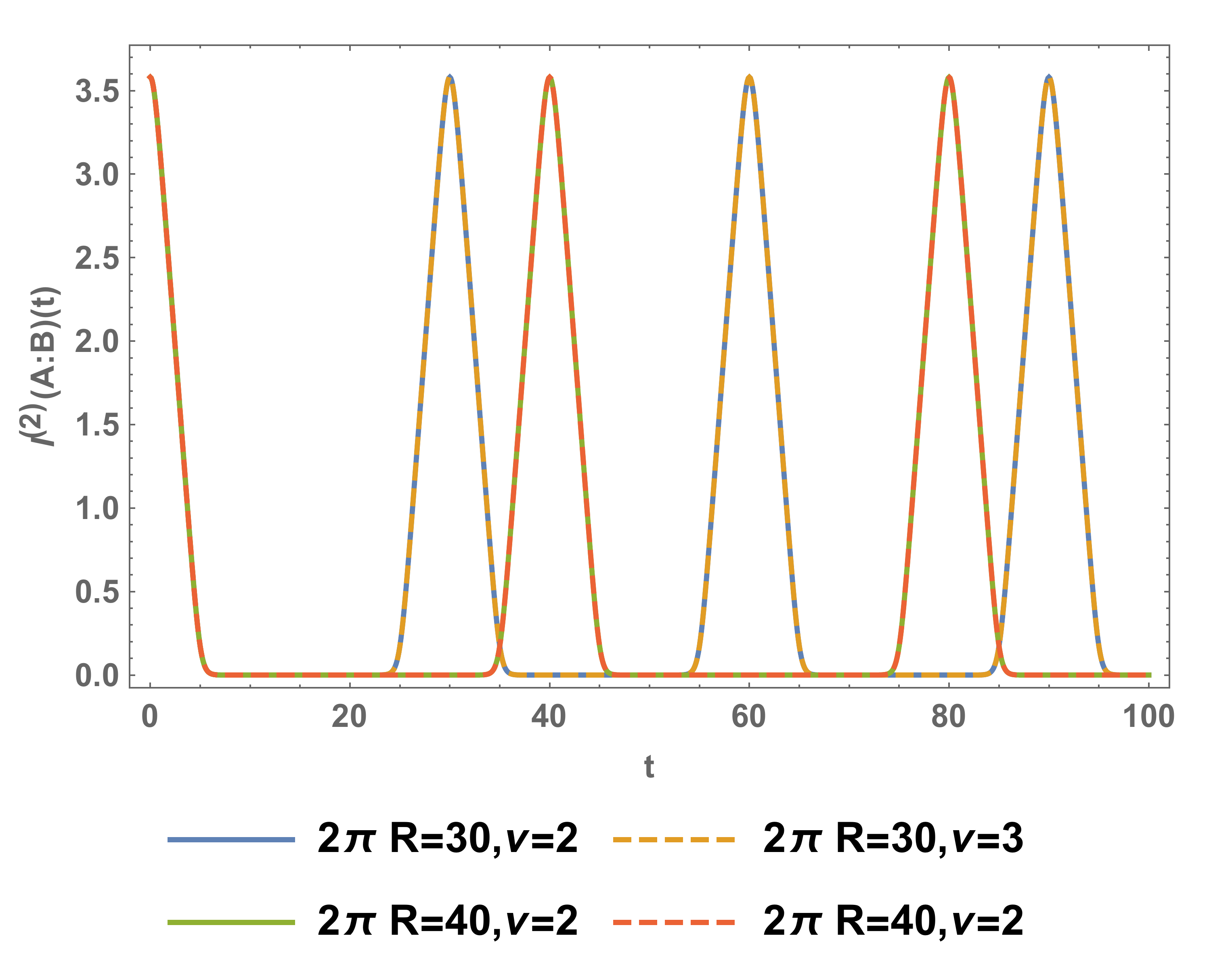}
    \includegraphics[trim={4cm 0 28cm 0},clip,valign=t,width=0.32\textwidth]{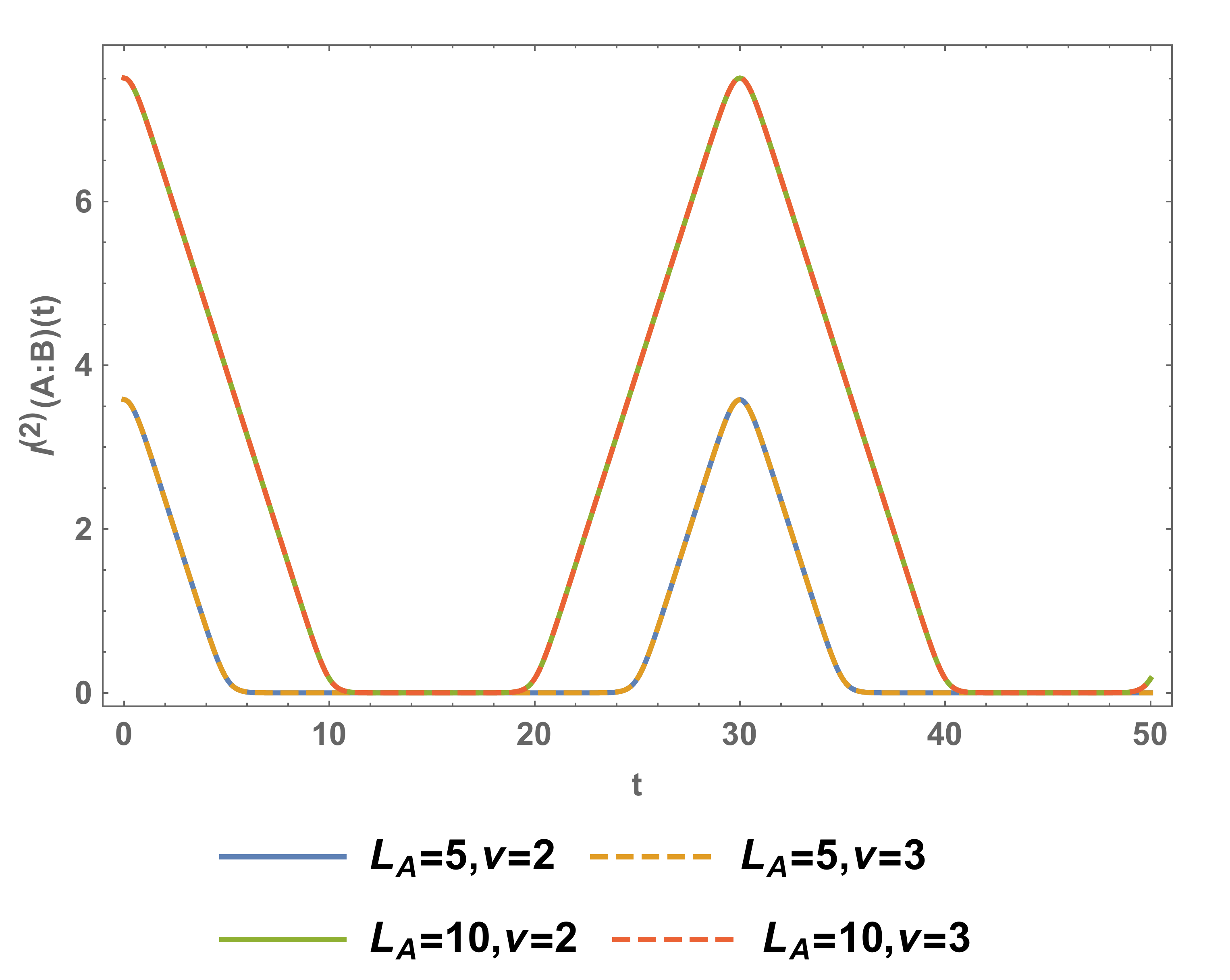}
    \includegraphics[trim={4cm 0 28cm 0},clip,valign=t,width=0.32\textwidth]{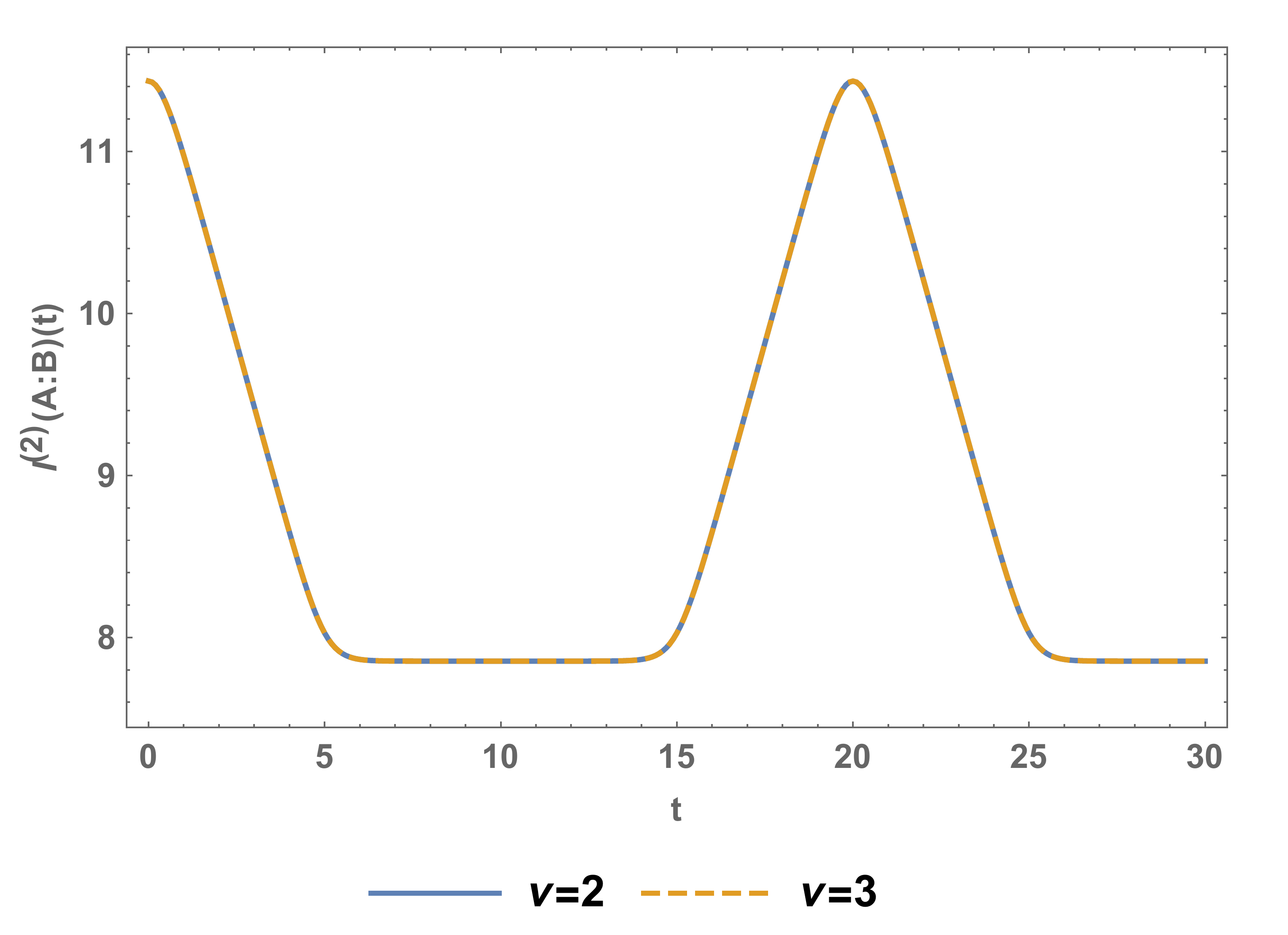}
    \caption{Plots of BOMI for the $c=1$ Dirac fermion with $n=2$ and $\epsilon=1$ with symmetric intervals. (\textbf{Left}:) Set $X_2=Y_2=0$ and $X_1=Y_1=5$ for two different spatial radius $R = \frac{15}{\pi},\frac{20}{\pi}$. (\textbf{Middle}:) Fixed the total system radius to be $R=\frac{15}{\pi}$ and considered two choices of input subsystem $[X_2,X_1]=[20,25],[20,30]$. (\textbf{Right}:) Set $X_1=Y_1=15$, $X_2=Y_2=0$ and $R=\frac{10}{\pi}$. This corresponds to the case where the union $A\cup B$ has a length greater than $2\pi R$ so there are always some quasiparticles in $B$ and hence the BOMI is never zero.
    }
    \label{FiniteSystemBOMI_SymmetricIntervals_Physical}
\end{figure}

First, consider two subsystems $A$ and $B$ that are spatially identical. Plots of the resulting BOMI are shown in Figure \ref{FiniteSystemBOMI_SymmetricIntervals_Physical}.
In following plots of BOMI for the free fermion CFT, the regulator will be set to $\epsilon=1$. In the leftmost plot, the subsystems have a fixed length of $L_A = L_B = 5$ while the total system size $2\pi R = 30,40$ is varied. Since both subsytems are spatially identical with length $L_A = L_B = 5$, all quasi-particles leave the subsystem by $t=5$ and the BOMI vanishes at this time. The quasi-particles re-enter the subsystem $B$ at $t= 25$ and $t = 35$ for a total system size of $2\pi R = 30$ and $2\pi R = 40$ respectively, causing the BOMI to increase and return to its original value. In the middle plot, the total system size is fixed at $2\pi R = 30$ while the subsystem sizes are $L_A = L_B = 5,10$. When the subsystem size is doubled, it takes twice as long for the quasi-particles to exit the subsystem initially, so it takes twice as long for the BOMI to first vanish. However, since the total system is spatially compact, the quasi-particles for the larger subsystem will re-enter the subsystem at an earlier time hence the BOMI starts to increase at an earlier time. In the rightmost plot, the sum of the lengths of the two subsystems $L_A+L_B$ is greater than the total system size $2\pi R$. Therefore, there will always be some quasi-particles contained in subsystem $B$ and the BOMI never vanishes.

\subsubsection{Asymmetric Intervals}
\begin{figure}
    \centering
    \textbf{Asymmetric Intervals}\par\medskip
    \includegraphics[trim={4cm 0 28cm 0},clip,valign=t,width=0.45\textwidth]{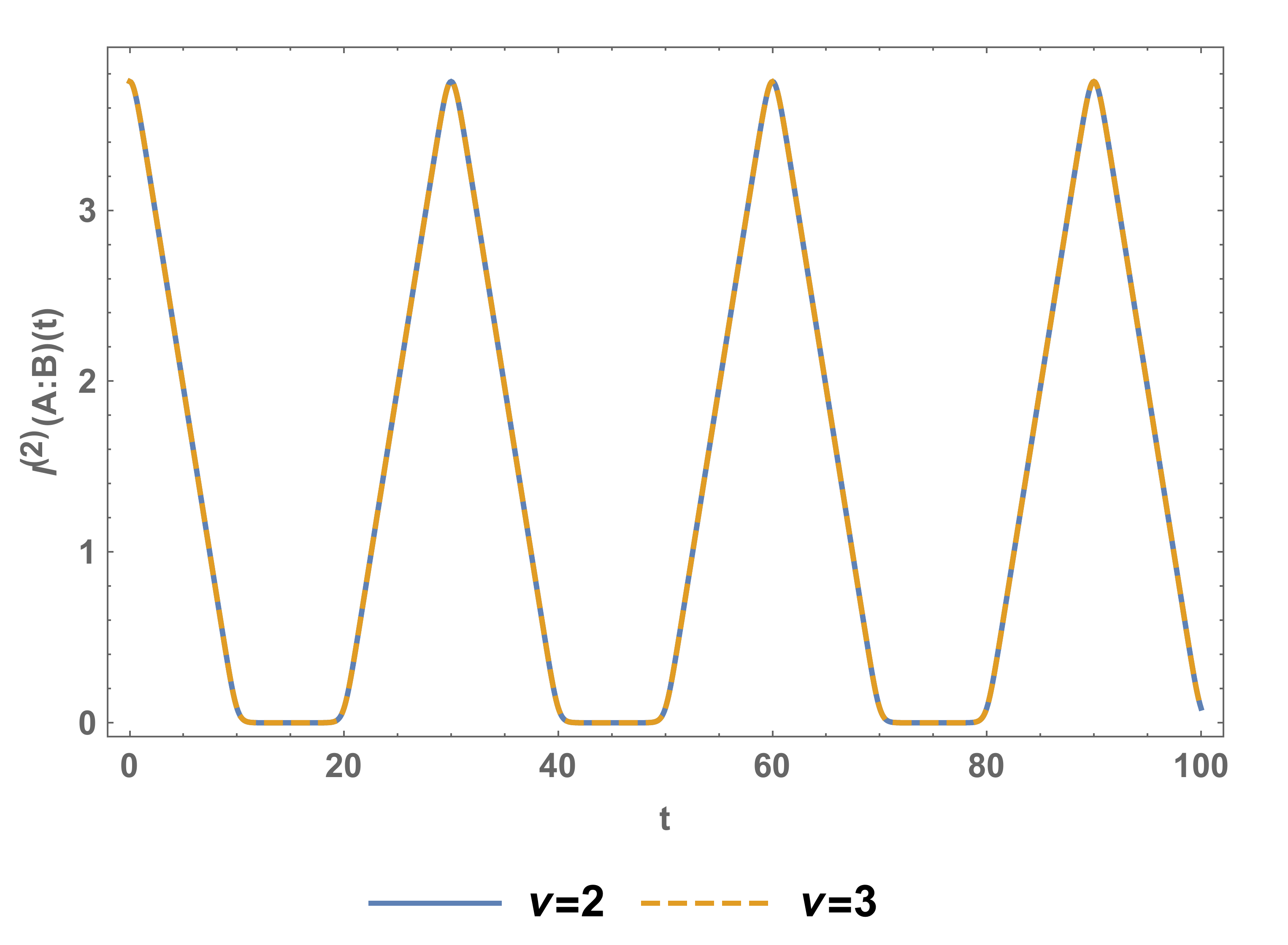}
    \includegraphics[trim={4cm 0 28cm 0},clip,valign=t,width=0.45\textwidth]{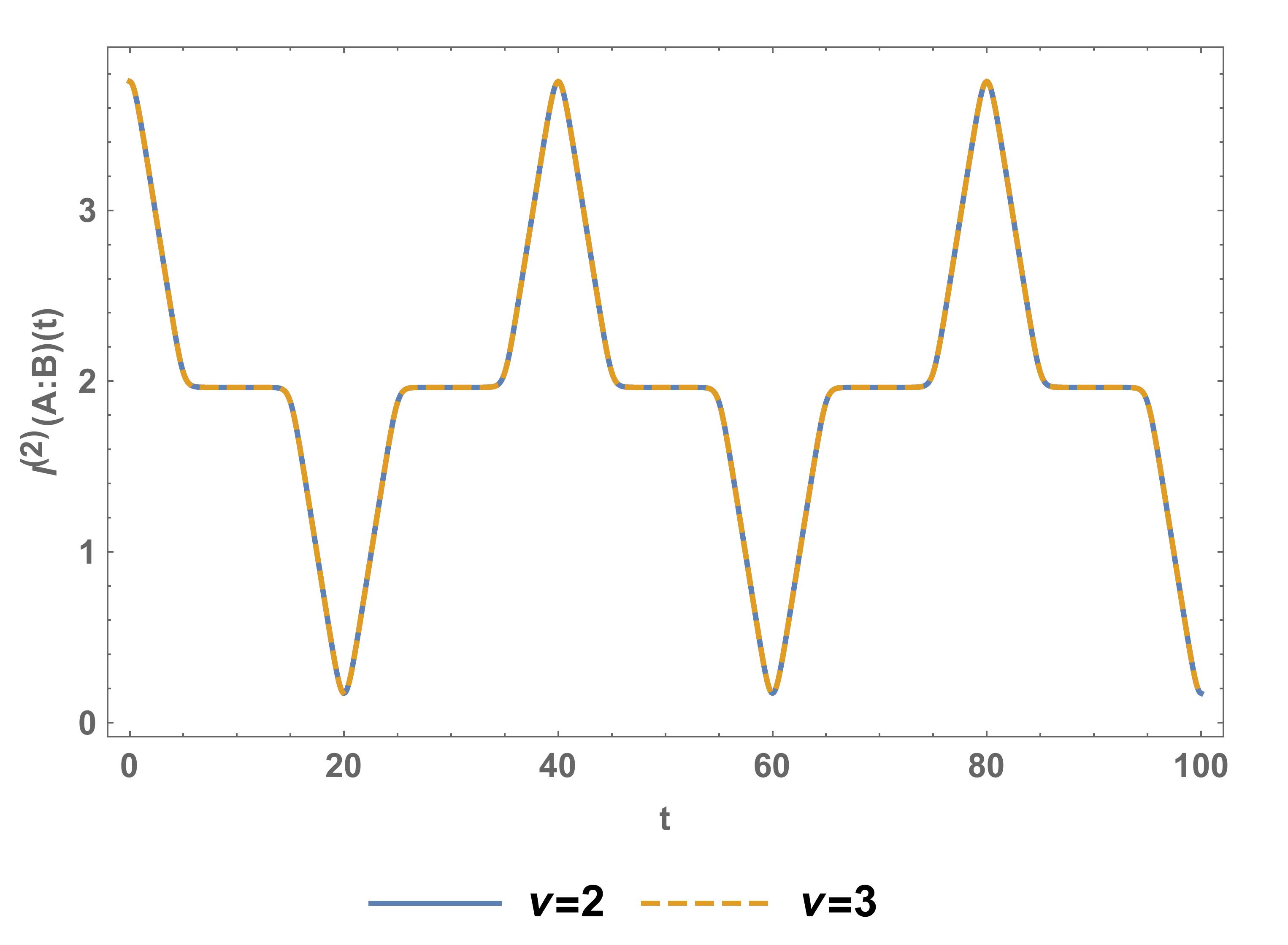}
    \caption{Plots of BOMI between two non-identical spatially overlapping subsystems $A$ and $B$. $(\textbf{Left}:)$ The subsystems are $A = [X_2,X_1]=[0,10]$ and $B=[Y_2,Y_1]=[0,5]$ and the total system size is $2\pi R = 30$. $(\textbf{Right}:)$ The subsystems are $A = [X_2,X_1]=[5,25]$ and $B=[Y_2,Y_1]=[5,10]$ and the total system size is $2\pi R = 40$.}
    \label{FiniteSystemBOMI_AsymmetricIntervals}
\end{figure}

Now, consider the case where the two intervals are no longer spatially identical but still have a non-zero intersection. Plots of BOMI for some of these configurations are shown in Figure \ref{FiniteSystemBOMI_AsymmetricIntervals}. First, consider the leftmost plot where $L_A = 10$ and $L_B = 5$. From $t=0$ to $t=5$, there is a loss of right-moving quasi-particles from the output subsystem while from $t=5$ to $t = 10$, there is a net loss of left-moving quasi-particles. Therefore, the BOMI decreases linearly to zero until $t = 10$. At $t = 20$, the quasi-particles re-enter the subsystem due to the periodicity of space and the BOMI returns to its original value. In the rightmost plot, the subsystems have lengths $L_A = 20$ and $L_B = 5$ while the total system size is $2\pi R = 40$. There is now a horizontal plateau between $t = 5$ and $t = 15$ because the rate of left-moving quasi-particles entering and leaving the output subsystem as this time are equal. At $t = 20$, all the left-moving quasi-particles have just left the output subsystem while the right-moving quasi-particles are beginning to re-enter the subsystem, leading to a rise in BOMI. Eventually, the BOMI returns to its original value and the process repeats itself indefinitely.

\subsubsection{Disjoint Intervals}
\begin{figure}
    \centering
    \textbf{Disjoint Intervals}\par\medskip
    \includegraphics[trim={4cm 0 28cm 0},clip,valign=t,width=0.45\textwidth]{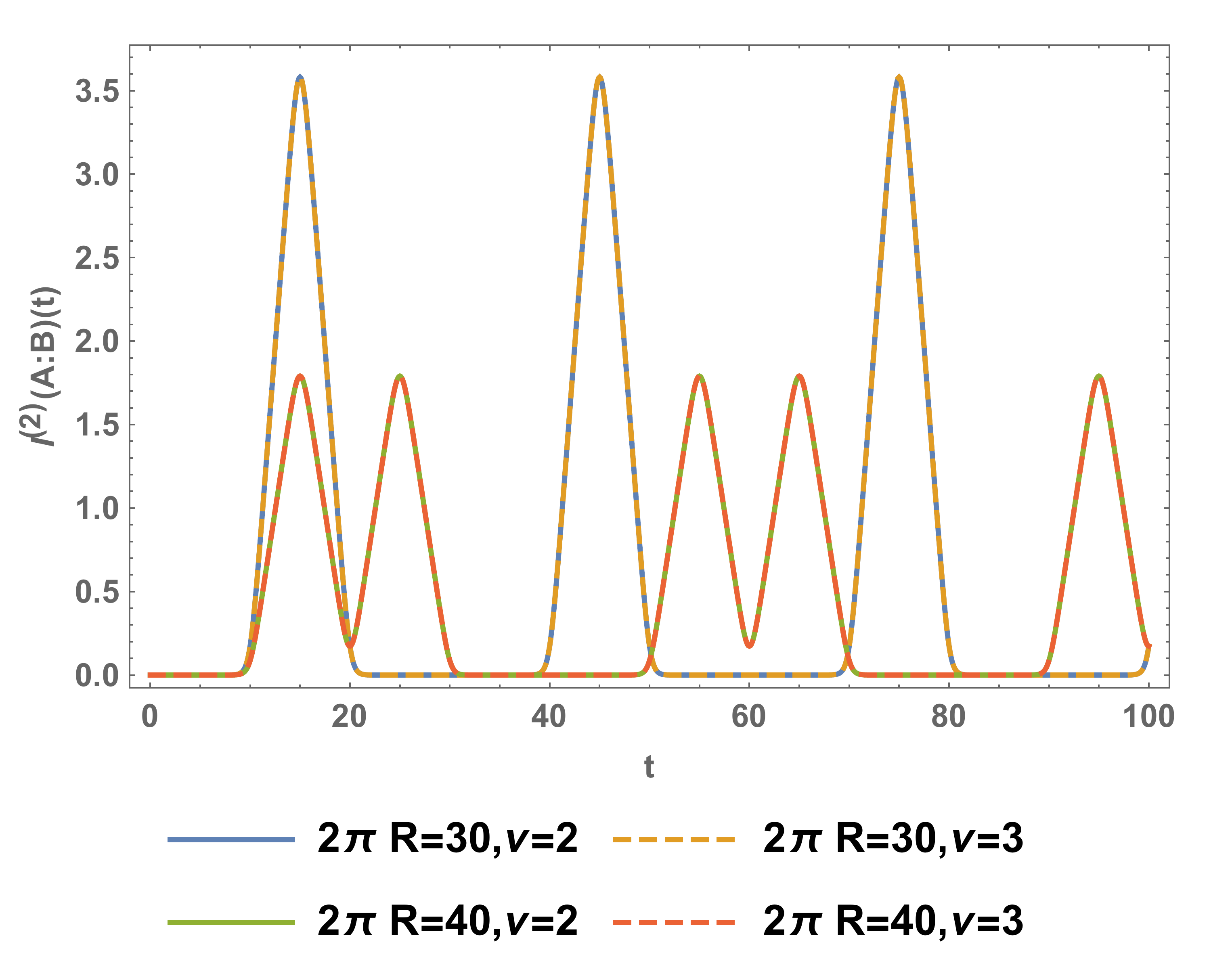}
    \includegraphics[trim={4cm 0 28cm 0},clip,valign=t,width=0.45\textwidth]{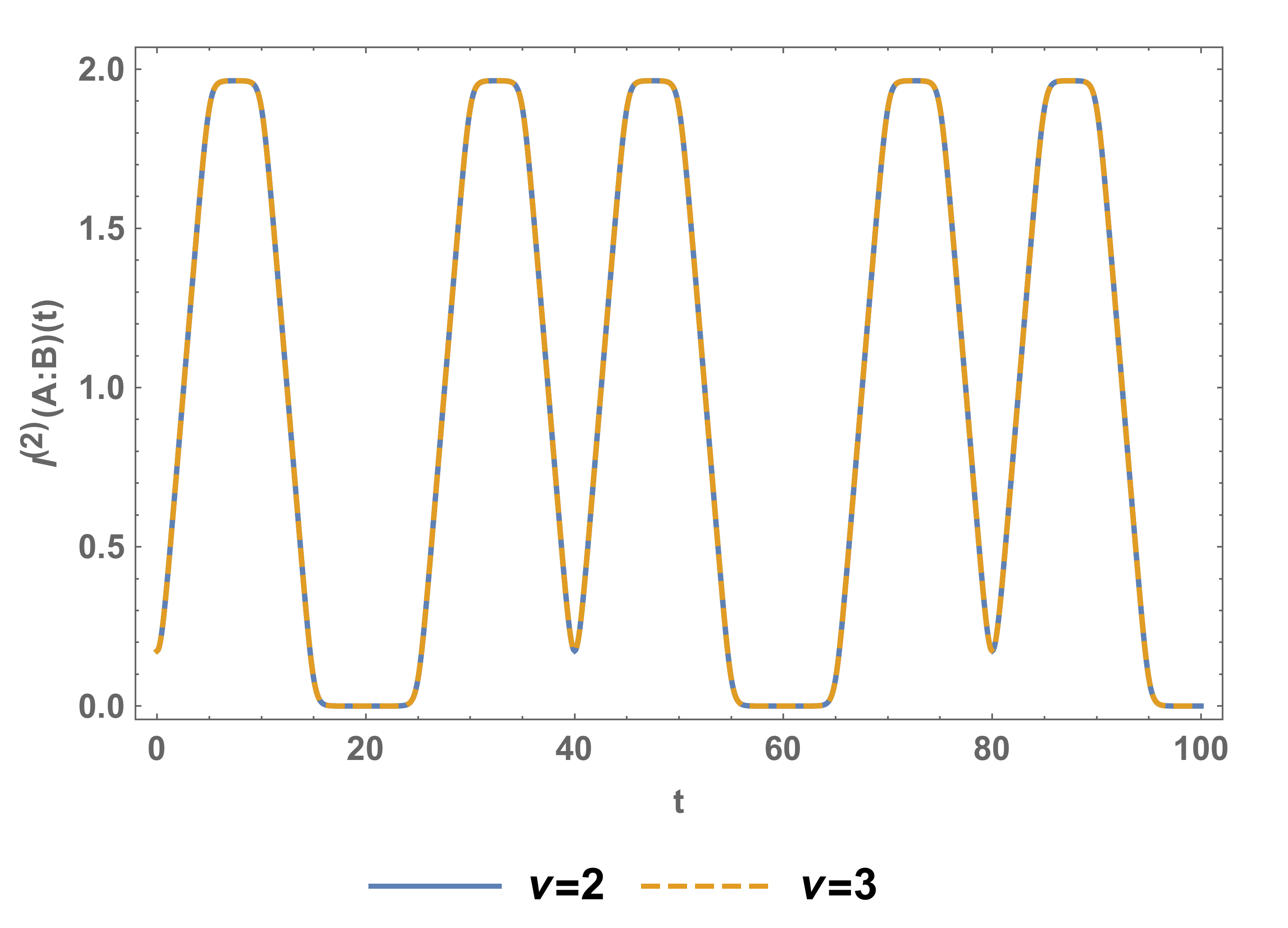}
    \caption{Plots of BOMI between two disjoint intervals $A$ and $B$. (\textbf{Left}:) The input and output subsystems are fixed as $A = [X_2,X_1]=[20,25]$ and $B = [Y_2,Y_1]=[5,10]$ while the total system size $2\pi R=30, 40$ is varied. (\textbf{Right}:) The input and output subsystems are $A = [X_2,X_1]=[10,20]$ and $B = [Y_2,Y_1]=[5,10]$ while the total system size $2\pi R=40$.}
    \label{FiniteSystemBOMI_DisjointIntervals}
\end{figure}

Lastly, consider the case where the input and output subsystems are disjoint. Plots of the BOMI when the input and output subsystems have no overlap are shown in Figure \ref{FiniteSystemBOMI_DisjointIntervals}. In the left plot, the total system size is varied. When $2\pi R = 30$, both the left-moving and right-moving quasi-particles enter and leave the output subsystem at the same time but when the total system size is $2\pi R = 40$, the right-moving quasi-particles enter the output subsystem immediately after all the left-moving quasi-particles have left it. This explains why there are two bumps in the latter case and why the single bump in the former case is twice as tall. In the right plot, the two subsystems have no overlap and yet are spatially adjacent so the BOMI begins to increase immediately. Since the input and output subsystems are not of equal length, plateaus arise in the BOMI.

\subsubsection{Tripartite Operator Mutual Information}
With the equation for the BOMI for free fermions \eqref{BOMI} at hand, the TOMI is readily computed. The free fermion BOMI can be divided into a universal spin-structure independent piece $I_{\text{univ.}}^{(n)}(A:B)$ and a spin-structure dependent piece $I_{\text{non-univ.},\nu}^{(n)}(A:B)$:
\begin{equation}
    I_\nu^{(n)}(A:B) = I_{\text{univ.}}^{(n)}(A:B)+I_{\text{non-univ.},\nu}^{(n)}(A:B)
\end{equation}
where $I_{\text{univ.}}^{(n)}(A:B)$ is the first term of \eqref{BOMI} while $I_{\text{non-univ.},\nu}^{(n)}(A:B)$ is the second. Let $A = [X_2,X_1]$ be an interval in the input Hilbert space and $B_1 = [Y_2,Y_1]$ and $B_2 = [Y_3,Y_2]$ be subsystems of the output Hilbert space where $Y_3<Y_2<Y_1$. The universal part of the TOMI cancels out exactly since
\begin{equation}
    I_{\text{univ.}}^{(n)}(A:B_1:B_2) = I_{\text{univ.}}^{(n)}(A:B_1)+I_{\text{univ.}}^{(n)}(A:B_2)-I_{\text{univ.}}^{(n)}(A:B_1\cup B_2) = 0.
\end{equation}
Note that this cancellation is exact and does not require any particular limits to be taken. On the other hand, the non-universal piece of TOMI, which is a linear combination of the spin-structure terms, does not cancel out exactly. The spin-structure part of the OEE is given by the last term of \eqref{TwistOperatorLogarithm},
\begin{equation}
    S_{C,\nu}^{(n)} = \frac{1}{1-n}\sum_{k=-\frac{n-1}{2}}^{\frac{n-1}{2}} \log \left|\frac{\theta_\nu\left(\frac{k}{n}\frac{z}{2\pi R}\bigg|\tau\right)}{\theta_\nu(0|\tau)}\right|^2
\end{equation}
where $z = X_2-X_1, Y_1-Y_2,X_2-X_1+ Y_1-Y_2 $ for $C=A,B,A\cup B$ respectively. Performing an S-modular transformation gives
\begin{equation}
    S_{C,\nu}^{(n)} =
    \frac{z^2}{24\epsilon R}\frac{n+1}{n}+
    \frac{2}{1-n}\sum_{k=-\frac{n-1}{2}}^{\frac{n-1}{2}} \log \left[\frac{\theta_\mu\left(i\frac{kz}{2n\epsilon}\bigg|i\frac{\pi R}{\epsilon}\right)}{\theta_\mu\left(0\bigg|i\frac{\pi R}{\epsilon}\right)}\right]
\end{equation}
for $(\nu,\mu)=(2,4),(3,3),(4,2)$. The theta functions are real and identical for both the holomorphic and anti-holomorphic piece. The argument of the logarithm is also postive because $z \in [-2\pi R,2\pi R]$. Following \cite{PhysRevD.77.064005, Herzog2013}, the sum of the logarithm can be re-written by applying the product representation of the elliptic theta functions followed by a Mercator expansion of the logarithms which converges because $z \in [-2\pi R,2\pi R]$. For the spin-structures that are anti-periodic along the thermal cycle, $\nu=2,3$, this results in 
\begin{equation}
    S_{C,\nu}^{(n)} =
    \frac{z^2}{24\epsilon R}\frac{n+1}{n}+
    \frac{2}{n-1}\sum_{j=1}^\infty \frac{(-1)^{j\nu}}{j\sinh{\frac{\pi^2 R j}{\epsilon}}}\left[\frac{\sinh{\frac{\pi j z}{2\epsilon}}}{\sinh{\frac{\pi j z}{2\epsilon n}}}-n\right],\qquad \nu=2,3
\end{equation}
The replica limit can be taken by applying L'Hospitals rule which gives
\begin{equation}
    S_{C,\nu} = \frac{z^2}{12\epsilon R}+2 \sum_{j=1}^\infty \frac{(-1)^{j\nu}}{j\sinh{\frac{\pi^2 R j}{\epsilon}}} \left[
    \frac{\pi j z}{2\epsilon}\coth{\frac{\pi j z}{2\epsilon}}-1
    \right]
\end{equation}
When $\epsilon\rightarrow 0$, the first term dominates while the second term is exponentially suppressed. Therefore, to leading order in $\epsilon$, 
\begin{equation}
    I_{\text{non-univ.},\nu}(A:B)
    = \frac{(X_1-X_2)(Y_1-Y_2)}{6\epsilon R}+\ldots
\end{equation}
where the sub-leading terms are exponentially suppressed. This leading order term is also symmetric in the lengths of subsystem $A$ and $B$ as it should be. Applying this formula, the spin-structure part of the TOMI also cancels out at leading order, i.e.
\begin{equation}
    I_{\text{non-univ.},\nu}(A:B_1:B_2) = 0 +\ldots
\end{equation}
Therefore, the total TOMI vanishes up to some exponentially suppressed terms. In the small regulator limit, the quasi-particle picture holds perfectly for free fermions on the circle (when $\nu=2,3$) and there is no scrambling of information by the corresponding unitary channel.

\subsubsection{Summary of Free Fermion CFT results}
In this section, bosonization was employed to compute the OEE of the $c=1$ Dirac fermion CFT defined on the torus $S^1 \times S^1$. The BOMI was found to be in perfect agreement with the quasi-particle picture just like in \cite{Nie_2019}. Due to the periodicity of space, the BOMI showed perfect revivals as the quasi-particles traversed the spatial circle $S^1$ periodically and indefinitely. When the regulator $\epsilon$ was taken to be much smaller than the total system size, the TOMI vanished for $\nu=2,3$, in agreement with the quasi-particle picture.

\subsection{An application of standard quasiparticle picture  to operator entanglement.}
In the previous part of section 3, we explained the time evolution of BOMI and TOMI of free fermion on compact spacetime using the quasiparticle picture proposed in \cite{Nie_2019}, where the time evolution of BOMI and TOMI on non-compact spacetime follows.
In this section, we will extend the standard quasiparticle picture, which can describe the time evolution of state entanglement, to describe the time evolution of operator entanglement.
We consider a double copy of free fermion (and scalar) theory and study mutual information between the two copies. This is basically considering the TFD state in these theories and study mutual information between subregions on different Hilbert spaces. Since we are studying free theories, an interesting question is whether we can understand the dynamics of BOMI in terms of free streaming quasi-particles or not? We show that imposing a suitable adaptation on the standard quasi-particle picture perfectly agrees with our numerical simulations modulo the zero mode effect which is not captured by the quasi-particle picture. 

In the following we first describe how to find exact quasi-particle formulae for dynamics of mutual information in given configuration. Next we present our numerical results and compare them with the quasi-particle picture. 

\subsection{Quasi-particle picture}
In this section we show how a suitable adaptation leads to an exact predictive quasi-particle picture for BOMI and for standard mutual information between subsystems on different sides of the TFD state in generic configurations. We will follow the picture introduced by Alba and Calabrese in \cite{Alba:2016, Alba:2017lvc} where we assume integrability together with the knowledge of the final steady state (basically a generalized Gibbs ensemble in this case), in order to fix the contribution of each single mode to the propagation of entanglement all over the evolution. Let's forget about our more complex setup of interest for studying mutual information for a while in order to review how this picture leads to a predictive formula for entanglement evolution for the simplest case, namely a single connected interval in a single copy of the Hilbert space. In such a case, the entanglement entropy is given by
\begin{align}\label{eq:AC1}
S(t)
=2t\int\displaylimits_{2|v(k)|t<L_A}\, dk \,s(k)v(k)+L_A \int\displaylimits_{2|v(k)|t>L_A}\, dk \,s(k),
\end{align}
where $v(k)$ is the group velocity of modes with label $k$, which is basically its momentum in free theories, and $s(k)$ is the entropy density in the momentum space. $s(k)$ carries the information of the quench protocol and the initial (pre-quench) state, which can be fixed by reading off the spectrum of the final generalized Gibbs ensemble from the constants of motion, where the latter can be read from the initial state (for more details see \cite{Alba:2017lvc}). When we consider a global quench, each spatial point is assumed to be a source of a pair of free streaming quasi-particles which are maximally entangled with each other. At time $t$, those quasi-particles which one pair is in the entangling region  and the other pair is outside the region contribute to the time dependent part of the entropy.

\begin{figure}[h]
\begin{center}
\includegraphics[scale=0.9]{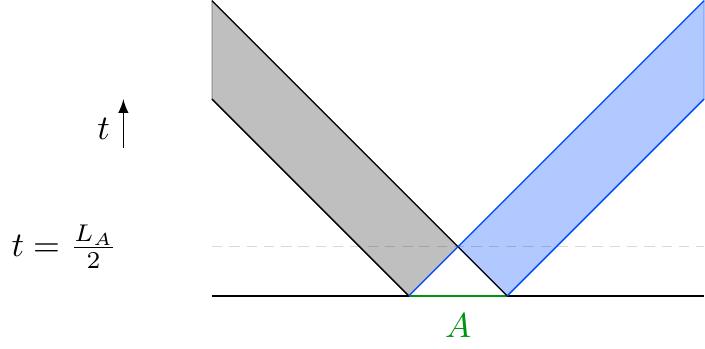}
\end{center}
\caption{Evolution of entanglement entropy of a single interval. For $0<t<L_A/2$, the intersection of constant time slices and the blue and gray shaded regions are increasing while time is increased. Afterwards for $L_A/2<t<\infty$ this intersections remain constant.}
\label{fig:QP1}
\end{figure}

For later convenience, we rephrase this formula \eqref{eq:AC1} pictorially. For simplicity of illustration, all our figures merely present the case that the group velocity is equal to unity, though in the analytic formulae we sum over all velocities corresponding to different momenta. We show left/right-moving quasi-particles in black/blue. The simplest case that is the entanglement entropy for a single interval has been shown in Figure \ref{fig:QP1}. The first term in \eqref{eq:AC1} corresponds to the black and the blue triangles in $0<t<L_A/2$ window which are responsible for the increase of the entropy, while the second term corresponds to the black and blue parallelograms extended over $L_A/2<t<\infty$, where the entropy has stopped increasing. 

In the following we will consider the quasi-particle picture introduced in \cite{Nie_2019} and used the aforementioned pictorial understanding to find an exact quasi-particle formulae for BOMI and mutual information in the TFD state in generic configurations. The key point of the picture introduced in \cite{Nie_2019} is that the region $A$ is the only source for free streaming quasi-particles and at a given time $t$, those quasi-particles which are inside $B$ contribute to the mutual information $I(A:B)$, regardless of the position of their pair. The final thing we need is to fix $s(k)$. Since our case of interest is basically the TFD state, the final state is nothing but two copies of thermal states, though $s(k)$ is simply two times the contribution of a thermal states, namely
\be\label{eq:ThEntDen}
s(k)
=2\left[\frac{\beta\,  \omega (k)\,e^{-\beta\,  \omega (k)}}{1\pm e^{-\beta\,  \omega (k)}}\pm\ln \left(1\pm e^{-\beta\,  \omega (k)}\right)\right],
\ee
corresponding to fermionic/bosonic theories.

We would like to also mention that the symmetric configuration where $L_A=L_B$ in TFD states has been previously discussed in \cite{Chapman:2018hou}, where it is possible to understand the $S_{A\cup B}$ in terms of the standard quasi-particle picture, i.e., where all spatial points are considered as sources for quasi-particles. In this case the entanglement between two copies is understood in terms of a linear combination of the quasi-particles propagating on each single side. Our results for generic configurations reduces to theirs for the specific case of symmetric configurations. It is also worth to note that a similar qualitative interpretation has been also discussed in \cite{es} assuming the left/ right-moving quasi-particles merely propagate on the left/right copies of the TFD state correspondingly. 

In the following we present examples of the three family of symmetric, asymmetric and disjoint configurations and show how to extract the corresponding quasi-particle picture on compact spatial direction. 

\subsection*{Symmetric}
\begin{figure}[h!]
\begin{center}
\includegraphics[scale=.88]{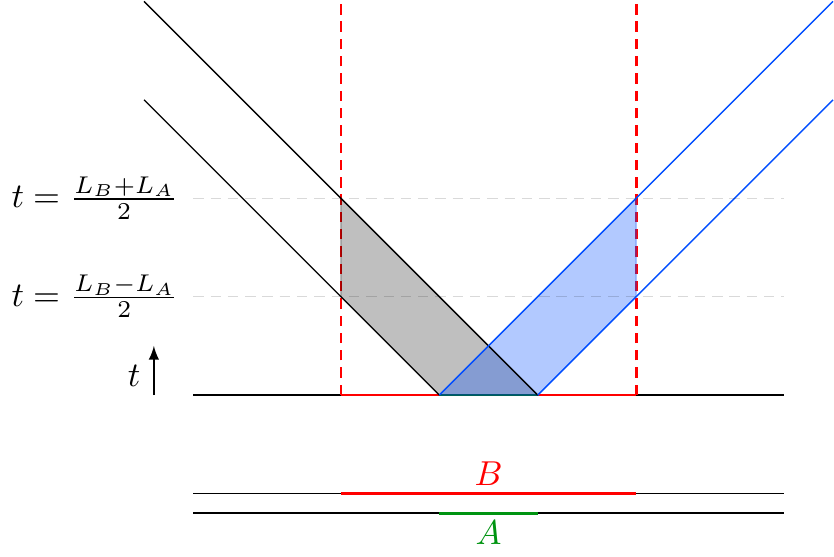}
\hspace*{10mm}
\includegraphics[scale=.88]{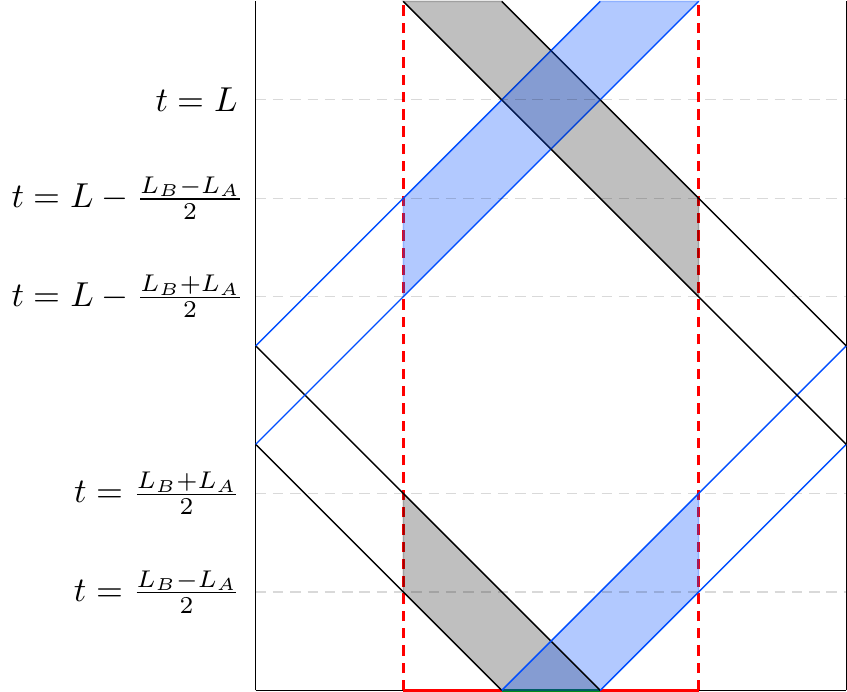}
\end{center}
\caption{Symmetric configuration. Left/Right: on infinite/compact system.}
\label{fig:QPsym}
\end{figure}

It is not hard to see that the quasi-particle picture on an infinite spatial direction is given by
\begin{align}\label{eq:symInf}
I(A:B)=
2L_A \int_{v(k)t \leq \frac{L_B-L_A}{2}}\, \frac{dk}{2\pi} \,s(k)
+ \int_{v(k)t \leq \frac{L_B+L_A}{2}}\, \frac{dk}{2\pi} \,s(k)\left[L_B+L_A - 2v(k)t\right].
\end{align}
It is easy to understand how we came to this formula for the infinite system case. In the right panel of Figure \ref{fig:QPsym}, it is clear that the first term in \eqref{eq:symInf} is contributing for $0<t<\frac{L_B-L_A}{2}$, where the intersection of a time slice with both the blue and black parallelograms have constant length $L_A$. While time increase, in $\frac{L_B-L_A}{2}<t<\frac{L_B+L_A}{2}$, the intersection on a time-slice with the parallelograms is decreasing and hence each color contributes $-s(k)v(k)t$ to the integrand. It is straightforward to find the constant part of the integrand in the second term of \eqref{eq:symInf} geometrically, which also ensures the continuity of the expression at $t=\frac{L_B-L_A}{2}$. For $t>\frac{L_B+L_A}{2}$ there is no quasi-particle in region $B$, hence mutual information vanishes. 

On the finite system with size $L$ shown in Figure \ref{fig:QPsym}, we follow exactly the same logic but also keep in mind the the quasi-particles reenter the region due to the periodicity of the spatial direction. A careful follow up of this reentrance has been illustrated in the right panel of Figure \ref{fig:QPsym}. It is not hard to see that in this case one finds
\begin{align}
\begin{split}
I(A:B)
&=
2L_A \int_{\left\{\frac{v(k)t}{L}\right\} \leq \frac{L_B-L_A}{2L}}\, \frac{dk}{2\pi} \,s(k)
+ \int_{\frac{L_B-L_A}{2L}<\left\{\frac{v(k)t}{L}\right\} \leq \frac{L_B+L_A}{2L}}\, \frac{dk}{2\pi} \,s(k)\left[L_B+L_A - L\left\{\frac{2v(k)t}{L}\right\}\right]
\\&
+2L \int_{1-\frac{L_B+L_A}{2L}<\left\{\frac{v(k)t}{L}\right\} \leq 1-\frac{L_B-L_A}{2L}}\, \frac{dk}{2\pi} \,s(k)\left\{2v(k)\left(\frac{t}{L}-1+\frac{L_B+L_A}{2L}\right)\right\}
\\&
+2L_A \int_{1-\frac{L_B-L_A}{2L}<\left\{\frac{v(k)t}{L}\right\} \leq 1}\, \frac{dk}{2\pi} \,s(k)
\end{split}
\end{align}
where $\left\{\bullet\right\}$ is the fractional part of $\bullet$. 
\subsection*{Asymmetric}
\begin{figure}[t!]
\begin{center}
\includegraphics[scale=.55]{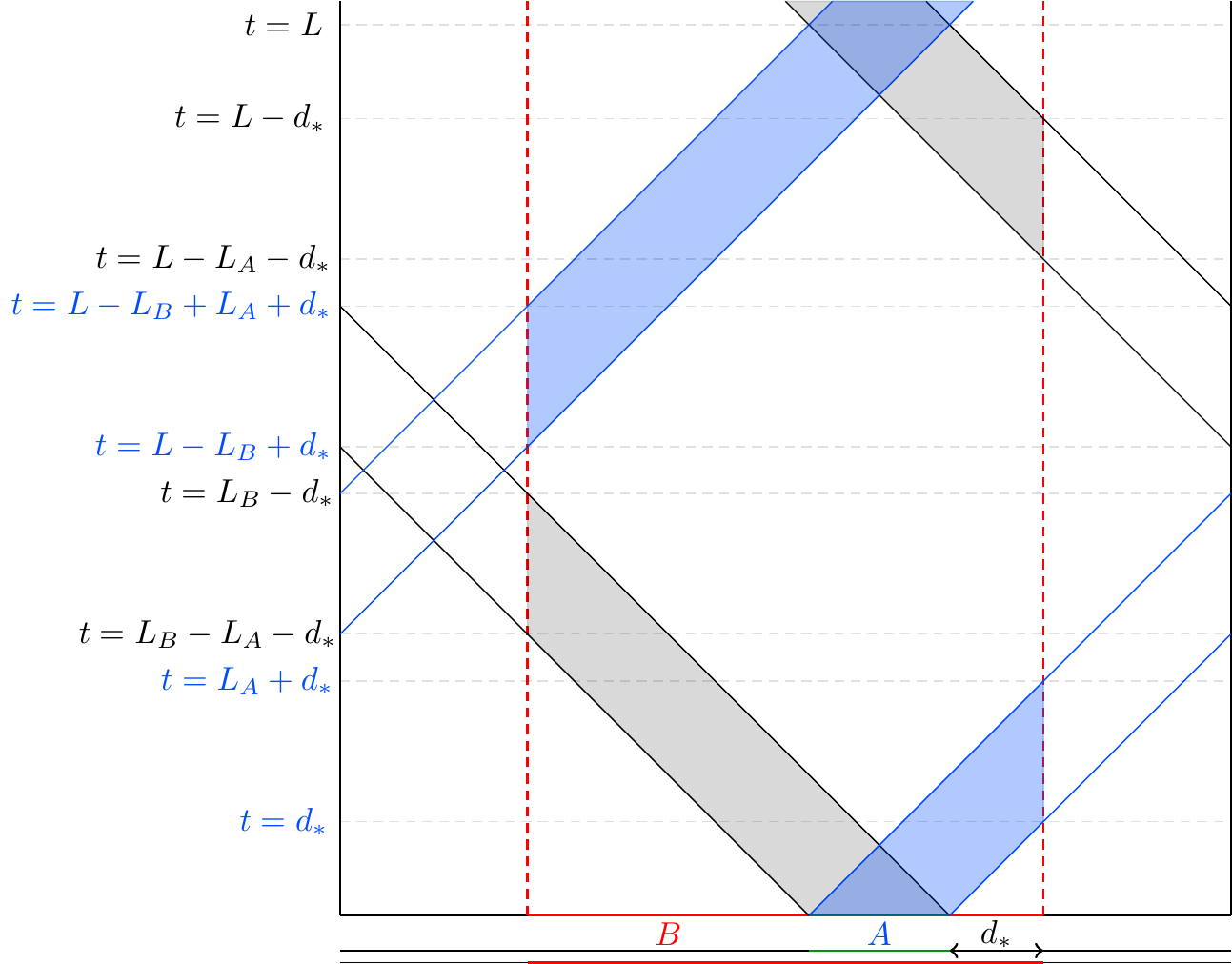}
\hspace*{5mm}
\includegraphics[scale=.55]{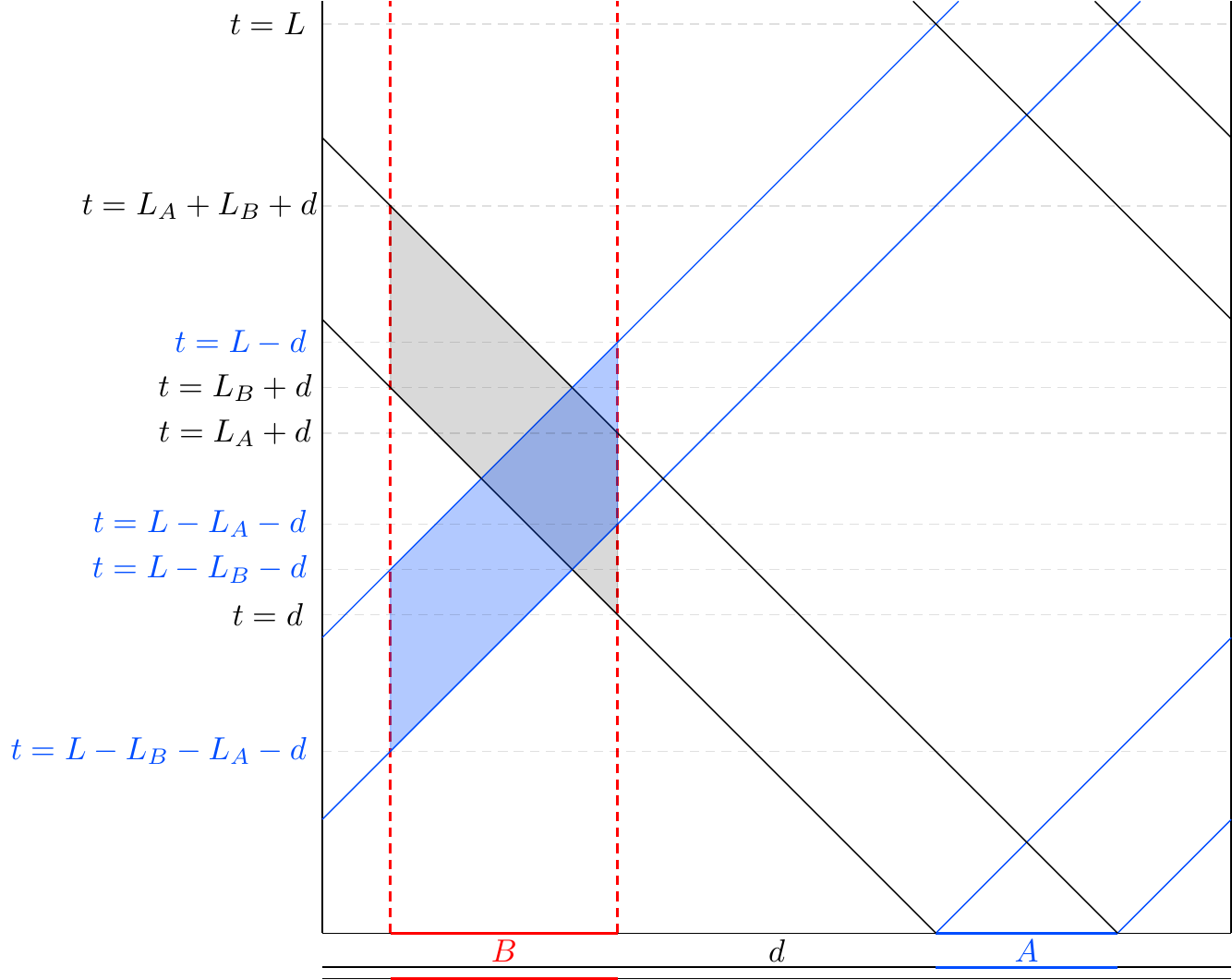}
\end{center}
\caption{Examples of asymmetric (left) and disjoint (right) configurations.}
\label{fig:QPAsymDis}
\end{figure}

For the asymmetric configurations, several cases may happen depending on the length of $L_A$, $L_B$, the asymmetric parameter $d_*$, and the length of the total system $L$. An example is shown in the left panel of Figure \ref{fig:QPAsymDis}. For the case shown in this figure, following the same logic described in the symmetric case, one can easily verify that for an infinite system mutual information is given by
\begin{align}
\begin{split}
I(A:B)
&=
2L_A \int_{t \leq \frac{d_*}{v(k)}}\, \frac{dk}{2\pi} \,s(k)
+\int_{\frac{d_*}{v(k)} < t \leq \frac{L_A+d_*}{v(k)}}\, \frac{dk}{2\pi} \,s(k)\left[2L_A + d_* - v(k)t\right]
\\&
+L_A\int_{\frac{L_A + d_*}{v(k)} < t \leq \frac{L_B-L_A-d_*}{v(k)}}\, \frac{dk}{2\pi} \,s(k)
+\int_{\frac{L_B-L_A-d_*}{v(k)} < t \leq \frac{L_B-d_*}{v(k)}}\, \frac{dk}{2\pi} \,s(k)\left[L_B - d_* - v(k)t\right],
\end{split}
\end{align}
and on the finite system we find
\begin{align}
\begin{split}
I(A:B)
&=
2L_A \int_{\left\{\frac{v(k)t}{L}\right\} \leq \frac{d_*}{L}}\, dk \,s(k)
+ \int_{\frac{d_*}{L}<\left\{\frac{v(k)t}{L}\right\} \leq \frac{L_A+d_*}{L}}\, \frac{dk}{2\pi} \,s(k)\left[2L_A+d_* - L\left\{\frac{v(k)t}{L}\right\}\right]
\\&\hspace*{-1cm}
+L_A \int_{\frac{L_A+d_*}{L}<\left\{\frac{v(k)t}{L}\right\} \leq \frac{L_B-L_A-d_*}{L}}\, \frac{dk}{2\pi} \,s(k)
+ \int_{\frac{L_B-L_A-d_*}{L}<\left\{\frac{v(k)t}{L}\right\} \leq \frac{L_B-d_*}{L}}\, \frac{dk}{2\pi} \,s(k)\left[L_B-d_* - L\left\{\frac{v(k)t}{L}\right\}\right]
\\&\hspace*{-1cm}
+L \int_{1-\frac{L_B-d_*}{L}<\left\{\frac{v(k)t}{L}\right\} \leq 1-\frac{L_B-L_A-d_*}{L}}\, \frac{dk}{2\pi} \,s(k)\left\{v(k)\left(\frac{t}{L}-1+\frac{L_B-d_*}{L}\right)\right\}
\\&\hspace*{-1cm}
+L_A \int_{1-\frac{L_B-L_A-d_*}{L}<\left\{\frac{v(k)t}{L}\right\} \leq 1-\frac{L_A+d_*}{L}}\, \frac{dk}{2\pi} \,s(k)
\\&\hspace*{-1cm}
+\int_{1-\frac{L_A+d_*}{L}<\left\{\frac{v(k)t}{L}\right\} \leq 1-\frac{d_*}{L}}\, \frac{dk}{2\pi} \,s(k)\left[L_A+L\left\{v(k)\left(\frac{t}{L}-1+\frac{L_A+d_*}{L}\right)\right\}\right]
\\&\hspace*{-1cm}
+2L_A \int_{1-\frac{d_*}{2L}<\left\{\frac{v(k)t}{L}\right\} \leq 1}\, \frac{dk}{2\pi} \,s(k)\,.
\end{split}
\end{align}
The same procedure leads to all other possible choice of parameters which we skip to mention here.
\subsection*{Disjoint}
This configuration also may happen in several cases regarding the parameters $L_A$, $L_B$, $d$, and $L$. We have shown an example in the right panel of Figure \ref{fig:QPAsymDis}. In this case on an infinite system we find
\begin{align}
\begin{split}
I(A:B)
&=
\int_{\frac{d}{v(k)} < t \leq \frac{L_A+d}{v(k)}}\, dk \,s(k)\left[v(k)t -d\right]
+L_A\int_{\frac{L_A + d}{v(k)} < t \leq \frac{L_B+d}{v(k)}}\, \frac{dk}{2\pi} \,s(k)
\\&
+\int_{\frac{L_B+d}{v(k)} < t \leq \frac{L_B+L_A+d}{v(k)}}\, \frac{dk}{2\pi} \,s(k)\left[L_B +L_A + d - v(k)t\right],
\end{split}
\end{align}
and on a finite system we find
\begin{align}
\begin{split}
I(A:B)
&=
L\int_{1-\frac{L_B+L_A+d}{L}<\left\{\frac{v(k)t}{L}\right\} \leq \frac{d}{L}}\, \frac{dk}{2\pi} \,s(k)
\left\{v(k)\left(\frac{t}{L}-1+\frac{L_A+L_B+d}{L}\right)\right\}
\\&\hspace*{-1cm}
+L\int_{\frac{d}{L}<\left\{\frac{v(k)t}{L}\right\} \leq 1-\frac{L_B+d}{L}}\, \frac{dk}{2\pi} \,s(k)
\left\{2v(k)\frac{t-d}{L}\right\}
\\&\hspace*{-1cm}
+
L\int_{1-\frac{L_B+d}{L}<\left\{\frac{v(k)t}{L}\right\} \leq 1-\frac{L_A+d}{L}}\, \frac{dk}{2\pi} \,s(k)
\left[\left\{v(k)\left(\frac{t}{L}-1+\frac{L_B+d}{L}\right)\right\}+L-L_B+L_A-2d\right]
\\&\hspace*{-1cm}
+\left(L-2d\right) \int_{1-\frac{L_A+d}{L}<\left\{\frac{v(k)t}{L}\right\} \leq \frac{L_A+d}{L}}\, \frac{dk}{2\pi} \,s(k)
\\&\hspace*{-1cm}
 -L\int_{\frac{L_A+d}{L}<\left\{\frac{v(k)t}{L}\right\} \leq \frac{L_B+d}{L}}\, \frac{dk}{2\pi} \,s(k)
\left[\left\{v(k)\frac{t-L_A-d}{L}\right\}+L-2d\right]
\\&\hspace*{-1cm}
-2L\int_{\frac{L_B+d}{L}<\left\{\frac{v(k)t}{L}\right\} \leq 1-\frac{d}{L}}\, \frac{dk}{2\pi} \,s(k)
\left[\left\{v(k)\frac{t-L_B-d}{L}\right\}+L+L_A-L_B-2d\right]
\\&\hspace*{-1cm}
-L\int_{1-\frac{d}{L}<\left\{\frac{v(k)t}{L}\right\} \leq \frac{L_A+L_B+d}{L}}\, \frac{dk}{2\pi} \,s(k)
\left[\left\{v(k)\left(\frac{t}{L}-1+\frac{d}{L}\right)\right\}-L+L_A+L_B+2d\right]
\end{split}
\end{align}

\subsubsection{Numerical results}\label{sec:numericalmethod}
\begin{figure}[t!]
\begin{center}
\includegraphics[scale=.3]{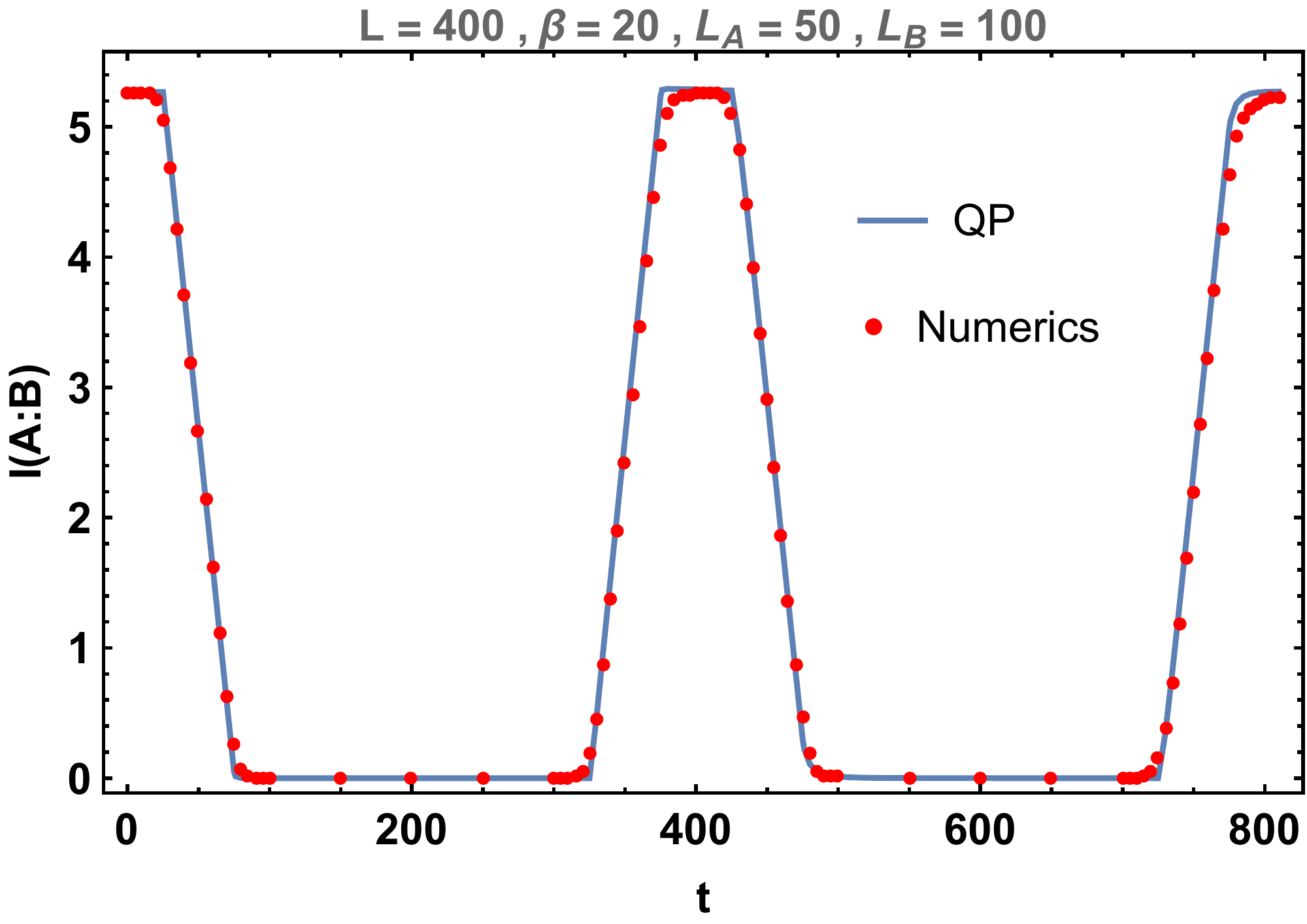}
\hspace*{5mm}
\includegraphics[scale=.3]{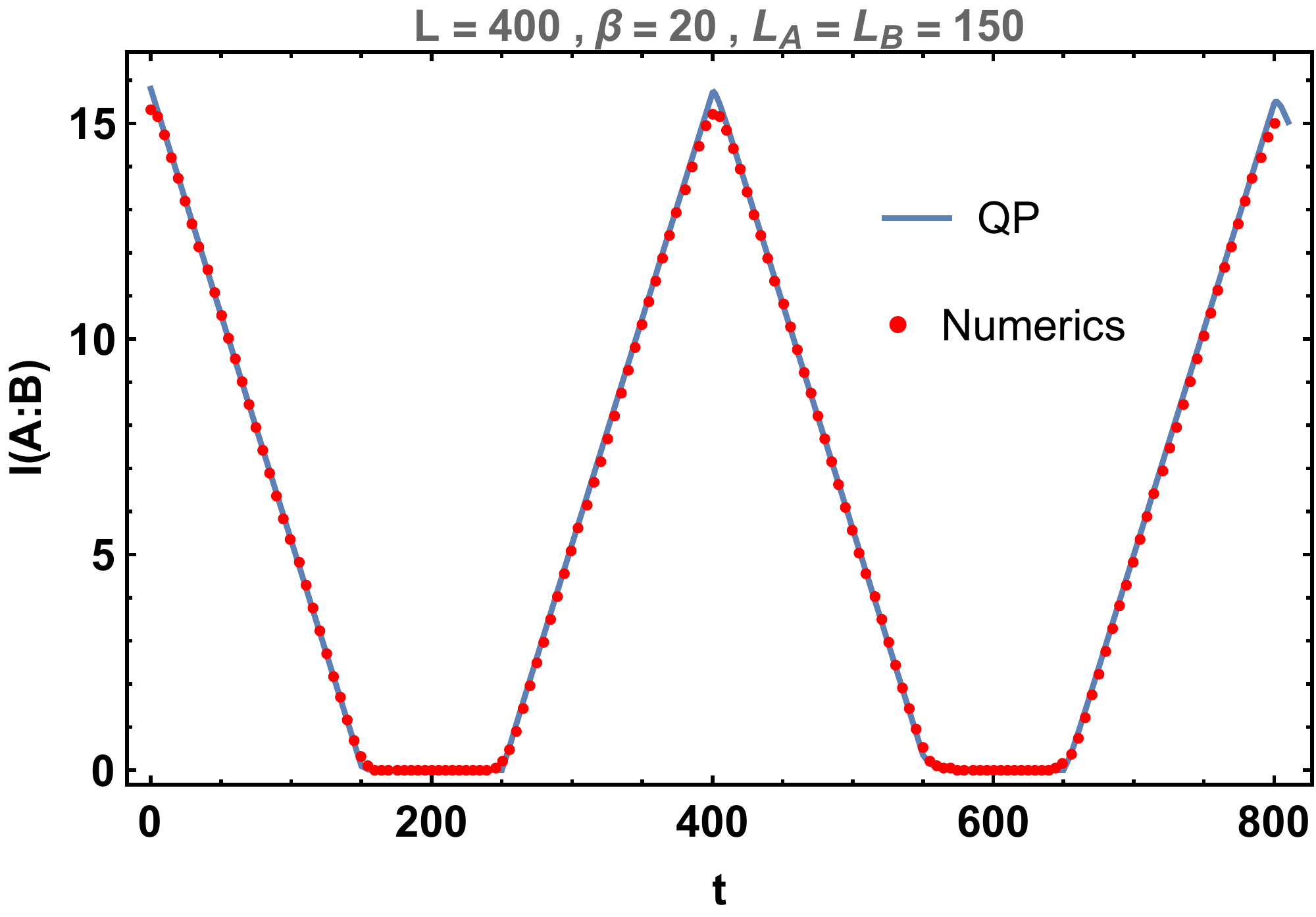}
\includegraphics[scale=.3]{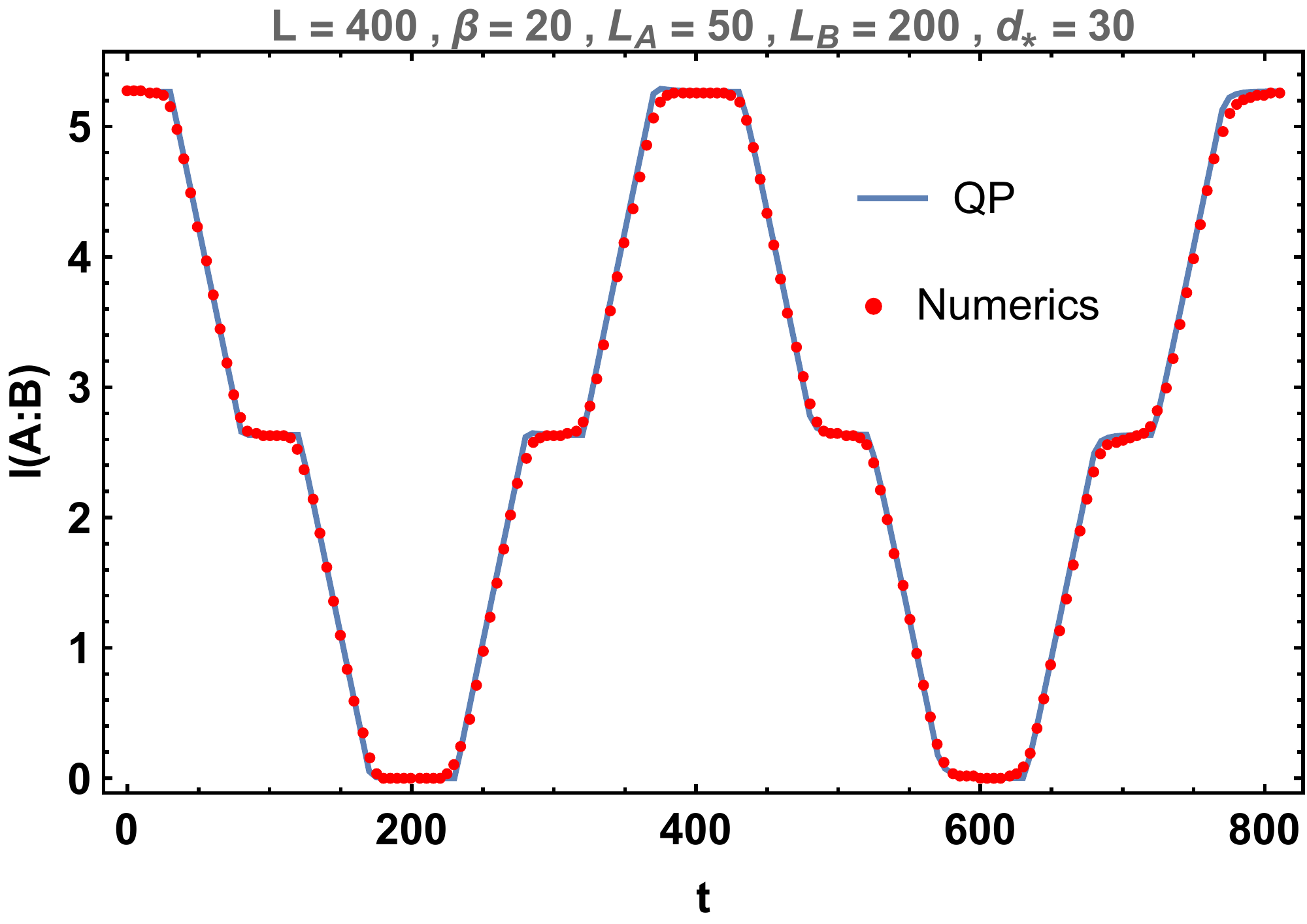}
\hspace*{5mm}
\includegraphics[scale=.3]{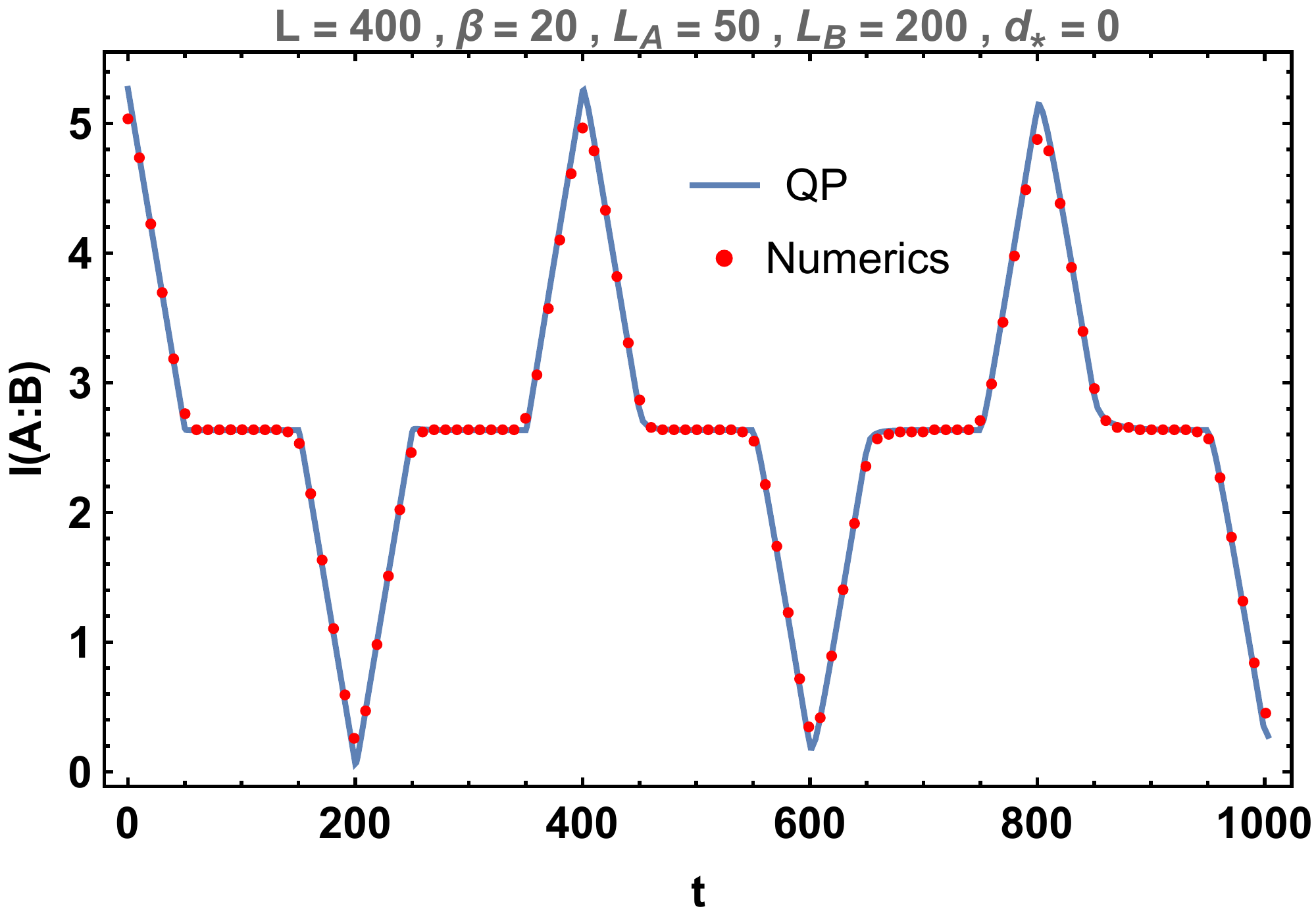}
\includegraphics[scale=.325]{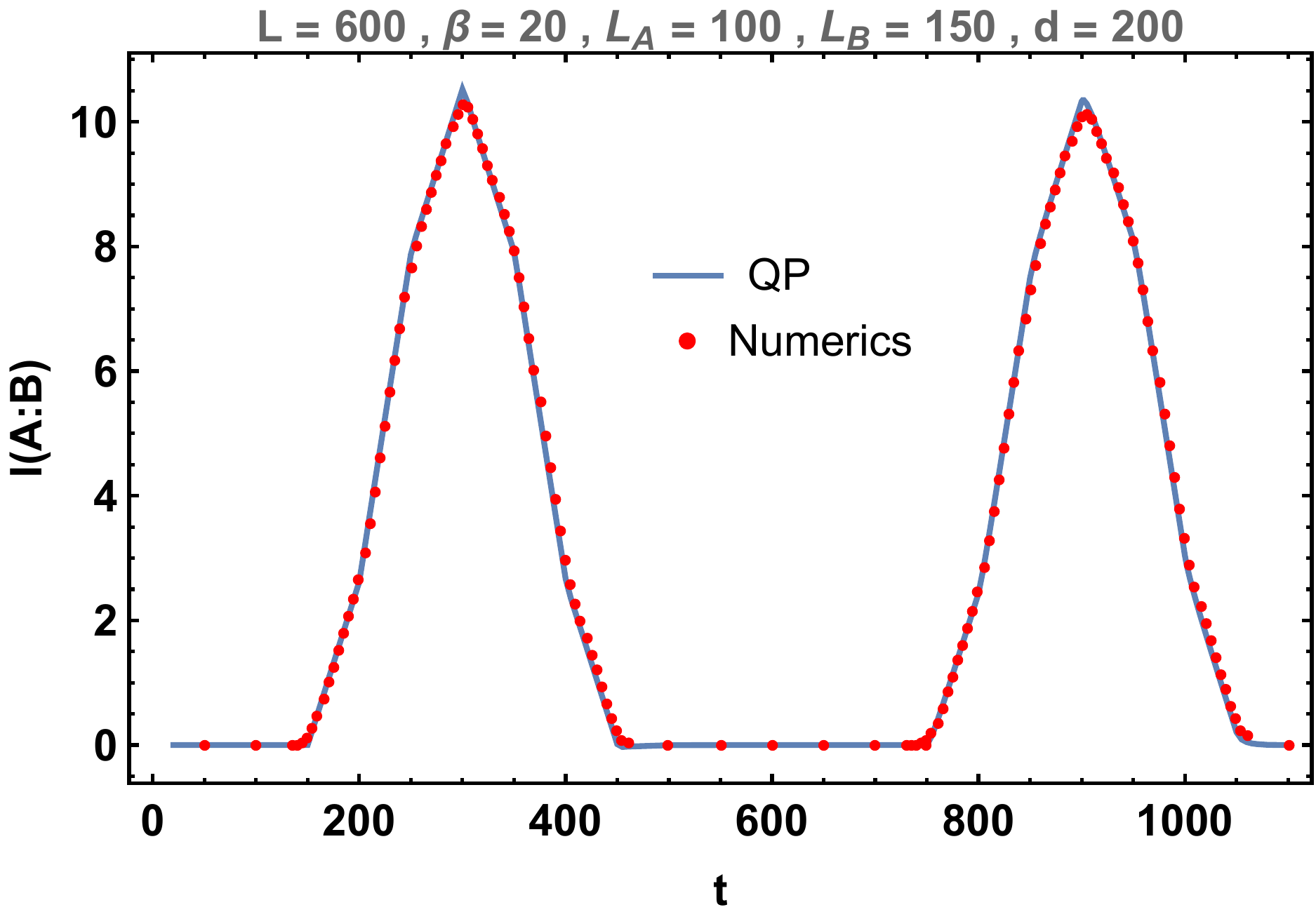}
\hspace*{5mm}
\includegraphics[scale=.325]{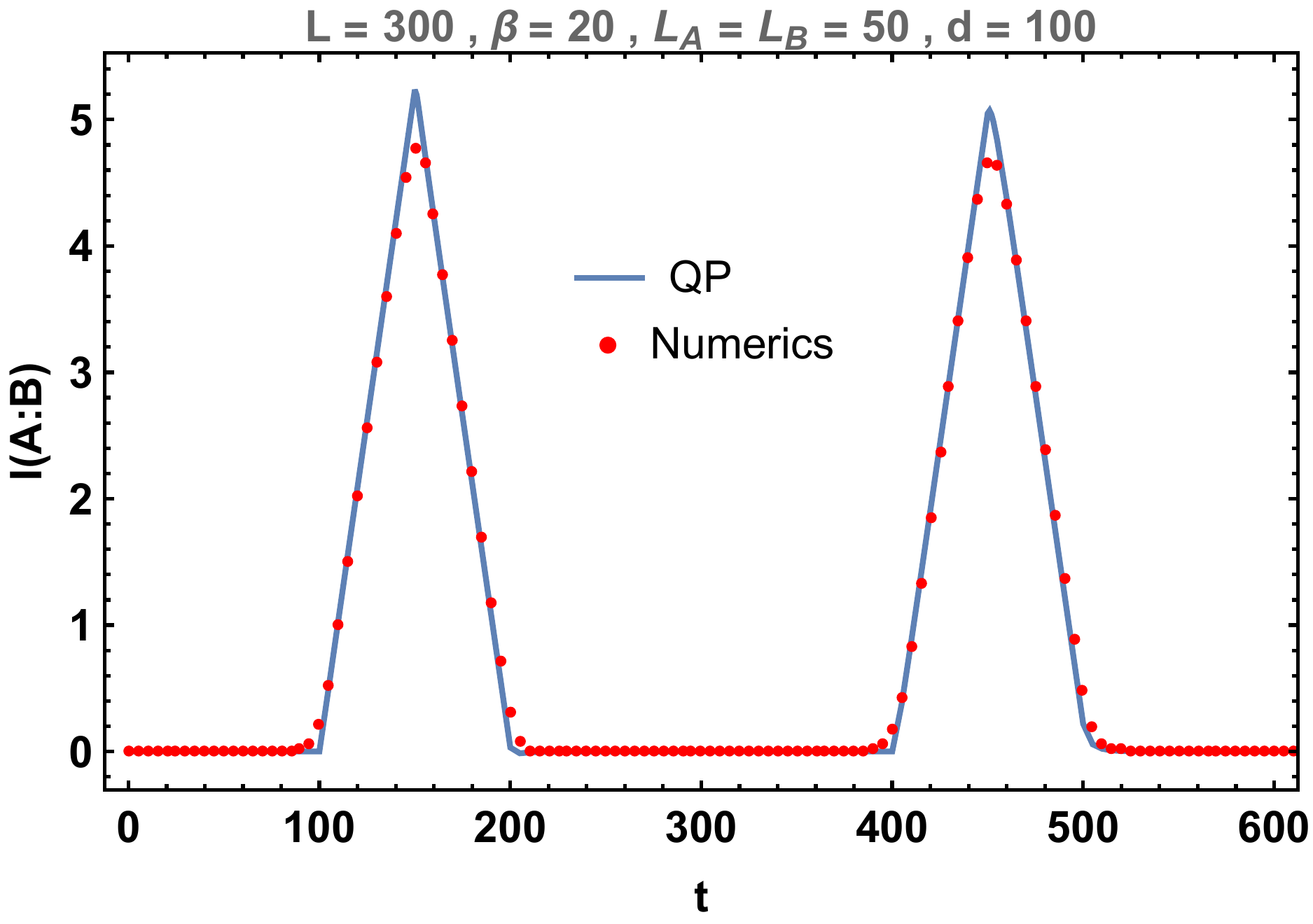}
\end{center}
\caption{Numerical results for free fermion theory (red dots) versus quasi-particle picture (solid blue curves). In the first row we present examples of symmetric configurations with $L_A=L_B$ (right) and $L_A<L_B$ (left). In the second row we present examples of the asymmetric configurations with $d_*=0$ (right) and $d_*>0$ (left). In the third row we show examples of disjoint configurations with $L_A=L_B$ (right) and $L_A<L_B$ (left).}
\label{fig:NQP}
\end{figure}

In this section, we numerically study mutual information for free fermionic TFD states and compare the numerical results with the predictions of quasi-particle picture. The Hamiltonian is given by  
\be
H=\sum_{k=0}^{N-1} \left(H^{(1)}_k+H^{(2)}_k\right)
\;\;\;\;\;,\;\;\;\;\;
H^{(i)}_k=\omega_k \hat{n}^{(i)}_k
\;\;\;\;\;,\;\;\;\;\;
\omega_k=\frac{1}{\delta}\cos\left(\frac{2\pi k}{N}\right),
\ee
where $\delta$ is the lattice spacing hence $N\,\delta=L$, $\hat{n}_k=c_k^\dagger c_k$ and the fermion operators obey standard anti-commutation relations $\{c_k,c_{k'}^\dagger\}=\delta_{kk'}$. $N$ is the number of sites in the lattice.  The fermionic TFD state with the inverse temperature $\beta$ is defined as
\be
|\beta\rangle=\bigotimes_{k=1}^N \left(1+e^{-\beta\omega_k}\right)^{-\frac{1}{2}}\left(|0\rangle_1|0\rangle_2+e^{-\beta\omega_k/2}e^{-i\omega_k t}|1\rangle_1|1\rangle_2\right).
\ee
We use the covariance matrix method formalism to compute the entanglement and mutual information. Considering a union of two subregions which we denote by $A$ and $B$ containing $N_A$ and $N_B$ sites. We define the following $2(N_A+N_B)$ dimensional vector
\be\label{eq:covvec}
r=\left(q_1,q_2,\cdots,q_{N_A},q_{N_A+1},\cdots,q_{N_A+N_B},p_1,p_2,\cdots,p_{N_A},p_{N_A+1},\cdots,p_{N_A+N_B}\right),
\ee
where $q_A$ to $q_{N_A}$ corresponds to $c_i$'s in region $A$ and from $q_{N_A+1}$ to $q_{N_A+N_B}$ correspond to those in $B$ followed by the $p$'s corresponding to $c^\dagger$ operators in the same ordering. To compute the spectrum of the reduced density matrix, we need to construct the covariance matrix
\be
\Gamma=\begin{pmatrix}
Q&R\\
{R}^T&P
\end{pmatrix}
\ee
where $Q$ and $P$ and $R$ are $(N_A+N_B)$-dimensional square matrices given by
\be\label{eq:Gamma}
\Gamma_{ij}=\frac{1}{2}\langle\beta|\left\{r_i,r_j\right\}|\beta\rangle.
\ee
The covariance matrix $\Gamma$ contains different blocks as 
\be
\Gamma=\begin{pmatrix}
\Gamma^{(cc)}&\Gamma^{(c c^\dagger)}\\
-{\Gamma^{(c c^\dagger)}}^T&\Gamma^{(c^\dagger c^\dagger)}
\end{pmatrix}
\ee
where each block has the following structure
\be
\Gamma^{(cc)}=\begin{pmatrix}
\Gamma^{(cc)}_{11}&\Gamma^{(cc)}_{12}\\
-{\Gamma^{(cc)}}_{12}^T&\Gamma^{(cc)}_{22}
\end{pmatrix}
\ee
with the similar structure for other blocks. It is easy to check that the components of these blocks for the diagonal blocks are given by
\begin{align}
\begin{split}
\Gamma^{(cc)}_{11}&=0
\;\;\;\;\;\;\;\;\;\;\;\;\;\;\;\;\;\;,\;\;\;\;\;\;\;
\Gamma^{(c^\dagger c^\dagger)}_{11}=0
\\
{\Gamma^{(c c^\dagger)}_{11}}_{ij}
&=
\frac{1}{L}\sum_{k=0}^{N-1}\tanh\frac{\beta\omega_k}{2}\cos\left(\frac{2\pi k(i-j)}{L}\right)
\end{split}
\end{align}
and for the off-diagonal blocks are given by 
\begin{align}
\begin{split}
{\Gamma^{(cc)}_{12}}_{ij}&=-{\Gamma^{(c^\dagger c^\dagger)}_{12}}_{ij}=
\frac{1}{L}\sum_{k=0}^{N-1}\mathrm{sech}\frac{\beta\omega_k}{2}\cos(\omega_kt)\cos\left(\frac{2\pi k(i-j)}{L}\right)
\\
{\Gamma^{(c c^\dagger)}_{12}}_{ij}&=
\frac{1}{L}\sum_{k=0}^{N-1}\mathrm{sech}\frac{\beta\omega_k}{2}\sin(\omega_kt)\cos\left(\frac{2\pi k(i-j)}{L}\right)
\end{split}
\end{align}
With these in hand, the spectrum of $i\Gamma$ denoted by $\{\pm\nu_i\}$, gives a double copy of the spectrum of the reduced density matrix that gives the entropy as \be
S_{A_1\cup A_2}=-\sum_{i=1}^{N_{A_1}+N_{A_2}}\left[\frac{\nu_i+1}{2}\ln\left(\frac{\nu_i+1}{2}\right)+\frac{\nu_i-1}{2}\ln\left(\frac{\nu_i-1}{2}\right)\right]
\ee
In Figure \ref{fig:NQP} we have presented the quasi-particle results versus numerical results corresponding to six examples showing perfect match. 

Using the same formalism, namely implementing the corresponding entropy density for bosonic theories in \eqref{eq:ThEntDen} into the same quasi-particle formulae leads to the quasi-particle prediction for bosonic theories. There are subtleties in this case due to the extremely large effect of the zero mode which is not captured by the quasi-particle picture. More details about this case is discussed in appendix \ref{app:boson}.

\subsection{Holographic CFT}
The goal of this section is to study OMI in holographic channels. To this end, we perform holographic calculations of entanglement entropy which form the main subject of this section. 
\subsubsection{Warmup: Holographic entanglement entropy}
The purpose of the following section is to review these calculations in a pedagogical fashion. The reader who is familiar with these holographic calculations may skip to the next subsection. 

\subsubsection*{BTZ blackhole in embedding coordinates}
In two-dimensional holographic CFTs, minimal surfaces in the dual geometry are simply minimal geodesics. Let us first demonstrate a useful way to calculate these geodesic lengths using the embedding formalism.

It is well-known that vacuum solutions in three-dimensional AdS are locally equivalent to empty AdS. Based on this fact, the geodesic length can be calculated via the embedding coordinates, 
\begin{align}
\mathrm{d}s^2&=-\mathrm{d}u^2-\mathrm{d}v^2+\mathrm{d}x^2+\mathrm{d}y^2, \\
    U^2&=\eta_{AB}U^AU^B=-u^2-v^2+x^2+y^2=-1.
\end{align}
We are particularly interested in the high-temperature or high-energy limit; hence, it is sufficient to consider the BTZ black hole \cite{Banados:1992wn}. 
In these embedding coordinates, the BTZ black hole can be expressed as
\begin{align}
    u&=\frac{r}{r_+}\cosh\left(r_+\phi\right),\\
    v&=\frac{\sqrt{r^2-r_+^2}}{r_+}\sinh\left(r_+t\right),\\
    x&=\frac{r}{r_+}\sinh\left(r_+\phi\right),\\
    y&=\frac{\sqrt{r^2-r_+^2}}{r_+}\cosh\left(r_+t\right).
\end{align}
with the identification $\phi\sim\phi+2\pi$, namely, 
\begin{align}
    u&\sim u\cosh\left(2\pi r_+\right)+x\sinh\left(2\pi r_+\right), \label{eq:identification_BTZ1}\\
    x&\sim x\cosh\left(2\pi r_+\right)+u\sinh\left(2\pi r_+\right). \label{eq:identification_BTZ2}
\end{align}
Then, the induced metric is given by the standard BTZ coordinates,
\begin{equation}
    ds^2=-(r^2-r_+^2)dt^2+\frac{dr^2}{r^2-r_+^2}+r^2d\phi^2.
\end{equation}
We need to identify the inverse temperature of the BTZ black hole as $ \beta=2\pi/r_+$ in order to obtain a smooth geometry near $r=r_+$ when we analytically continue Lorentzian time $t$ to Euclidean time. 

It is worth noting that the above embedding coordinates are invariant under the parameter rescaling $\phi\rightarrow R \phi$, $r\rightarrow r/R$ and $r_+\rightarrow r_+/R$. Hence, the inverse temperature can be rescaled as $\beta\rightarrow R\beta$. This rescaling corresponds to a boundary scale transformation. Therefore, if we consider a CFT on a torus with spatial circumference $2\pi R$, the inverse temperature should be identified with the blackhole radius as $\beta=2\pi R/r_+$\footnote{We also have to rescale the cutoff $r_\infty\rightarrow r_\infty/R$.}. 

The lengths of spacelike geodesics in the embedding coordinates have a simple expression
\begin{align}
   \sigma(U_1, U_2)= \cosh^{-1}(-U_1\cdot U_2),
\end{align}
where $U_{1,2}$ are the end points of the geodesic (the derivation can be seen in \cite{Bengtsson:1992, Anegawa:2020lzw}, for example). The above expression may correspond to  non-minimal geodesics because we have non-trivial identifications \eqref{eq:identification_BTZ1} and \eqref{eq:identification_BTZ2}. In order to compute the area of the minimal surface, the geodesic with minimal length has to be selected. 
\subsubsection*{Single interval}
First, let us look at the single interval case. We take the subsystem $A$ on the boundary on a fixed time slice to be $[X_2,X_1]$. The corresponding minimal surfaces should be anchored at two points $U_2$ and $U_1$ which are located on the cutoff surface at $r=\tilde{r}_\infty$ with the corresponding $X$ and $t$ coordinates determined by
\begin{align}
-U_1\cdot U_2=\dfrac{2\tilde{r}^2_\infty}{\tilde{r}_+^2}\sinh\left[\frac{\tilde{r}_+}{2}(w_1-w_2+2\pi nR)\right]\sinh\left[\frac{\tilde{r}_+}{2}(\bar{w}_1-\bar{w}_2+2\pi nR)\right]+\mathcal{O}(\tilde{r}_\infty^0),
\end{align}
where we defined $w_i=X_i+t_i, \bar{w}_i=X_i-t_i$ so that $X_i\sim X_i+2\pi R$. Here, $\tilde{r}_\infty$ and  $\tilde{r}_+$ are the cutoff and horizon radius in the rescaled coordinates. Note that here we have bunch of local minimum labeled by integer $n$. As mentioned previously, this is a consequence of the non-trivial identifications \eqref{eq:identification_BTZ1} and \eqref{eq:identification_BTZ2}. The magnitude of $n$ corresponds to the winding of the geodesics around the blackhole.

\begin{figure}[t]
\begin{center}
\includegraphics[scale=0.25]{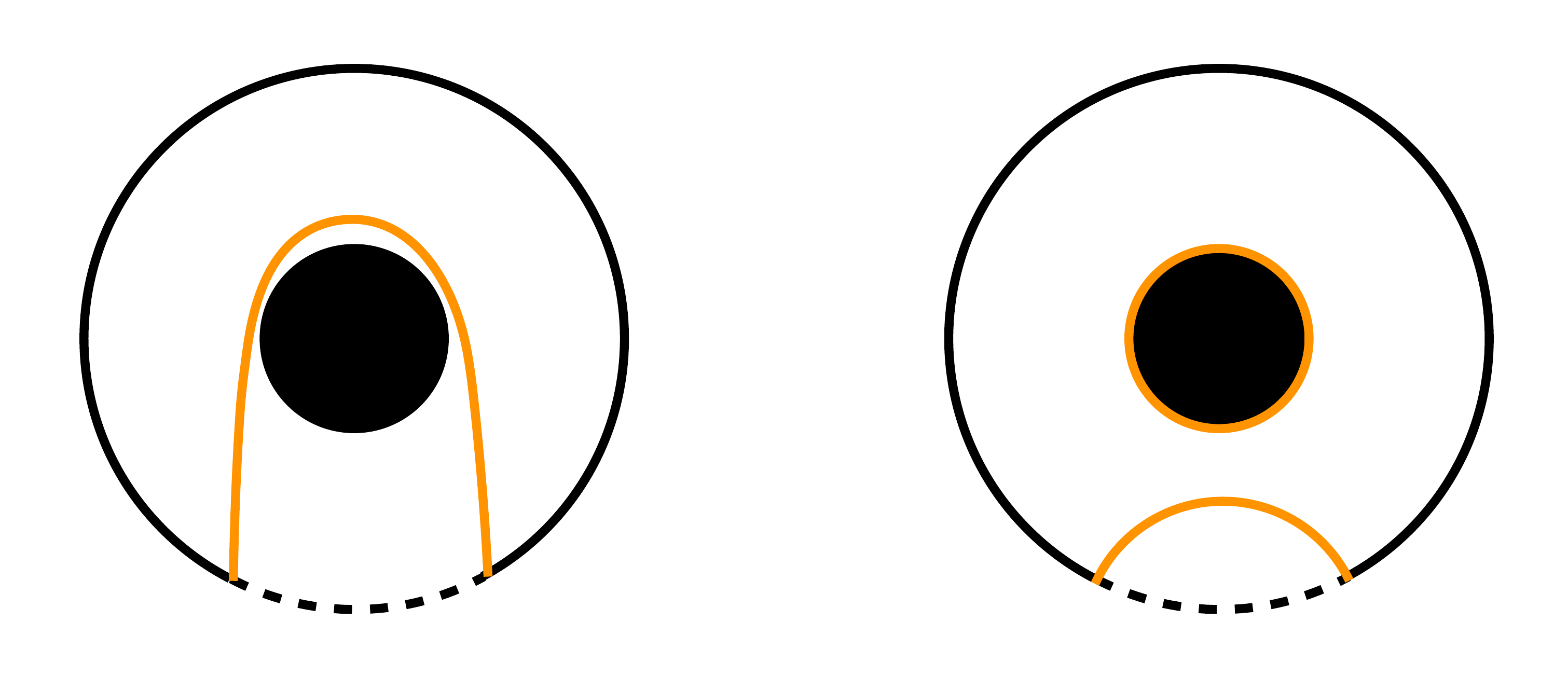}
\end{center}
\caption{Candidate Ryu-Takayanagi surfaces (orange curves) in the calculation of holographic entanglement entropy. The dashed lines on each boundary circle represent subsystem $A$. Each black circle on the center describes the black hole. In particular, in the right panel, we have the minimal surface wrapping the black hole. The contribution from this surface explains the black hole entropy $S_{BH}$. }\label{fig:btz_single}
\end{figure}

We rewrite the formula in terms of CFT parameters \cite{Brown:1986nw} as
\begin{align}
\dfrac{\sigma(U_2, U_1)}{4G_N}&\simeq\frac{c}{6}\log[-2U_2\cdot U_1]\\
&=\frac{c}{6}\log\left[\frac{\beta^2}{a^2\pi^2}\sinh\left[\frac{\pi(w_1-w_2+2\pi nR)}{\beta}\right]\sinh\left[\frac{\pi(\bar{w}_1-\bar{w}_2+2\pi nR)}{\beta}\right]\right],
\end{align}
 where $a$ is a lattice cutoff in the $2$d CFT with a spatial circumference of $2\pi R$.
Therefore, the holographic entanglement entropy is
\begin{align}
    S_{A}&=\min\left[\dfrac{\sigma(U_2, U_1)}{4G_N}\right]\nonumber\\
    &=\begin{cases}
     \frac{c}{3}\log\left[\frac{\beta}{a\pi}\sinh\left[\frac{\pi(X_1-X_2)}{\beta}\right]\right] & (X_1-X_2<\ell_{\text{cl.}}), \vspace{3mm}\\
     \frac{c}{3}\log\left[\frac{\beta}{a\pi}\sinh\left[\frac{\pi(2\pi R-(X_1-X_2))}{\beta}\right]\right]+S_{BH}  & (X_1-X_2>\ell_{\text{cl.}}),
    \end{cases}
\end{align}
where $S_{BH}$ is Bekenstein-Hawking entropy,
\be
S_{BH}=\dfrac{c\pi}{3}\dfrac{2\pi R}{\beta},
\ee
and $\ell_{\text{cl.}}$ is a critical length \cite{Bao:2017guc} given by
\be
\ell_{\text{cl.}}=\dfrac{\beta}{2\pi}\log\left[\dfrac{1+\mathrm{e}^{\frac{2\pi }{\beta}2\pi R}}{2}\right].
\ee
See Figure \ref{fig:btz_single} for an illustration of the geodesics. 
Later, we will take $\beta=2\epsilon$ and focus on the high temperature limit $\epsilon\rightarrow0$. In this limit, the second phase becomes negligible as $\ell_{\text{cl.}}\rightarrow2\pi R$. Namely, the contribution of $S_{BH}$ is a consequence of a mixed state and does not show up if we see pure states \cite{Asplund:2014coa,Kusuki:2019rbk,Kusuki:2019evw}. It is also worth noting that if we consider a (typical) pure state, the high temperature limit of $S_{A}$ follows the Page curve with respect to the subsystem size $L_A=X_1-X_2$. The calculation for the subsystem $B=[Y_2,Y_1]$ proceeds in an identical fashion. 
\subsubsection*{Two disjoint intervals on different boundaries}
\begin{figure}[t]
\begin{center}
\includegraphics[scale=0.25]{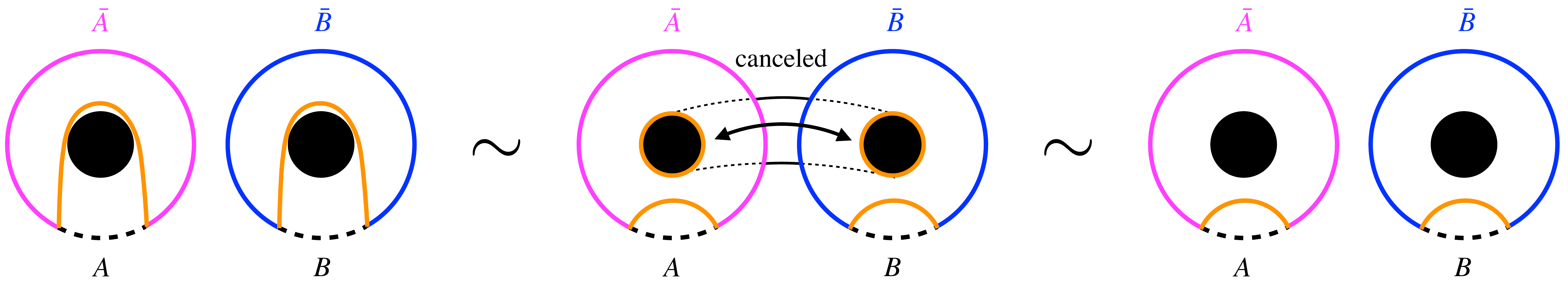}
\end{center}
\caption{We illustrate a deformation of RT surfaces in the disconnected phase $S_{\text{dis.}}$  (orange curves) where each of these surfaces belong to an equivalent homology class. The dashed line on the boundary circles correspond to subsystems $A$ and $B$. In particular, the geodesics winding around the black hole can be canceled out through the wormhole (dotted tube in the middle panel). Note that we cannot do this for the calculation of $S_A$ or $S_B$ as we have no access to the black hole interior. Each face represents the surface of the wormhole cylinder as seen from the outside, and the orientation of the RT surfaces in these figures is determined with respect to that case.}\label{fig:ch}
\end{figure}

We are interested in the operator mutual information which can be holographically computed as the area of minimal surfaces in a two-sided AdS black hole. In addition to the minimal surfaces we have seen so far, there is a new phase that requires spacelike geodesics that connect the two different asymptotic boundaries \cite{HM}. As in the field theory side, we can obtain the coordinates of the other boundary by shifting the time coordinate $t\rightarrow t+i\beta/2$.
As mentioned previously, the holographic entanglement entropy for two disjoint interval has two phases,
\begin{align}
    S_{AB}=\min[S_{\text{con.}}, S_{\text{dis.}}],
\end{align}
where 
\begin{align}
    S_{\text{con.}}&=S(X_1,Y_1)+S(X_2,Y_2),\\
    S(X,Y)&=\frac{c}{6}\log\left[\frac{\beta^2}{\pi^2a^2}\cosh \left[\frac{\pi  (X-Y+t_X-t_Y)}{\beta}\right] \cosh \left[\frac{\pi  (X-Y-(t_X-t_Y))}{\beta }\right]\right]\\
    &=\frac{c}{6}\log\left[\frac{\beta^2}{2\pi^2 a^2}\left(\cosh \left[\frac{2 \pi  (t_X-t_Y)}{\beta }\right]+\cosh \left[\frac{2 \pi  (X-Y)}{\beta }\right]\right)\right].
\end{align}
and $S_{\text{dis.}}$ is roughly given by $S_A+S_B$ but now we do not include any $S_{BH}$ contributions. This is because our entire TFD state is a pure state, therefore  $S_{A\cup B}=S_{\overline{A\cup B}}$. To be precise, this can be understood as a consequence of the homology condition of the RT formula (see Figure \ref{fig:ch} for details\footnote{Again, we can see the similar Page curve as in the single interval setup mentioned above since the total systems is a pure state. Interestingly, in the present setup, the appearance of a (spatial) wormhole inside the entanglement wedge is crucial to explain the decrease in $S_{AB}$. (The wormhole plays a role of ``purifier''.)}
. We will call $S_{A\cup B}=S_{\text{con.}}$ the connected phase and  $S_{A\cup B}=S_{\text{dis.}}$ the disconnected phase because the entanglement wedges between $A$ and $B$ in these phases are connected and disconnected respectively. Note that the minimal surface in the connected phase consists of geodesics connecting two different asymptotic boundary points. 
\subsubsection{Holographic Bipartite Operator Mutual Information}\label{subsec:HBOMI}
 We study the holographic BOMI and explore the extent to which the quasi-particle picture is valid in the holographic system. To this end, we will look at a series of progressively complicated setups. In what follows, we will mainly take the high-temperature limit, {\it i.e.} $\beta\equiv2\epsilon\rightarrow0$. 
 We will take input subsystem to be  $A=\{X|X_2<X<X_1\}$ and the output subsystem to be $B=\{X|Y_2<X<Y_1\}$.
\subsubsection*{Symmetric setup ($A=B$)}

Let us first set $X_1=Y_1$ and $X_2=Y_2$ for simplicity. First, we assume that $0<X_1-X_2<\pi R$ so that each subsystem is smaller than total system size $2\pi R$.

Then, the connected pieces are given by
\begin{align}
S_{\text{con.}}=\frac{c\pi t}{3\epsilon}+\frac{2c}{3}\log\left[\frac{2\epsilon}{a\pi}\right].
\end{align}
On the other hand, the disconnected phase gives
\begin{align}
   S_{\text{dis.}}= \frac{c\pi (X_1-X_2)}{3\epsilon}+\frac{2c}{3}\log\left[\frac{2\epsilon}{a\pi}\right]. 
\end{align}

Therefore, we obtain
\begin{align}
    I(A:B)=\begin{cases}
    \frac{c\pi}{3\epsilon}(X_1-X_2-t) & (0<t<X_1-X_2), \\
    0 & (X_1-X_2<t).
    \end{cases}
    \;\;\;(0<X_1-X_2<\pi R)
\end{align}
However, if $X_1-X_2>\pi R$, there are deviations between $S_A+S_B$ and $S_{AB}$ at late times. This happens because $S_A$ and $S_B$ remain in the original phase\footnote{Keep in mind that we have homology constraints for $S_A$ and $S_B$. We also took the high-temperature limit which makes the phase transition impossible.}, whereas the replacement of $X_1-X_2$ with $2\pi R-(X_1-X_2)$ should be made in $S_{A\cup B}$. Note that $S_{\text{con.}}$ has no contributions from the black hole entropy. Therefore, we obtain
\begin{align}
    I(A:B)=\begin{cases}
    \frac{c\pi}{3\epsilon}(X_1-X_2-t) & (0<t<2\pi R-(X_1-X_2)), \\
    \frac{2c\pi}{3\epsilon}(X_1-X_2-\pi R) & (2\pi R-(X_1-X_2)<t).
    \end{cases}
    \;\; (\pi R<X_1-X_2<2\pi R)
\end{align}

The results in both cases do not show any  quantum revival that is characterized by the scale of the total system size. 
\subsubsection*{Asymmetric setup ($A\subset B$ or $A\supset B$)}
Next, we consider the case where $Y_1>X_1>X_2=Y_2$; namely, the input system $A$ is a proper subset of the output system $B$, $A\subset B$ (after identification of the different Euclidean time slices). One can easily obtain results for the $X_1>Y_1>X_2=Y_2$ case (that is, $A\supset B$) by exchanging all $X$ and $Y$ coordinates in what follows.

We will once again take the high-temperature limit; hence, $S_A$ and $S_B$ are uniquely determined. 
Depending on the time evolution, we have to consider whether $S_{A\cup B}$ is in the connected or disconnected phase. If $S_{A\cup B}$ is in the disconnected phase, we have two choices of minimal surfaces as previously discussed. In the high temperature limit, this can be divided into two cases depending on the total length of our subsystem $A\cup B$. 

If $0<X_1-2X_2+Y_1<2\pi R$, we have
\begin{align}
S_{AB}=\dfrac{c\pi}{6\epsilon}\min\big[\max\left[t+Y_1-X_1,2t\right],X_1+Y_1-2X_2\big]+S_{\text{div.}},
\end{align}
where 
\begin{align}
S_{\text{div.}}=\dfrac{c}{3}\log\left(\frac{\epsilon^2}{2\pi^2a^2}\right).
\end{align}
The first two possibilities within $\max[\cdots]$ come from the same connected phase and are the result of the high temperature limit. Therefore, we obtain
\begin{align}
    I(A:B)=\begin{cases}\begin{cases}
    \frac{c\pi}{6\epsilon}(2L_A-t) & (0<t<2L_A), \\
    0 & (2L_A<t).
    \end{cases} & (L_B>3L_A),\vspace{3mm}\\
    \begin{cases}
    \frac{c\pi}{6\epsilon}(2L_A-t) &  (0<t<L_B-L_A),\\
    \frac{c\pi}{6\epsilon}(L_A+L_B-2t) & (L_B-L_A<t<\frac{L_A+L_B}{2}), \\
    0 & (\frac{L_A+L_B}{2}<t).
    \end{cases}& (L_B<3L_A), \label{eq:asm1}
    \end{cases}
\end{align}
where $L_A=X_1-X_2$ and $L_B=Y_1-X_2$. 

If $2\pi R<X_1-2X_2+Y_1<4\pi R$, we have
\begin{align}
S_{A\cup B}=\dfrac{c\pi}{6\epsilon}\min\big[\max\left[t+Y_1-X_1,2t\right], 4\pi R-(X_1+Y_1-2X_2)\big]+S_{\text{div.}}.
\end{align}
Therefore,
\begin{align}
    I(A:B)=\begin{cases}\begin{cases}
    \frac{c\pi}{6\epsilon}(2L_A-t) & (0<t<L), \\
    \frac{c\pi}{3\epsilon}(L_A-\bar{L}_B) & (2\bar{L}_B<t).
    \end{cases} & (Y_1-X_1>2\bar{L}_B),\vspace{3mm}\\
    \begin{cases}
    \frac{c\pi}{6\epsilon}(2L_A-t) &  (0<t<L_B-L_A),\\
    \frac{c\pi}{6\epsilon}(L_A+L_B-2t) & (Y_1-X_1<t<\frac{2\bar{L}_B+L_B-L_A}{2}), \\
    \frac{c\pi}{3\epsilon}(L_A-\bar{L}_B) & (\frac{2\bar{L}_B+L_B-L_A}{2}<t).
    \end{cases}& (Y_1-X_1<2\bar{L}_B),
    \end{cases} \label{eq:asm2}
\end{align}
where $\bar{L}_B\equiv 2\pi R-L_B$.

\subsubsection*{No overlap ($A\cap B=\emptyset$)}
Next, we consider the case where $X_2<X_1<Y_2<Y_1$. In other words, the two subsystems $A$ and $B$ have no overlap. 
This assumption leads to $0<X_1-X_2+Y_1-Y_2<2\pi R$ and we have
\begin{align}
S_{A\cup B}=\dfrac{c\pi}{6\epsilon}\min\bigg[\max[Y_1-X_1+Y_2-X_2, t+Y_1-X_1&, t+Y_2-X_2,2t] \nn\\
&,X_1-X_2+Y_1-Y_2\bigg]+S_{\text{div.}},
\end{align}
In this case, $S_{A\cup B}$ always lies in the disconnected phase and hence
\begin{align}
I(A:B)=0. \label{eq:disj}
\end{align}

\subsubsection{Holographic Tripartite Operator Mutual Information}\label{subsec:HTOMI}
Finally, we would like to evaluate the tripartite mutual information defined as
\begin{align}
    I_3(A:B_1:B_2)=I(A:B_1)+I(A:B_2)-I(A:B_1\cup B_2).
\end{align}
We will take $A=\{X|\,0<X<X_1=L_A\}$, $B_1=\{X|\,0<X<Y_1=L_{B_1}\}$, and $B_2=\{X|\,L_{B_1}=Y_1<X<2\pi R\}$. 
It will be useful later on to note that
\begin{align}
    I(A:B_1\cup B_2)&=\dfrac{c\pi L_A}{3\epsilon}.
\end{align}
For the remaining calculation of $I(A:B_1)$ and $I(A:B_2)$, we can reuse the results derived in section \ref{subsec:HBOMI}. 
First, we will focus on the simplest setup where $A$ and $B_2$ do not overlap. Then, we will briefly discuss the setup $A$ and $B_2$ do overlap. 
\subsubsection*{Small subsystems ($0<L_A+L_{B_1}<2\pi R$) with $A\cap B_2=\emptyset$}
If $X_1<Y_1$, \textit{i.e.} $A\cap B_2=\emptyset$, one can apply \eqref{eq:asm1} to $I(A:B_1)$ and \eqref{eq:disj} to $I(A:B_2)$. Therefore, we obtain
\begin{align}
    I_3(A:B_1:B_2)=\begin{cases}\begin{cases}
    \frac{c\pi}{6\epsilon}(-t) & (0<t<2L_A), \\
    \frac{c\pi}{3\epsilon}(-L_A) & (2L_A<t).
    \end{cases} & (L_{B_1}>3L_A),\vspace{3mm}\\
    \begin{cases}
    \frac{c\pi}{6\epsilon}(-t) &  (0<t<L_{B_1}-L_A),\\
    \frac{c\pi}{6\epsilon}(L_{B_1}-L_{A}-2t) & (L_{B_1}-L_A<t<\frac{L_A+L_{B_1}}{2}), \\
    \frac{c\pi}{3\epsilon}(-L_A) & (\frac{L_A+L_{B_1}}{2}<t).
    \end{cases}& (L_{B_1}<3L_A).
    \end{cases}
\end{align}

\subsubsection*{Large subsystems ($2\pi R<L_A+L_{B_1}<4\pi R$) with $A\cap B_2=\emptyset$}
 If $X_1<Y_1$,  \textit{i.e.} $A\cap B_2=\emptyset$, one can apply \eqref{eq:asm2} to $I(A:B_1)$ and \eqref{eq:disj} to $I(A:B_2)$. 
Therefore, we obtain
\begin{align}
    I_3(A:B_1:B_2)=\begin{cases}\begin{cases}
    \frac{c\pi}{6\epsilon}(-t) & (0<t<2L_{B_2}), \\
    \frac{c\pi}{3\epsilon}(-L_{B_2}) & (2L_{B_2}<t).
    \end{cases} & (\bar{L}_A>3L_{B_2}),\vspace{3mm}\\
    \begin{cases}
    \frac{c\pi}{6\epsilon}(-t) &  (0<t<L_{B_1}-L_A),\\
    \frac{c\pi}{6\epsilon}(-L_A+L_{B_1}-2t) & (L_{B_1}-L_A<t<\frac{2L_{B_2}+(L_{B_1}-L_A)}{2}), \\
    \frac{c\pi}{3\epsilon}(-L_{B_2}) & (\frac{2L_{B_2}+(L_{B_1}-L_A)}{2}<t),
    \end{cases}& (\bar{L}_A<3L_{B_2}),
    \end{cases}
\end{align}
where $\bar{L}_A=2\pi R-X_1$ and $L_{B_2}=2\pi R-Y_1>0$.

\subsubsection*{More general setup}
Finally, we discuss the case $X_1>Y_1$, \textit{i.e.} $A\cap B_2\neq\emptyset$ briefly. Since the case classification is complicated, we will only look at the late time limit. Let us consider the case with $0<Y_1<\pi R<X_1<2\pi R$ and $0<X_1+Y_1<2\pi R$ as an example. The late time limit of BOMI gives
\begin{align}
    I(A:B_1)&=0,\\ I(A:B_2)&=2S_A+2S_{B_2}-2S_{BH}=2S_A-2S_{B_1},\\ 
    I(A:B_1\cup B_2)&=2S_{A},
\end{align}
Therefore, we obtain
\begin{align}
    I_3(A:B_1:B_2)=-2S_{B_1},
\end{align}
as the late time value. Similarly, if $0<Y_1<\pi R<X_1<2\pi R$ with $2\pi R<X_1+Y_1<4\pi R$, we obtain
\begin{align}  I_3(A:B_1:B_2)=-2S_{\bar{A}}.
\end{align}
One can also consider the case with $\pi R<Y_1<X_1<2\pi R$ and we obtain the general expression as
\begin{align}
    I_3(A:B_1:B_2)=-2\min[S_A,S_{\bar{A}},S_{B_1},S_{B_2}]. 
\end{align}
%
\subsubsection{Summary of holographic CFT results}
In summary, we have found that OMI in holographic CFT does not display quantum revival. In particular, the time evolution of the holographic OMI cannot simply be explained by the quasiparticle picture. Strictly speaking, there are a few exceptions that show up in the early time behavior of BOMI in the symmetric and asymmetric setups. In these cases, the quasiparticle picture holds until the BOMI stops decreasing ({\it i.e.} before reaching the plateau regions), at which time the quasiparticle picture breaks down. The TOMI can be understood in the same way, since it is a linear combination of the BOMI. 

However, it does not mean that the OMI does not have finite size effects at all. Indeed, the BOMI $I(A:B)$ keeps a non-zero value when the total size of $A$ and $B$ is sufficiently large. Also, the TOMI $I_3(A:B_1:B_2)$ at late time is given by $-2S_X$ where $X$ is the smallest subsystem among $A$, $\bar{A}$, $B_1$, and $B_2$. This effect can be also understood as the Page curve argument, namely the ``finiteness'' of dimension of Hilbert space on the compact space. Although we are now considering field theories including UV divergences, it does not matter for regulated quantities such as the (operator) mutual information.  

It is also intriguing to note that this is in contrast with OTOC where we do not see such finite size effects. Therefore, the OMI is more sensitive to non-local effects in the information scrambling. 

In the next section, we will discuss the line-tension picture that describes the time-evolution of OMI in a chaotic system such as the holographic CFTs.

\subsection{Line-tension picture}\label{LineTensionPictureSection}
\begin{figure}[h]
\begin{center}
\includegraphics[scale=0.3]{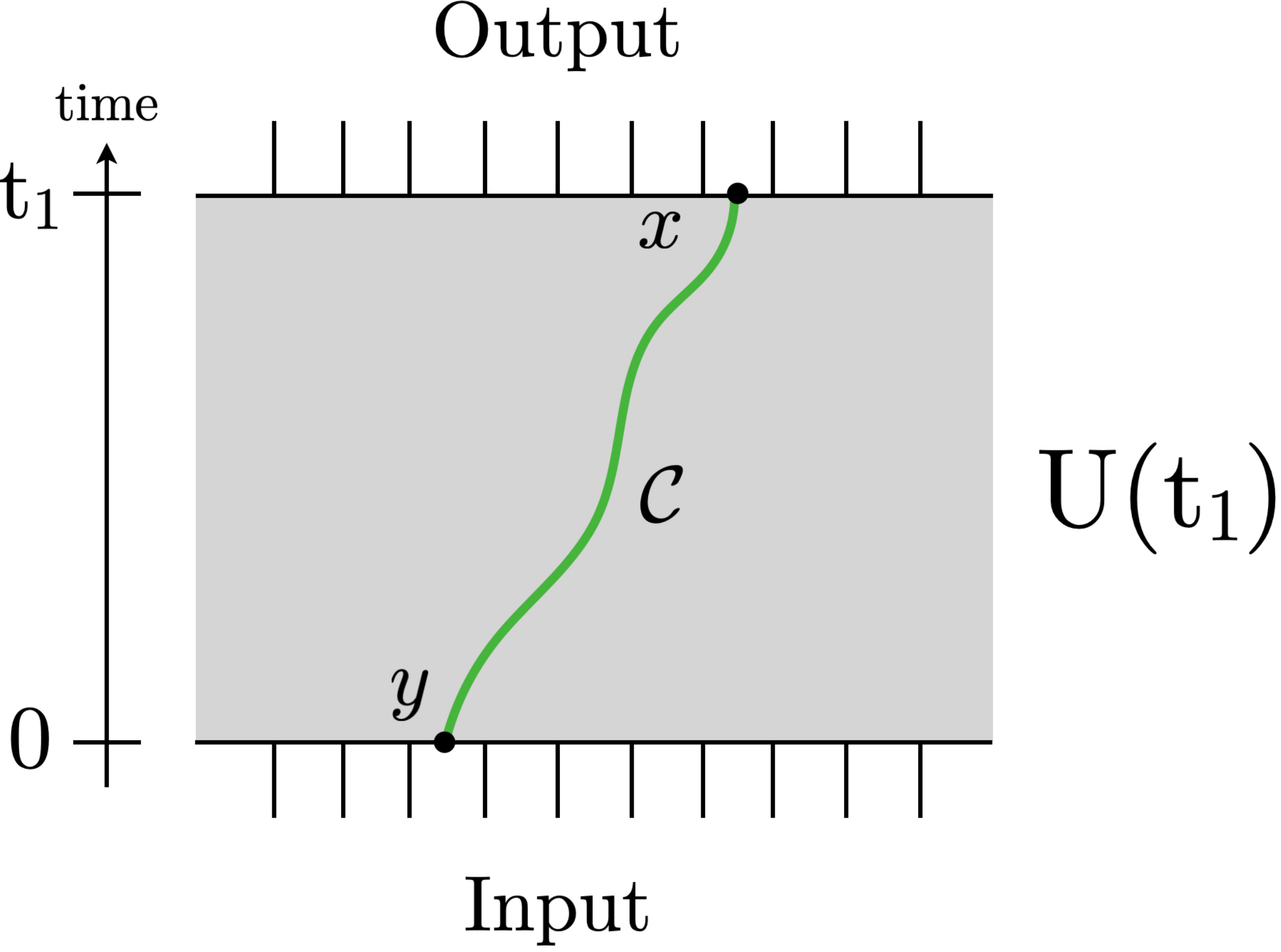}
\caption{The curve ${\cal C}$ that cuts the unitary circuit into two parts. In the line-tension picture, the entanglement entropy $S_U(x,y,t_1)$ is given by the integral of the line tension ${\cal T}(v)$ along the curve ${\cal C}$. }
\label{LT1}
\end{center}
\end{figure}
It is known that in an integrable system entanglement growth can be described by the spreading of quasiparticles. On the other hand, in a chaotic system, the hydrodynamics of entanglement production is well described by the ``line-tension picture'' introduced in \cite{2017PhRvX...7c1016N,
2018arXiv180300089J,Mezei:2018jco,PhysRevX.8.031058,PhysRevX.8.021013}.  We will briefly describe it in this section. 

The original line-tension picture was introduced in the (two-dimensional) chaotic system defined on an infinite line \cite{2017PhRvX...7c1016N}. We assume that the spatial direction is homogeneous. These assumptions can be relaxed, allowing for a compact spatial extent as well as spatial inhomogeneity \cite{SSD}, with the former being the main subject of this paper. For simplicity, we assume that the state is time-evolved by the unitary operator $U(t_1)$ from $t=0$ to $t=t_1$. We cut the infinite line where the system lives into two pieces at position $x$ at $t=t_1$. We also cut the line at position $y$ at $t=0$. The entanglement of the unitary operator $S_U(x,y,t_1)$ can be computed using the line tension picture.

The main ingredient of the line-tension picture is the ``line-tension'' ${\cal T}(v)$ associated to a curve ${\cal C}$ that connects the point $(x,t_1)$ and $(y,0)$ as in Figure \ref{LT1}. 
The curve ${\cal C}$  in spacetime has velocity $v=dx/dt$ and a line-tension ${\cal T}(v)$ that depends on $v$. In the coarse-grained limit, it counts the entanglement across the spacetime cut ${\cal C}$ through the unitary operator $U(t_1)$. This is related to the idea that the minimal cut through a tensor network provides an upper bound on the entanglement in the tensor network. Explicitly, the entanglement entropy $S_U(x,y,t_1)$ of the unitary operator is computed in  the line-tension picture as
\ba
S_U(x,y,t_1)={\rm min}_{\cal C}\int_{\cal C}dt\,  {\cal T}(v)\, ,
\ea
 where the minimization is taken over all the possible curves ${\cal C}$ that connects the point $(x,t_1)$ and $(y,0)$. In our case, where the spacetime is uniform, the minimal curve is given by a straight line with a constant velocity $v=(x-y)/t_1$.

The details of the function ${\cal T}(v)$ depend on the system but one can estimate it for a chaotic system using random unitary circuits. Random unitary circuits are toy models that illustrate the phenomena of quantum information scrambling in chaotic systems. In the scaling limit and in the limit of large bond dimension $q$, the line-tension is simply given by counting the number of bonds cut which is
\bal
\mathcal{T}(v)= \begin{cases}\log q & v<1 \\ v \log q & v>1\, .\end{cases}
\eal
To compute the entanglement of the unitary operator in a holographic CFT using the line-tension picture, we need to identify the bond dimension (the local Hilbert space dimension) $q$ in the holographic CFT. This can be accomplished by comparing the rates at which the information gets scrambled.
While the entanglement entropy grows at a rate of $\log q$ in random unitary circuits, it is known that in holographic CFTs the entanglement of the unitary operator (computed as the entanglement between two CFTs in the time-evolved thermofield double state) grows at a rate of $\f{c\pi}{3\beta}$. Here, $\beta$ is dimensionless with lattice spacing.
Therefore, we make the identification
\ba
q\sim e^{\f{c\pi}{3\beta}}\, .
\ea
Observe that $\log q$ simply corresponds to the entropy density given by the Cardy formula $S_{\rm Cardy}/(2\pi R)=\f{c\pi}{3\beta}$. Using this relation, one can correctly reproduce the growth of the entanglement in holographic CFTs which is
\ba
S_U(x,y,t_1)\sim \f{c\pi}{3\beta}t_1\, .
\ea
\begin{figure}[h]
\begin{center}
\includegraphics[scale=0.3]{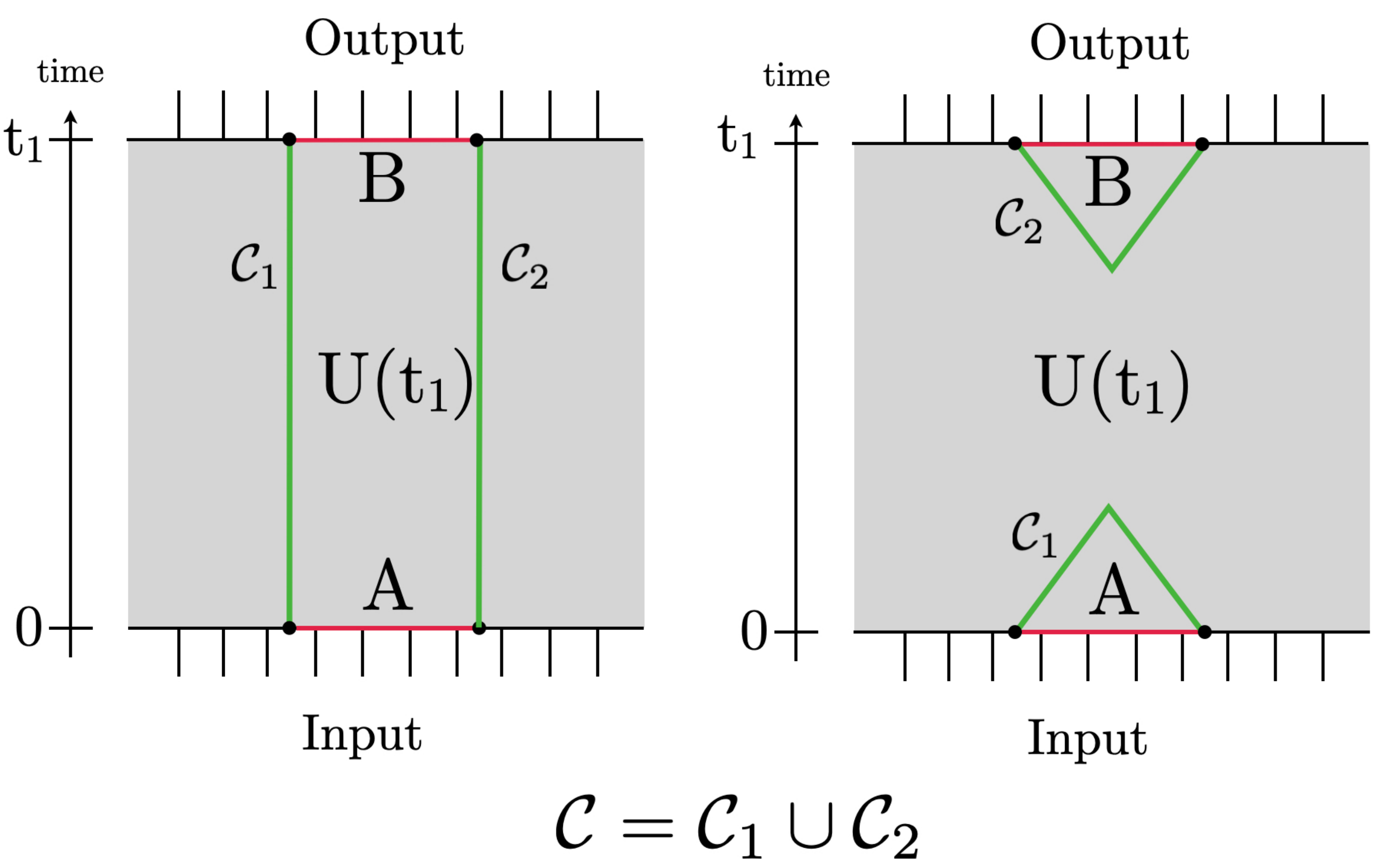}
\caption{Two candidates for the minimal curve that compute BOMI between the regions $A$ and $B$ that lie on the time slices at $t=0$ and $t=t_1$ respectively. The minimal curve ${\cal C}$ consists of two components ${\cal C}_1$ and ${\cal C}_2$. There are two possible configurations: a connected one (left) and a disconnected one (right). The connected configuration dominates at early times, while the disconnected one dominates at late times. This phase transition of the minimal curve is precisely the same kind of phase transition undergone by the Ryu-Takayanagi surface that computes the entanglement entropy in a two-dimensional holographic CFT.}
\label{LT2}
\end{center}
\end{figure}
A remarkable fact of the line-tension picture is that it can also reproduce a phase transition of the entanglement in chaotic systems. To see this, we consider a finite subregion $A$ with length $\ell$ at $t=0$ and and another subsystem $B$ at $t=t_1$. The minimal curve consists of two disconnected components that are anchored at the edges of the intervals $A$ at  $t=0$ and $B$ at $t=t_1$.
There are two possible configurations for the minimal curve, the connected phase where each curve stretches from $t=0$ to $t=t_1$, and the disconnected phase, where each curve is homologous to $A$ or $B$ respectively. Therefore the entanglement entropy is given by 
\ba
S_U(A,t_1)=\min\biggl\{ \f{c\pi}{3\beta}t_1,  \f{c\pi}{3\beta}\ell\biggl\}.
\ea

At early times $t_1<\ell$, the connected phase is dominant and the entanglement grows linearly in time. There is a phase transition at time $t_1=\ell$, after which the disconnected phase becomes dominant. This phase transition of the minimal curve is precisely the same phenomenon as the phase transition of the Ryu-Takayanagi surface that computes the entanglement entropy in a two-dimensional holographic CFT using holography. The line-tension picture correctly gives the leading order coarse-grained behavior of the entanglement entropy in a two-dimensional holographic CFT in the scaling limit.

\subsubsection{Line-tension picture on a compact space}
\begin{figure}[h]
\begin{center}
\includegraphics[scale=0.4]{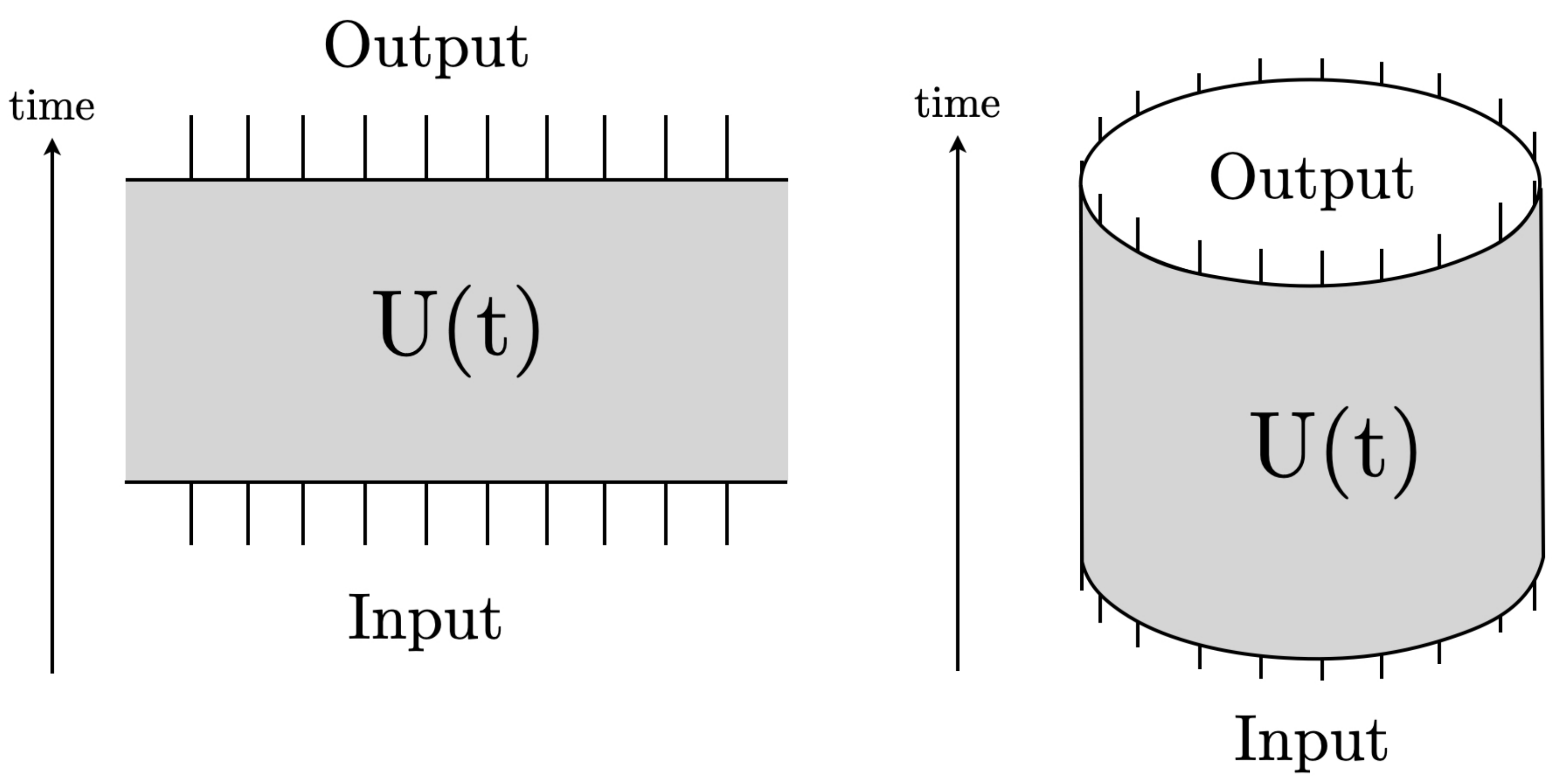}
\caption{Unitary circuit on a non-compact space (left) and a compact space (right).}
\label{UnitaryC}
\end{center}
\end{figure}
\begin{figure}[h]
\begin{center}
\includegraphics[scale=0.4]{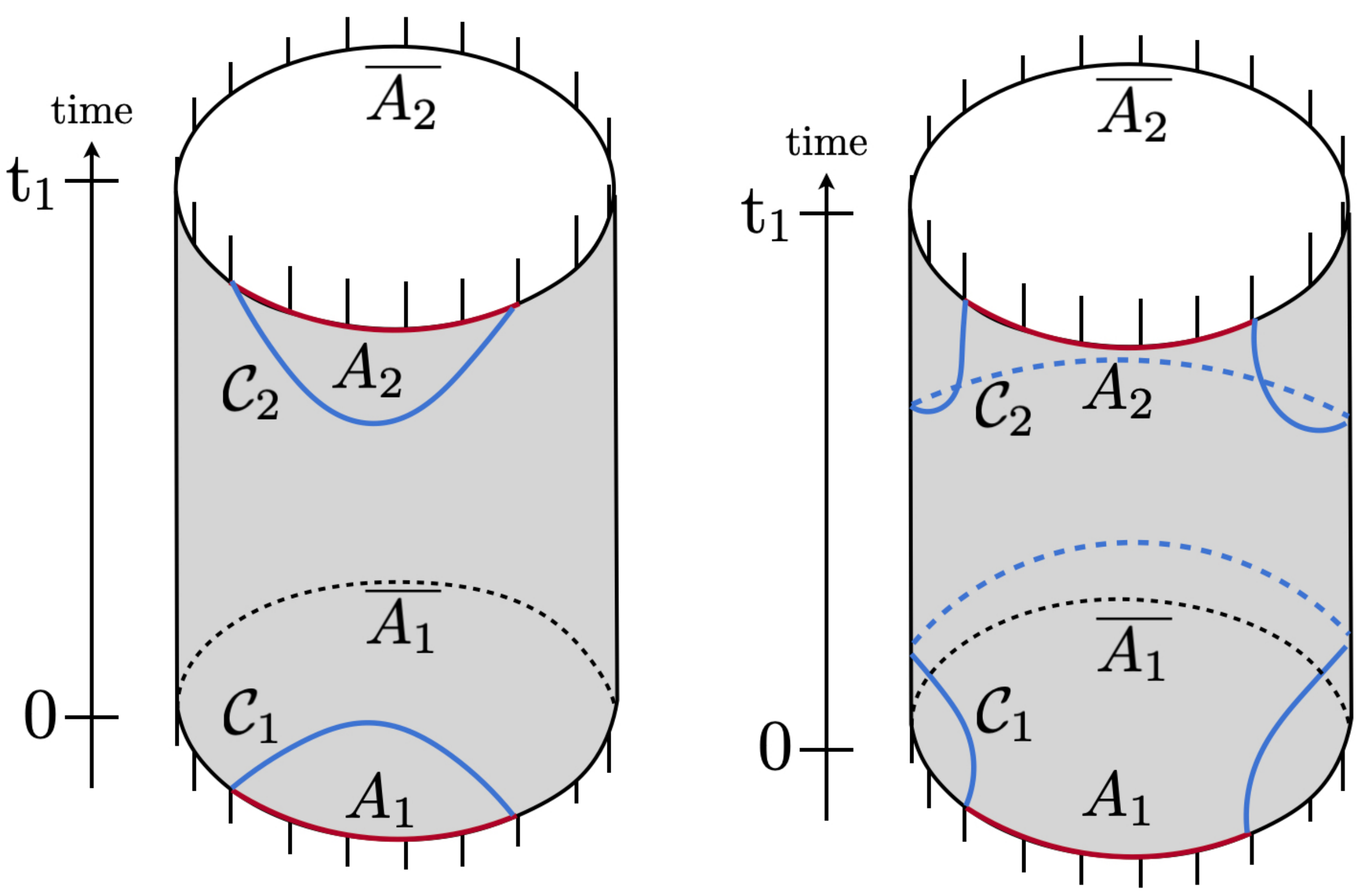}
\caption{Two candidates for the minimal curve in the disconnected phase: the homologically trivial configuration (left) and the homologically non-trivial configuration (right). }
\label{UnitaryLines}
\end{center}
\end{figure}

The original line-tension picture was proposed for a system on an infinite line. In this section, we generalize it to systems on compact spaces. In this case, the unitary circuits that represents the unitary operator $U(t)$ are defined on the compact space depicted in Figure\ref{UnitaryC}. We consider the curves ${\cal C}$ that cut the unitary circuit and compute the line-tension ${\cal T}(v)$ associated to that curve just as in the non-compact case. Since ${\cal T}(v)$ counts the {\it local} density of the entanglement across the cut ${\cal C}$ through the unitary operator $U(t)$, the basic prescription is {\it locally} the same as in a system on a non-compact space; we integrate the line tension over the curve ${\cal C}$ to obtain the entanglement. On the other hand, the minimality condition imposed for the curve ${\cal C}$ depends on the global structure. To explain that, let us turn our attention to the left panel of Figure \ref{UnitaryLines}. We take the intervals $A_1$ and $A_2$ on the time slices at $t=0$ and $t=t_1$ respectively, and denote $A=A_1\cup A_2$. We  compute the operator entanglement for the region $A$ in the line-tension picture. The minimal curve consists of two disconnected components ${\cal C}_1$ and ${\cal C}_2$ that are anchored at the edges of the intervals $A_1$ and $A_2$ respectively. We focus on the disconnected phase, where one component is anchored at the edges of $A_1$ and the other at the edges of $A_2$ respectively. In the case of a compact space, one can consider two possible configurations ${\cal C}={\cal C}_1\cup{\cal C}_2$ as shown in Figure \ref{UnitaryLines}. One possible configuration is that ${\cal C}_1$ is homologous to $A_1$, and   ${\cal C}_2$ is homologous to $A_2$ (left figure in Figure \ref{UnitaryLines}). Another configuration is that  ${\cal C}_1$ is homologous to $\bar{A}_1$, the complement region at $t=t_1$, and ${\cal C}_2$ is homologous to $\bar{A}_2$, the complement region at $t=0$, respectively (right figure in Figure \ref{UnitaryLines}). One might also consider the case where ${\cal C}_1$ is homologous to $A_1$ while the other is homologous to $\bar{A_2}$, or vice versa, but in our prescription, we do not consider such configurations as candidates for the minimal curve. 

More precisely, our definition of the unitary operator entanglement on a compact space  is 
\begin{align}
S_A=\underset{\substack{ {\cal C}\underset{\rm hom}{\sim} A
}}{\min}\int_{{\cal C}} dt\, {\cal T}(v)\, ,
\end{align}
with the line-tension given by
\begin{align}
\mathcal{T}(v)=\left\{\begin{array}{ll}\log q & v<1 \\ v \log q & v>1\end{array}\right.
\end{align}
where $q=c\pi /3\beta$ which is the same as the non-compact case.
Here, we impose the homology condition of the minimal curve: ${\cal C}\underset{\rm hom}{\sim} A$, which means that ${\cal C}=\bigcup_{i=1}{\cal C}_i$, which consists of several components, is homologous to the union of the regions $A=\bigcup_{i=1}A_i$ at $t=0$ and $t=t_1$. In the case where $A$ has two components one of which is defined on $t=0$ and the other at $t=t_1$, all candidates for the minimal curves that give the entanglement in the disconnected phase are shown in Figure \ref{UnitaryLines}.
\begin{figure}[h]
\begin{center}
\includegraphics[scale=0.4]{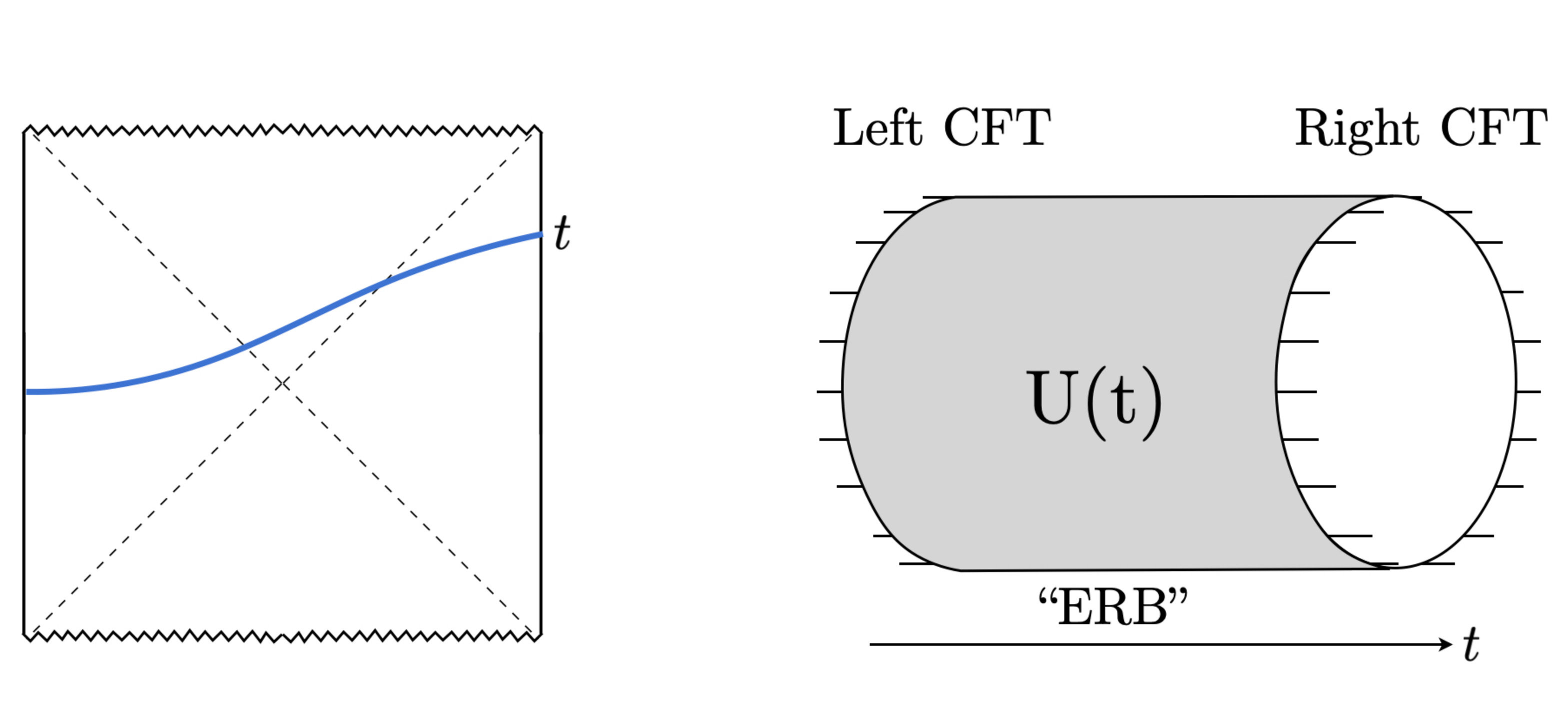}
\caption{Left: the eternal black hole dual to the thermofield double state., where two black holes are connected via the Einstein-Rosen bridge.   Right: Unitary circuit defined on a compact spacetime.  These are topologically equivalent, and the their lengths grow linearly in time. }
\label{UnitaryERB}
\end{center}
\end{figure}

Our prescription correctly reproduces the behavior of the entanglement entropy in holographic CFTs computed by the Ryu-Takayanagi surface on the AdS spacetime. To see this, let us remind ourselves that the unitary operator can be mapped by the channel-state map to the dual state, which is represented by the thermofield double state as
\ba
U(t)=\sum_{a} e^{-i t E_{a}}|a\rangle\langle a|\rightarrow| U(t)\rangle=\mathcal{N}\sum_{a} e^{-i t E_{a}}|a\rangle_{\text {out }}\left|a^{*}\right\rangle_{\text {in }}\, ,
\ea
i.e., the entangled state between the doubled Hilbert space ${\cal H}_{\rm in}$ and ${\cal H}_{\rm out}$. Here $|a\rangle$ is an eigenstate of the Hamiltonian of the system, and $|\cdot ^*\rangle$ is CPT conjugate to the state $|\cdot \rangle$. In the AdS/CFT correspondence, the thermofield double state is dual to the eternal black hole spacetime, i.e., the double-sided black hole, where two black holes are connected via the Einstein-Rosen bridge.  It is intriguing that the unitary circuits defined on a compact space and the Einstein-Rosen bridge in the AdS spacetime are {\it topologically} equivalent (i.e., cylindrical topology), and moreover both lengths grow linearly in time. This topological correspondence becomes clearer when we consider the line-tension picture.
\begin{figure}[h]
\begin{center}
\includegraphics[scale=0.4]{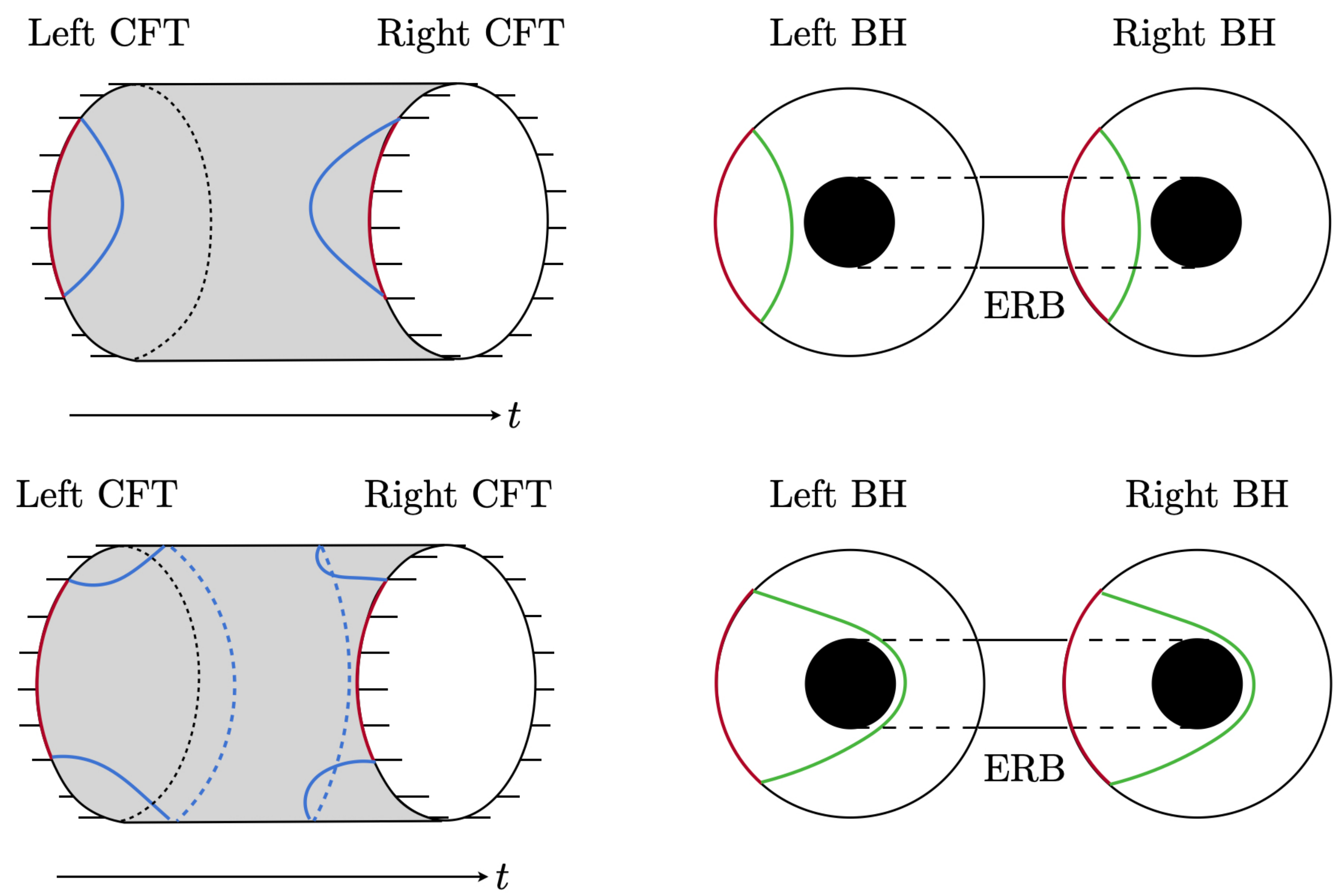}
\caption{ Two configurations in the disconnected phase: homologically trivial one (top) and the homologically non-trivial one (bottom). Left: the minimal curve in the line-tension picture. Right: Ryu-Takayanagi surface in the holographic calculation. }
\label{UnitaryBH}
\end{center}
\end{figure}

We compute the operator entanglement by taking the subregions $A_1$ with length $L$ at a initial time ($t=0$) and $A_2$ with the same length at time $t$.  By the state-channel map, the  region $A_1$ is put on ${\cal H}_{\rm in}$ and $A_2$ on ${\cal H}_{\rm out}$. To compute the entanglement entropy in the line tension picture, we need to compute the line-tension ${\cal T}(v)$ integrated along the minimal curve ${\cal C}$ anchored at the edges of $A=A_1\cup A_2$ that cuts the unitary circuits. As we explained, in the disconnected phase, we have two candidates for the minimal surface that computes the entanglement entropy: (a) the homologically trivial curve (top panel) and (b) the homologically non-trivial curve that encloses the cylinder where the unitary circuit is defined (bottom panel). Therefore, the disconnected candidates for the entanglement entropy  $S_{A}^{\rm disc}$ is computed as
\begin{align}\label{S_disc}
S_{A}^{\rm disc}={\rm min}\bigg\{(a):\f{2\pi c}{3\beta} L ,\ (b): \f{2\pi c}{3\beta}(2\pi R- L)\biggl\}\, .
\end{align}
Since  the entanglement entropy satisfies 
\bal\label{Sdisc}
S_{A}^{\rm disc}&=2S_{A_1} \quad L<\pi R\, ,\no &<2S_{A_1}\quad \  L>\pi R\, ,
\eal
the BOMI does not decay to zero when $L>\pi R$. This does not happen when the system is defined on a non-compact space since there are no homologically non-trivial curves in the disconnected phase.

Interestingly, these configurations of the minimal curve topologically correspond to the Ryu-Takayanagi surfaces in the AdS spacetime that compute the entanglement entropy holographically. In the right panel of Figure \ref{UnitaryBH}, we draw the two candidates for the Ryu-Takayanagi surfaces extending to the bulk double-sided black hole.
In the disconnected phase, there are two possibilities as shown in the right panel of Figure \ref{UnitaryBH}. One candidate is (a) the homologically trivial curve (top panel) and the other is (b) the homologically non-trivial curve that encloses the black hole horizons (bottom panel), and they nicely correspond to the candidates for the minimal curves in the line-tension picture. Indeed, the holographic calculations of the entanglement entropy using the Ryu-Takayangi surfaces (a) and (b) equivalent to the result (\ref{S_disc}) at the leading order of the coarse-grained limit ($\beta$ being much smaller than the other scales of the system).

\begin{figure}[h]
\begin{center}
\includegraphics[scale=0.35]{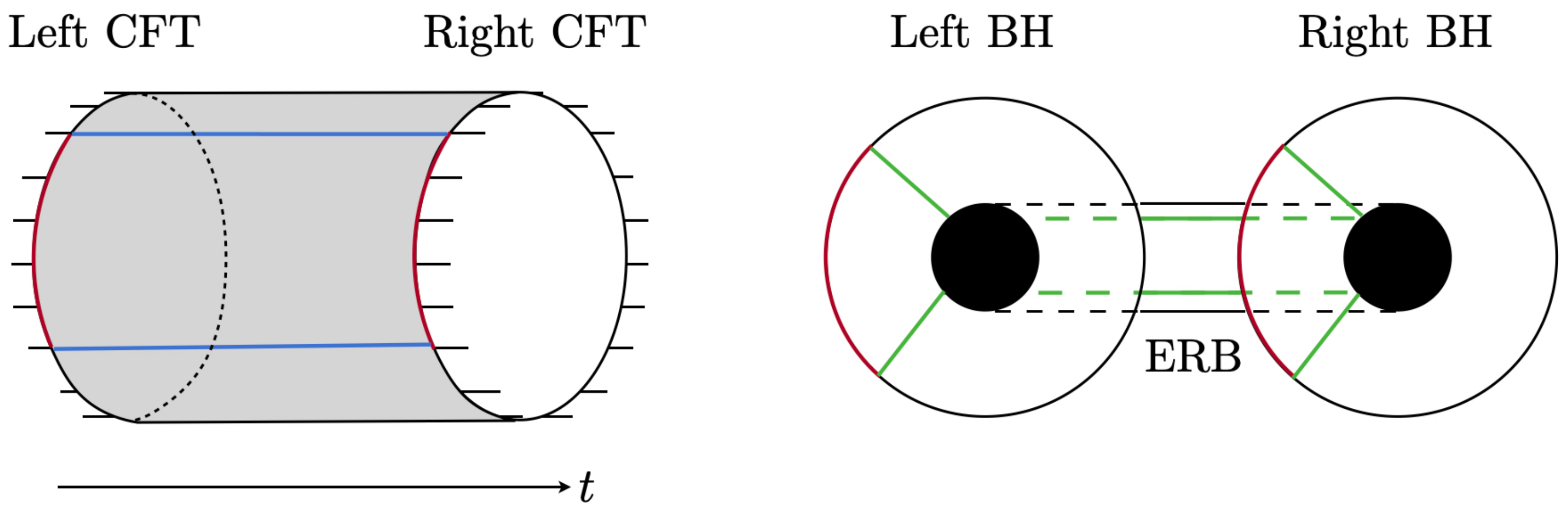}
\caption{Connected phase. Left: the minimal curve in the line-tension picture. Right: Ryu-Takayanagi surface in the holographic calculation. }
\label{UnitaryBH2}
\end{center}
\end{figure}

\begin{figure}[h]
\begin{center}
\includegraphics[scale=0.4]{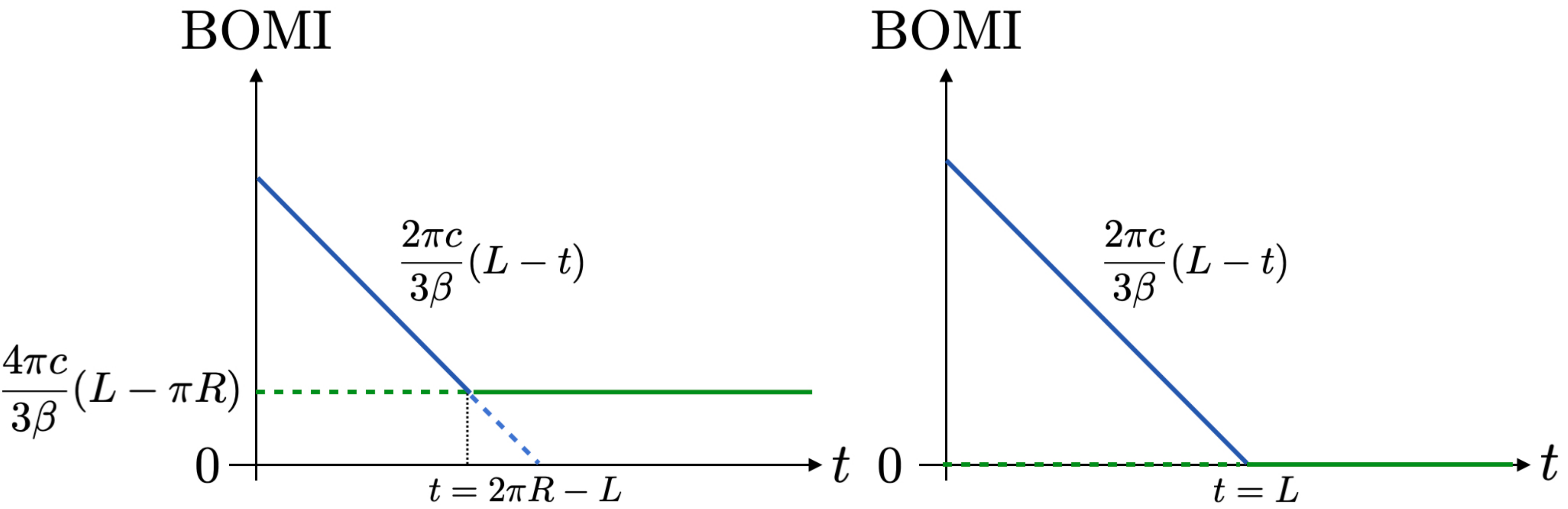}
\caption{Time dependence of BOMI in the symmetric case with $L_A=L_B=L$ in a compact-space (left) and a non-compact-space (right). The decaying part depicted by the orange line comes from the connected configuration of the minimal curve. The constant part  depicted by the blue line comes from the disconnected configuration of the minimal curve. In a compact space, there is a homologically non-trivial configuration as shown in the bottom of Figure \ref{UnitaryBH} and it prevents BOMI from decaying to zero.}
\label{mutual_information}
\end{center}
\end{figure}

In the connected phase, the configurations of the minimal surface in the line-tension picture and the corresponding Ryu-Takayanagi surface in the holographic calculation are shown in Figure \ref{UnitaryBH2}. In the symmetric case, the entanglement entropy computed by the line-tension picture is given by 
\begin{align}
S_{A }^{\mathrm{con}}=\frac{2 c \pi}{3 \beta} t\, .
\end{align}
The entanglement entropy for the region $A$ is given by the smaller of the connected configuration and the disconnected configuration as
\ba
S_{A }={\rm min}\bigg\{S_{A }^{\mathrm{disc}}, S_{A }^{\mathrm{con}}\biggl \}\, .
\ea
To summarize, the time-dependence of BOMI computed by the line-tension picture  is given by Figure \ref{mutual_information}.

\newpage
\subsubsection{Comments on complexity in the line-tension picture}
In this section, we will estimate the ``computational complexity'' of the unitary operator $U(t)$ using the line-tension picture. The operator entanglement for the unitary operator $U(t)$ in a chaotic system saturates at a certain time due to the thermalization. This implies that the details of the quantum state evolved by the unitary operator $U(t)$ become indistinguishable due to the thermalization. On the other hand, the cylindrical geometry that represents the unitary $U(t)$ used in the line-tension picture continues to grow even after the thermalization time. This suggests that the cylinder representing $U(t)$ itself captures the details of the $U(t)$ that cannot be distinguished by the operator entanglement after thermalization.

Computational complexity is known as a quantity more sensitive to the details of quantum states than the entanglement. In this section, we propose that the volume of the cylinder representing the unitary operator $U(t)$ computes the complexity of $U(t)$.

Computational complexity is a quantum information-theoretic quantity that measures the difficulty of producing a given operator $U$ by a sequence of simple unitary operators called ``gates". 
A sequence of gates is called a qunatum circuit. As a simple operator, we usually consider a operator that involves one or two qubits. The complexity of a given (generically $K$-qubit) operator $U$ can be produced by a quantum circuit as $U=g_ng_{n-1}\cdots g_1$. The complexity of the operator $U$ is defined as the minimal number of gates that is necessary to produce the operator $U$.



\begin{figure}[h]
\begin{center}
\includegraphics[scale=0.3]{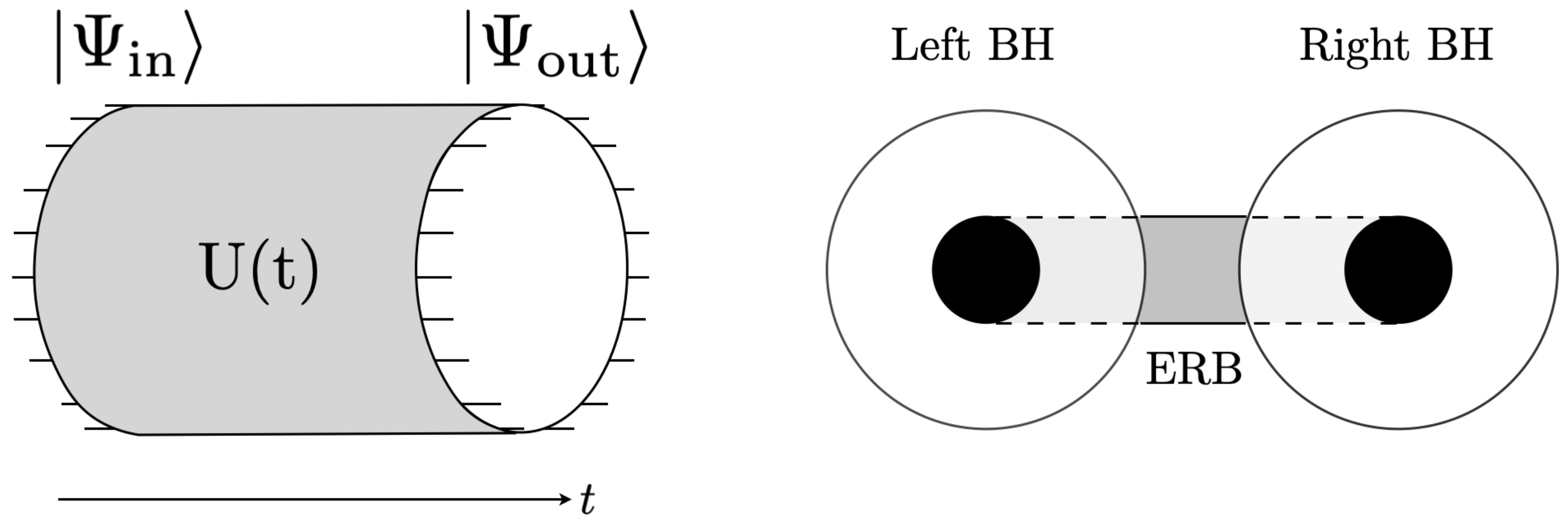}
\caption{The geometrical correspondence between the complexity i the line-tension picture (left) and the holographic complexity (right).}\label{fig:complexity}
\end{center}
\end{figure}

We propose how we can estimate the complexity of the unitary operator $U(t)$ parametrized by the time $t$ using the line-tension picture. We can represent the quantum circuit defined on a compact space as the cylinder Fig.\ref{fig:complexity}. One can estimate the complexity of this circuit defined as the number of gates as 
\begin{equation}
\text{Number of gates} =
    \frac{\text{Volume of the cylinder}}{\text{lattice cut-off}}\times(\text{number of d.o.f at each lattice site})\, .
\end{equation}
The volume of the cylinder is $2\pi R\times t$, and the lattice cut-off is set by $\ep$\footnote{One might wonder why the optimization of the quantum circuit is not considered here. We expect that the random unitary circuit filling up the cylinder already is the optimized quantum circuit representing a chaotic system. We leave this as a future work.}. The number of the degrees of freedom at each lattice site is given by the bond dimension (the local Hilbert space dimension) $q$.
Therefore, we obtain the complexity of the unitary operator as
\ba
{\cal C}=\f{2\pi Rt}{\ep^2}\log q\, .
\ea
Hitherto we have been considering the complexity of the operator, but it can also be interpreted as the complexity of the state given by the channel-state map from the operator. The above estimation shows that the complexity of the state dual to the unitary operator $U(t)$ is proportional to the volume of the cylinder representing $U(t)$ and  grows linearly in time. This is generally expected in a chaotic system. This is because the two different operations generically produces two distinct states in chaotic systems. On the other hand, in an integrable system, one can generically expect that we end up producing the same state even after different operators are applied to a state. In particular since $q=e^{\frac{c\pi}{3\beta}}$ for a holographic CFT, the growth rate is estimated as
\ba
\f{d{\cal C}}{dt}=\f{\pi c}{6\ep^2}(2\pi R)=S_{\rm BH}T,
\ea
i.e., a product of the entropy and the temperature of the black hole dual to the CFT state given by the channel-state map from $U(t)$. 

This nicely fits the recent proposal for the holographic dual to the computational complexity in \cite{Susskind:2014moa}. In \cite{Susskind:2014moa} they considered the time-evolution of the thermofield double state 
\be
|{\rm TFD}(t)\rangle \propto \sum_{a}  e^{-it E_a} \ket{a}_{L} \ket{a^*}_{R},
\ee
which is dual the double-sided black holes connected by a wormhole. They proposed that the computational complexity in a CFT (called holographic complexity) computes the volume of the wormhole in the dual gravity side. Since the volume of the wormhole grows linearly in $t$, it reproduces the expected behavior of the computational complexity in a chaotic system. In more detail, one can find that its growth rate is computed as
\ba
\f{d{\cal C}_{\rm hol}}{dt}\propto S_{\rm BH}T,
\ea
which nicely reproduces the result in the line-tension picture discussed above. 
As pointed out above, the cylinder in the line-tension picture topologically equivalent to the dual wormhole geometry (Fig.\ref{fig:complexity}). It is also intriguing that there is a geometrical correspondence between the volume of the cylinder representing $U(t)$ that estimates the complexity of the unitary operator $U(t)$ and the volume of the wormhole that computes the holographic complexity. 

\section{Discussions and Future directions}
Let us conclude the paper with a discussion of the main results as well as future directions. In this paper, we studied the operator entanglement of time evolution operators in order to study the scrambling effect of non-equilibrium processes in $2$d CFTs defined on spaces with finite extent.
We studied the time evolution of the BOMI and TOMI of the unitary time evolution operator for free fermion and holographic CFTs. The BOMI and TOMI for the free fermion time evolution operators were studied numerically and analytically, while those for the holographic CFT time evolution operator were studied analytically.
We observed quantum revival
in the free fermion theory in which the value of the BOMI returned to its initial value, but no such revivals were observed in the holographic CFTs. We also computed the TOMI which is a  measure of the scrambling effect of the system's dynamics and found that it is zero for the free fermion, regardless of whether the size of the system is finite or infinite \cite{2018arXiv181200013N}. This is because the time evolution of the BOMI for free fermions is perfectly described by the relativistic propagation of quasiparticles which are localized packets of information. On the other hand, the TOMI in holographic CFTs depends on whether or not the system is finite or infinite. We found that the absolute value of the TOMI at late times can be smaller when the system is finite than when the system is infinite depending on the choice of subsystems. This is consistent with the results obtained in numerical analysis of spin chains \cite{mascot2021local}.

\subsection*{Discussion in terms of density matrix}

In this section, we will discuss quantum revival in terms of density matrices.
If we expand the density matrix of the dual state of the time evolution operator in the energy eigenstate $\ket{n}$, it is given by 
\begin{equation}
\rho= \f{1}{\Tr e^{-2\epsilon H}} e^{2\epsilon H}+ \f{1}{\Tr e^{-2\epsilon H}} \sum_{n\neq m } \left(\ket{n}\bra{m}\otimes  \ket{n}\bra{m}\right)e^{- \epsilon (E_n+E_m)-it (E_n-E_m)}.
\end{equation}
Let $\rho_{\text{diag}}$ and $\rho_{\text{non-diag}}(t)$ denote the diagonal and non-diagonal components of the density matrix, respectively. The diagonal component $\rho_{\text{diag}}$ is the density matrix of the thermal state, and the time evolution of the physical quantity in the non-equilibrium process originates from the non-diagonal terms. 
When the contribution from this non-diagonal component is sufficiently small, the state can be well approximated by a thermal equilibrium state  \cite{PhysRevA.43.2046,PhysRevE.50.888,2008Natur.452..854R}.
In $2$d CFTs with system size $2\pi R$, the Hamiltonian is given by $H=\f{1}{R}\left(L_0+\overline{L}_0\right)-\f{c}{12R}$, where $L_0$ and $\overline{L}_0$ are Virasoro generators. Thus, in a $2$d CFT on finite spacetime, the spectral gap of energy is proportional to the size of the system, and thus quantum revival can occur with a period determined by the size of the system.
\subsubsection*{Free fermion}
First, let's consider the case of free fermion.
In the case of free fermion, since the eigenvalues are $E_n=\f{n}{R}-\f{c}{12R}$, the off-diagonal term $\rho_{\text{non-diag}}(t)$ of the density matrix is given by
\begin{equation}
\rho_{\text{non-diag}}(t)=\f{1}{\Tr e^{-2\epsilon H}} \sum_{n\neq m } \left(\ket{n}\bra{m}\otimes  \ket{a}\bra{b}\right)e^{- \epsilon \left(\f{ (n+m)}{R}-\f{c}{6R}\right)-\f{it}{R} (n-m)}.
\end{equation}
Here, $n$ and $m$ are integers.
Thus, the off-diagonal term of the density matrix satisfies the periodicity $\rho_{\text{non-diag}}(t)=\rho_{\text{non-diag}}(t+2\pi R)$, and the density matrix has a period of $2\pi R$. Therefore, the free fermion BOMI has a periodic behavior and quantum revival occurs.
\subsubsection*{Holographic CFT}
In general CFTs, the energy eigenstate $\ket{n}$ can be expanded in terms of conformal dimensions $(h_i,\overline{h}_j)$ and levels $(N^n_i,\overline{N}^n_j)$ as
\begin{equation} \label{eigenstate}
\ket{n}=\sum_{ij}C^n_{ij}\ket{h_i,N^n_i} \otimes \ket{\overline{h}_j,\overline{N}^n_j},
\end{equation}
where $\ket{h_i,N^n_i}$ and $\ket{\overline{h}_j,\overline{N}^n_j}$ are eigenstates of $L_0$ and $\overline{L}_0$ respectively, $N^n_i$ and $\overline{N}^n_j$ are integers, and we assume that $h_i$ and $\overline{h}_j$ are irrational numbers.
The product state consisting of these states is given by
\begin{align} \label{product_state}
H\left(\ket{h_i,N^n_i} \otimes \ket{\overline{h}_j,\overline{N}^n_j}\right) =&E_n\ket{h_i,N^n_i} \otimes \ket{\overline{h}_j,\overline{N}^n_j}  \\ \nonumber
=&\f{1}{R}\left[h_i+N^n_i+\overline{h}_j+\overline{N}^n_j-\f{c}{12}\right]\ket{h_i,N^n_i} \otimes \ket{\overline{h}_j,\overline{N}^n_j}. 
\end{align}
Expanding the density matrix in terms of (\ref{product_state}), the off-diagonal components of the density matrix are
\begin{equation}
\begin{split}
&\rho_{\text{non-diag}}(t)=\rho_{h_i+\overline{h}_j=h_I+h_{\overline{J}}}(t)+\rho_{h_i+\overline{h}_j\neq h_I+h_{\overline{J}}}(t),\\
&\rho_{h_i+\overline{h}_j=h_I+h_{\overline{J}}}(t)=\f{1}{\Tr e^{-2\epsilon H}} \sum_{n\neq m }\sum_{\substack{ ijkl;IJKL,\\ h_i+\overline{h}_j=h_I+h_{\overline{J}} }}e^{-\f{ \epsilon}{R}\left(2h_i+2\overline{h}_j+N^n_i+\overline{N}^n_j+N^m_I+\overline{N}^m_J-\f{c}{6}\right)}e^{-\f{i t}{R}\left(N^n_i+\overline{N}^n_j-N^m_I-\overline{N}^m_J\right)}\\
&\times C^n_{ij}C^n_{kl}C^{m*}_{IJ}C^{m*}_{KL}\ket{h_i,N^n_i,\overline{h}_j,\overline{N}^n_j}\bra{h_I,N^m_I,\overline{h}_J,\overline{N}^m_J}\otimes \ket{h_k,N^n_k,\overline{h}_l.\overline{N}^n_l}\bra{h_K,N^m_K,\overline{h}_L,\overline{N}^m_L},\\
&\rho_{h_i+\overline{h}_j\neq h_I+h_{\overline{J}}}(t)\\
&=\f{1}{\Tr e^{-2\epsilon H}} \sum_{n\neq m }\sum_{\substack{ ijkl;IJKL,\\ h_i+\overline{h}_j\neq h_I+h_{\overline{J}} }}e^{-\f{\epsilon}{R}\left(h_i+\overline{h}_j+h_I+\overline{h}_J+N^n_i+\overline{N}^n_j+N^m_I+\overline{N}^m_J-\f{c}{6}\right)}e^{-\f{it}{R}\left(h_i-h_I+\overline{h}_j-\overline{h}_J+N^n_i+\overline{N}^n_j-N^m_I-\overline{N}^m_J\right)}\\
&\times C^n_{ij}C^n_{kl}C^{m*}_{IJ}C^{m*}_{KL}\ket{h_i,N^n_i,\overline{h}_j,\overline{N}^n_j}\bra{h_I,N^m_I,\overline{h}_J,\overline{N}^m_J}\otimes \ket{h_k,N^n_k,\overline{h}_l.\overline{N}^n_l}\bra{h_K,N^m_K,\overline{h}_L,\overline{N}^m_L}.\\
\end{split}
\end{equation}
Here, the component of the density matrix $\rho_{h_i+\overline{h}_j=h_I+h_{\overline{J}}}(t)$ is periodic, i.e. $\rho_{h_i+\overline{h}_j=h_I+h_{\overline{J}}}(t)=\rho_{h_i+\overline{h}_j=h_I+h_{\overline{J}}}(t+2\pi R)$.
On the other hand, $\rho_{h_i+\overline{h}_j\neq h_I+h_{\overline{J}}}(t)$ is not periodic.
The upshot is that BOMI and TOMI do not exhibit quantum revival in holographic CFTs.
The time-dependent decrease of BOMI can most likely be attributed to the  $\rho_{h_i+\overline{h}_j\neq h_I+h_{\overline{J}}}(t)$ terms as these are the 
 components of $\rho_{\text{non-diag}}(t)$ that are not periodic. Also, from the point of view of the AdS/CFT correspondence, the time evolution of these terms are expected to be related to the growth of wormholes.
Therefore, we expect the information about the wormhole growth to be encoded in $\rho_{h_i+\overline{h}_j\neq h_I+h_{\overline{J}}}(t)$.

\subsubsection*{Why do not the BOMI and TOMI of holographic CFT exhibit quantum revival?}
In this section, we use the twist operator formalism to explain why the time evolution of BOMI and TOMI in holographic CFT does not exhibit a quantum revival.
Here, the subsystems are the intervals defined by $A=[X_2,X_1]$ and $B=[Y_2, Y_1]$, where $A$ is defined on the initial time slice and $B$ is defined on the time slice at time t.
The OEE for $A\cup B$ in twist operator formalism is given by
\be
\begin{split}
&S_{A\cup B} = \text{Min}\left[S_{\text{dis.}}, S_{\text{con.}}\right], \\
&S_{\text{dis.}} = \lim_{n\rightarrow 1} \f{1}{1-n}\log{\left \langle \sigma_n(X_1)\overline{\sigma}_n(X_2)\right \rangle}+\lim_{n\rightarrow 1} \f{1}{1-n}\log{\left \langle \sigma_n(Y_2,t)\overline{\sigma}_n(Y_1,t)\right \rangle},\\
&S_{\text{con.}} = \lim_{n\rightarrow 1} \f{1}{1-n}\log{\left \langle \sigma_n(X_1)\overline{\sigma}_n(Y_1,t)\right \rangle}+\lim_{n\rightarrow 1} \f{1}{1-n}\log{\left \langle \sigma_n(Y_2,t)\overline{\sigma}_n(X_2)\right \rangle}.
\end{split}
\ee
Since $S_{\text{dis.}}$ is given by the thermal two-point correlation function of the operator defined on the same time slice, this is independent of time. On the other hand, $S_{\text{con.}}$ continues to grow with time. For this reason, after a time of at most the subsystem size, $S_{A\cup B}$ is given by $S_{\text{dis.}}$, which is independent of time.
Since the time for this phase transition to occur is short compared to the size of the system,  the time evolution of BOMI and TOMI  in holographic CFT does not exhibit quantum revival.
\subsection*{Future directions}
Here, we will discuss some future directions.
\begin{itemize}
\item {\bf Quantum revival and scrambling effect:} We studied the behavior of BOMI and found that quantum revival does not occur at all in holographic CFTs whereas complete quantum revival occurs in free fermions which follows from the relativistic propagation of quasiparticles.
It would be interesting to study the extent to which quantum revival can be suppressed by studying the time evolution of BOMI in  non-equilibrium processes with varying degrees of information scrambling as in compact bosons.
\item {\bf Adding symmetry:} In this study, we found that when a holographic CFT is placed in a system with finite size $2\pi R$, quantum revival does not occur, but the scrambling effect is smaller than in an infinite volume system.
One of the interesting future directions is to study what modifications to the holographic CFT dynamics can weaken the scrambling effect.
For example, it would be interesting to see if the inclusion of additional symmetries in holographic CFTs would lead to the appearance of quantum revival or a suppression of information scrambling. 
One of the studies in this direction is \cite{2021arXiv210704043K}.
\item {\bf Relationship between the density matrix and the gravity dual:} In this section, we discussed the correspondence between the components of the density matrix and the gravity dual in the AdS/CFT correspondence. It would be interesting to clarify this correspondence further.
\end{itemize}
\section*{Acknowledgments}
We thank Jonah Kudler-Flam, Tokiro Numasawa and Shinsei Ryu for
fruitful discussion.
K.G.~is supported by JSPS Grant-in-Aid for Early-Career Scientists 21K13930. M.N.~is supported by JSPS Grant-in-Aid for Early-Career Scientists 19K14724. K.T.~is supported by JSPS Grant-in-Aid for Early-Career Scientists 21K13920.
A.M. is supported by JSPS Grant-in-Aid for Challenging Research (Exploratory) 18K18766. A.M. thanks school of physics of institute for research in fundamental sciences (IPM) for their hospitality during the final stages of this work. 

\newpage
\appendix
\section{Free fermions with unphysical boundary condition}\label{app:4NS}
\begin{figure}
    \centering
    \textbf{Symmetric intervals}\par\medskip
    \includegraphics[trim={4cm 0 28cm 0},clip,valign=t,width=0.45\textwidth]{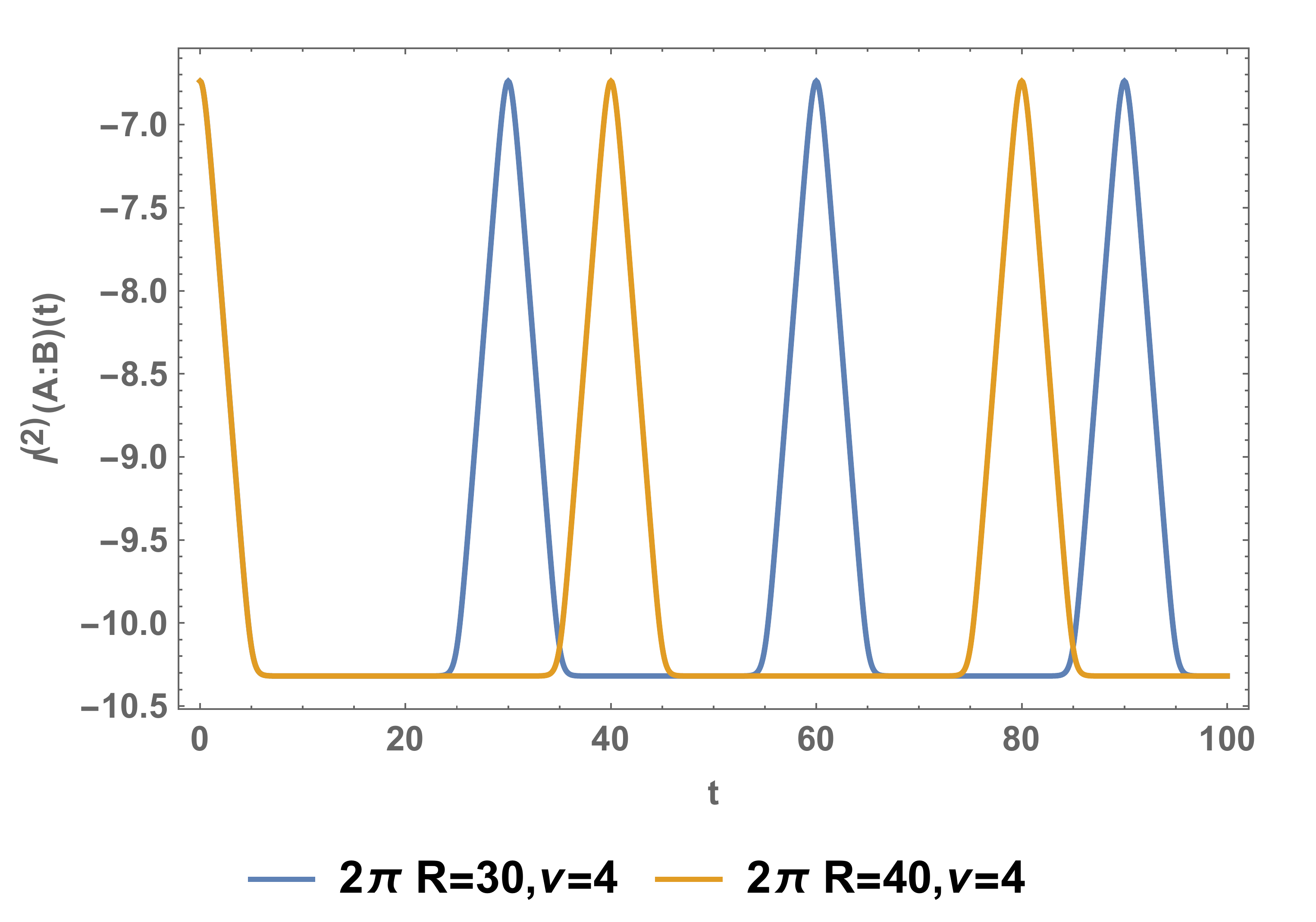}
    \includegraphics[trim={4cm 0 28cm 0},clip,valign=t,width=0.45\textwidth]{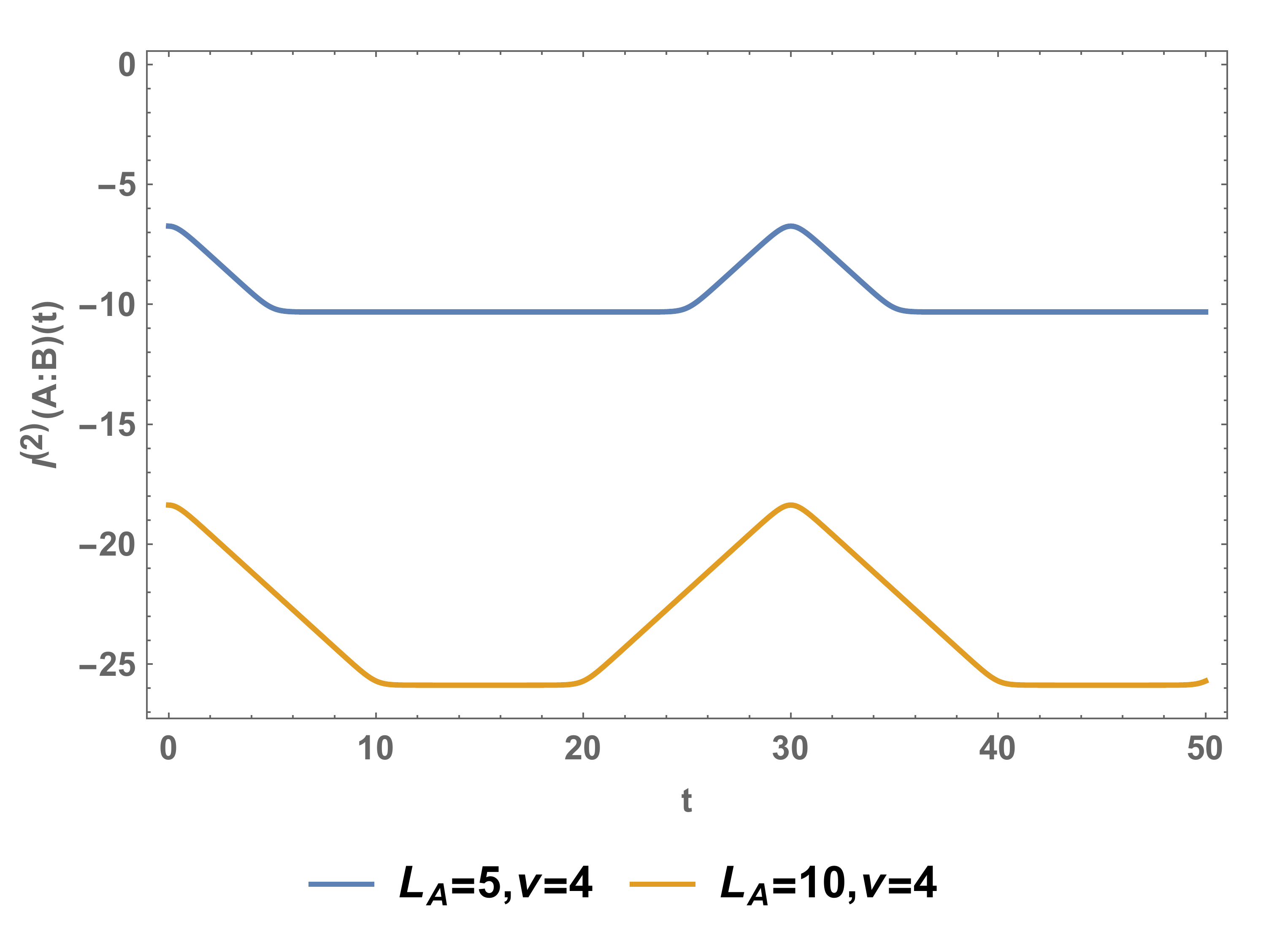}
    \textbf{Asymmetric intervals}\par\medskip
    \includegraphics[trim={4cm 0 28cm 0},clip,valign=t,width=0.45\textwidth]{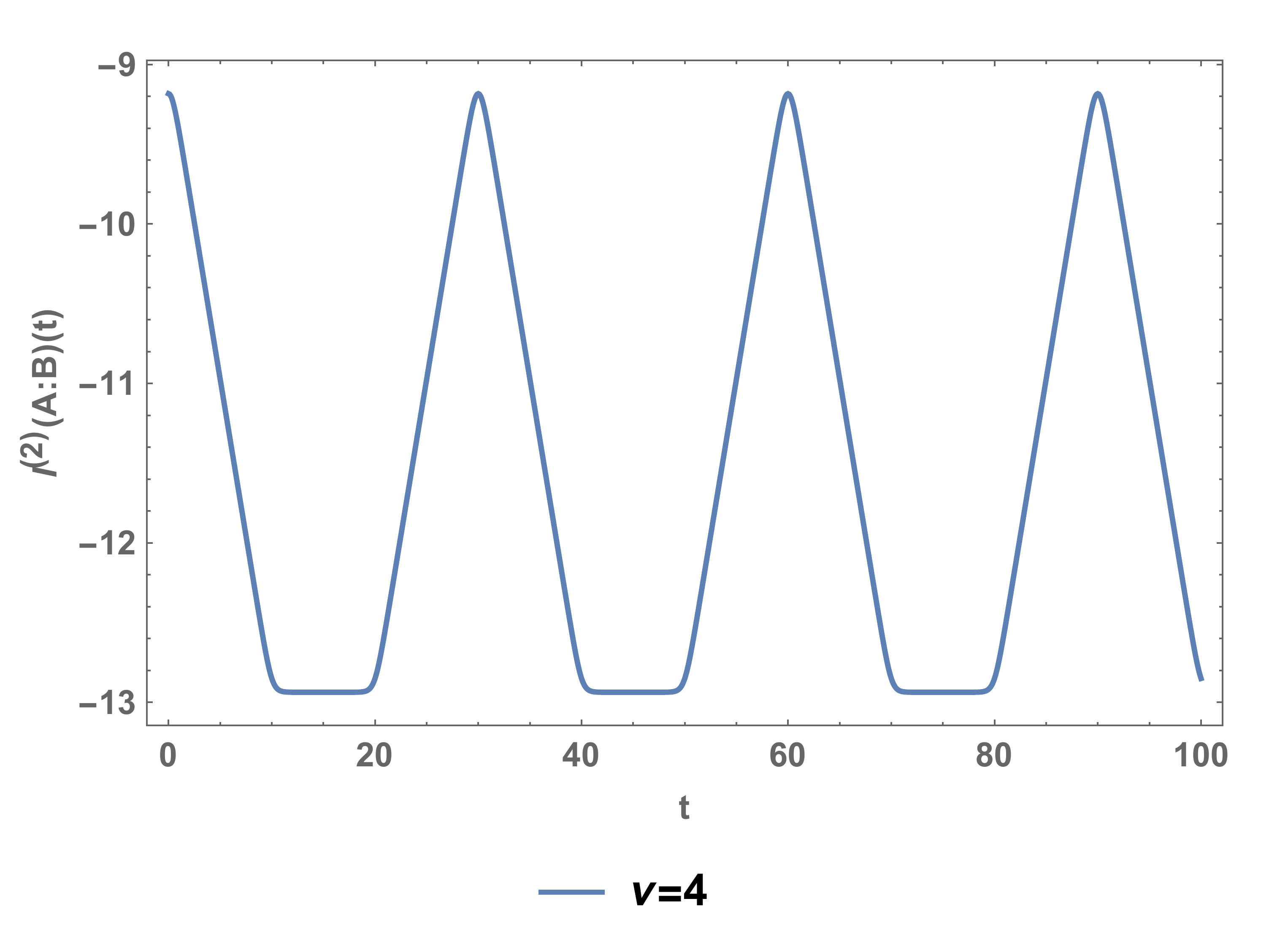}
    \includegraphics[trim={4cm 0 28cm 0},clip,valign=t,width=0.45\textwidth]{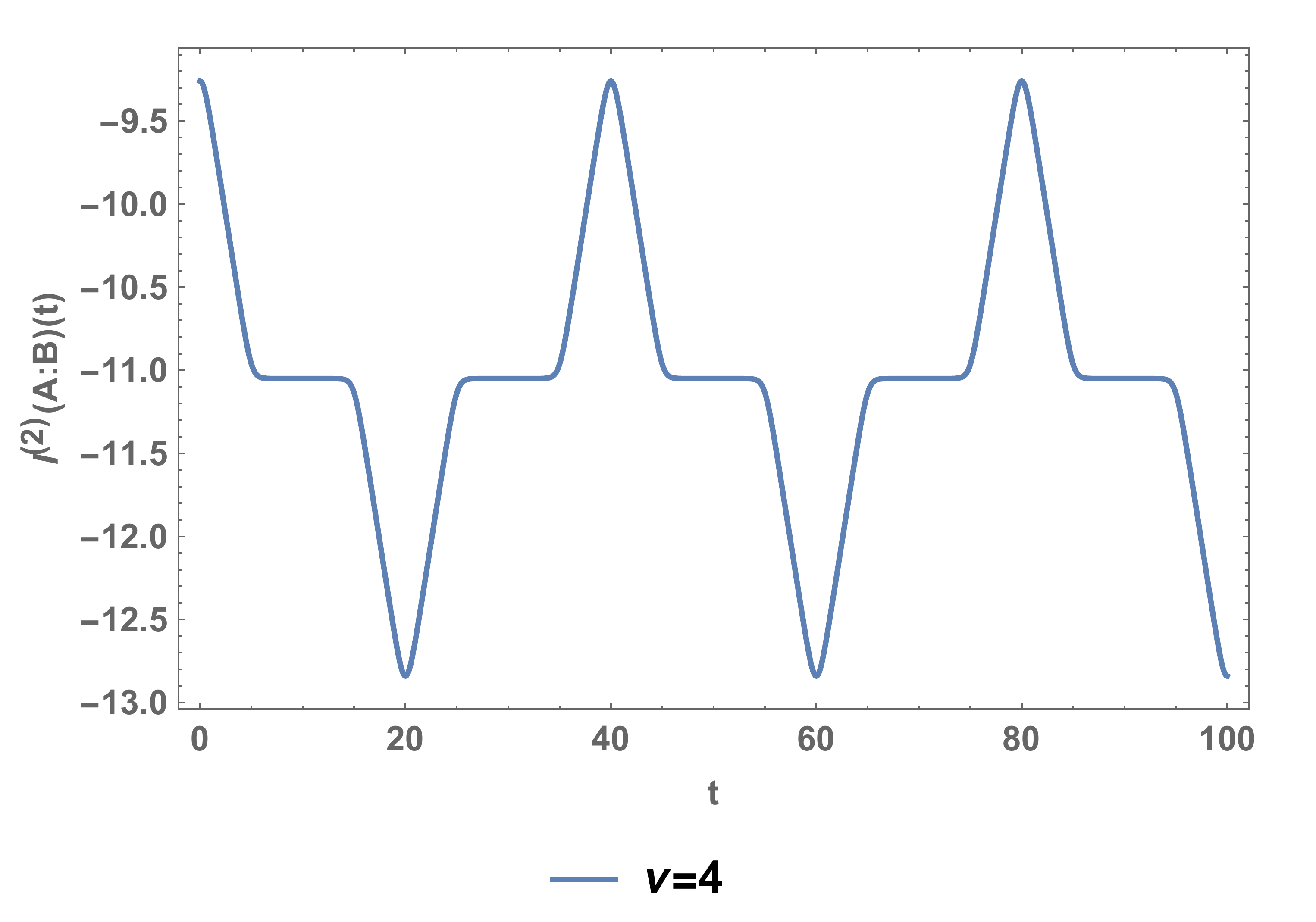}
    \caption{Plots of bipartite operator mutual information for symmetric and anti-symmetric intervals for the $c=1$ Dirac fermion with $n=2$, $\epsilon=1$ and periodic boundary conditions along the Euclidean time direction. (\textbf{Top row}:)
    The parameters are identical to the left and middle plots of figure \ref{FiniteSystemBOMI_SymmetricIntervals_Physical} but with a different spin structure $\nu=4$.
     $(\textbf{Bottom row}:)$ The total system size and the subsystems are the same as in figure \ref{FiniteSystemBOMI_AsymmetricIntervals} but with $\nu=4$.}
    \label{FiniteSystemBOMI_UnphysicalBC}
\end{figure}

The operator mutual information for free massless Dirac fermions with anti-periodic boundary conditions along the thermal cycle was discussed in section \ref{DiracFermionAnalyticalResultSection}. In this appendix, the boundary conditions along the thermal cycle are taken to be periodic instead. Plots of the corresponding mutual information are shown in Figure \ref{FiniteSystemBOMI_UnphysicalBC}. The mutual information shows exactly the same behavior as in the case where the boundary conditions along the thermal cycle were fixed to be anti-periodic. The only difference is that in this case, the mutual information has a time-independent offset that causes the 2\textsuperscript{nd} R\'{e}nyi mutual information to be negative. A possible explanation is that computing the mutual information with periodic boundary conditions imposed along the thermal cycle does not correspond to the mutual information of an operator state. The operator state lives on the doubled Hilbert space $\mathcal{H}=\mathcal{H}_1 \otimes \mathcal{H}_2$ where $\mathcal{H}_1$ and $\mathcal{H}_2$ are the input and output Hilbert spaces respectively. Tracing out the output Hilbert spaces gives
\begin{equation}\label{PathIntegralThermal}
\text{tr}_{\mathcal{H}} |U(t)\rangle\langle U(t) |=
    \text{tr}_{\mathcal{H}_1}e^{-\beta H}
\end{equation}
where $H$ is the Hamiltonian acting on the original Hilbert space $\mathcal{H}_1$. For fermions, the path integral \eqref{PathIntegralThermal} leads to anti-periodic boundary conditions along the Euclidean time direction. In order to have a fermionic path integral with periodic boundary conditions in the Euclidean time direction, a fermion parity operator has to be inserted into the path integral as follows:
\begin{equation}
    \text{tr}_{\mathcal{H}_1}\left(e^{-\beta H}(-1)^F\right)
\end{equation}
where $F$ is the fermion number operator. Such a path integral is not of the form \eqref{PathIntegralThermal} and therefore does not correspond to the unitary operator state. In fact, it might not even correspond to any operator state. Therefore, it is not surprising that the resulting mutual information behaves differently from the cases when anti-periodic boundary conditions are imposed along the Euclidean time direction.

\begin{figure}
    \centering
    \textbf{Disjoint intervals}\par\medskip
    \includegraphics[trim={4cm 0 28cm 0},clip,valign=t,width=0.45\textwidth]{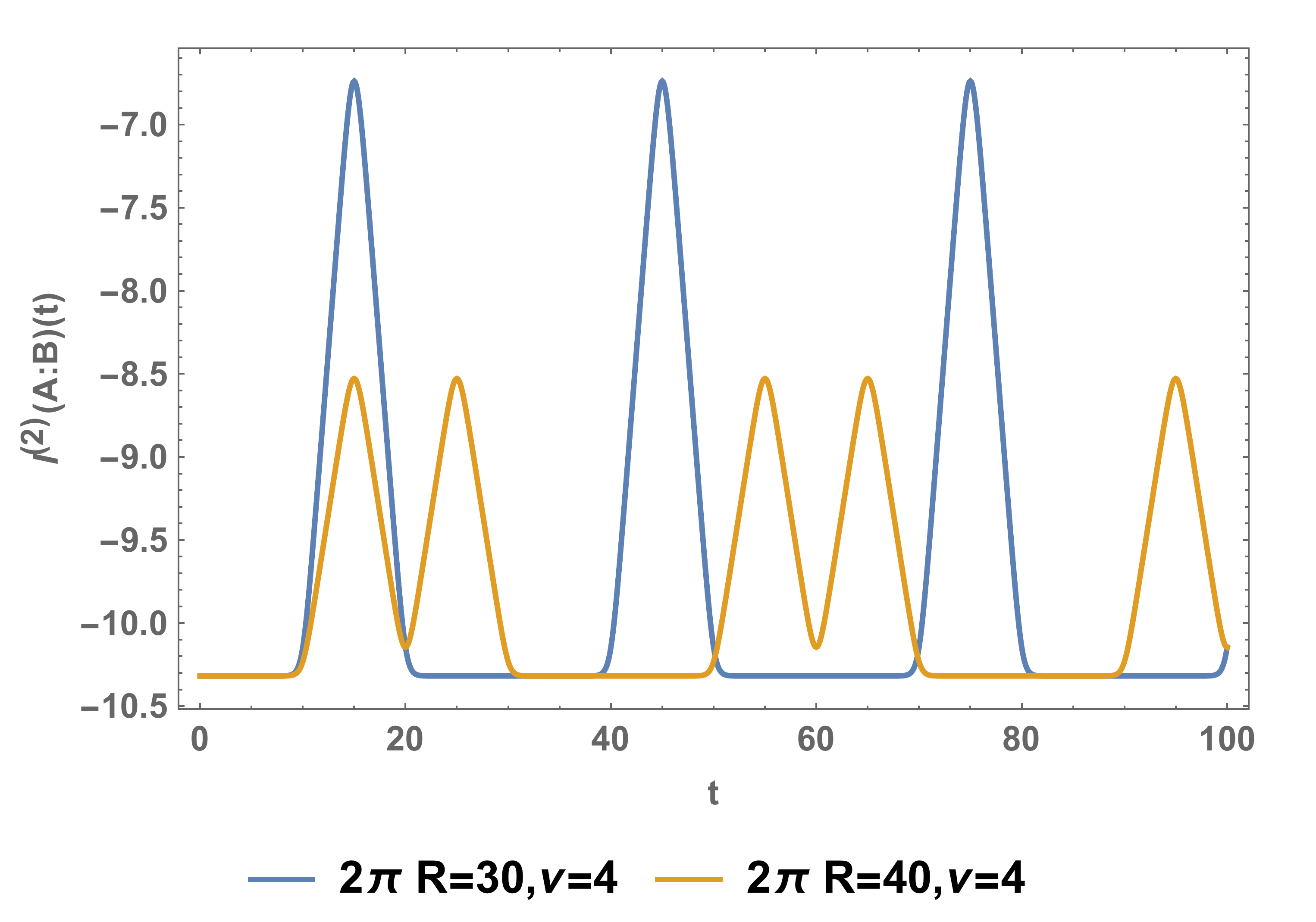}
    \includegraphics[trim={4cm 0 28cm 0},clip,valign=t,width=0.45\textwidth]{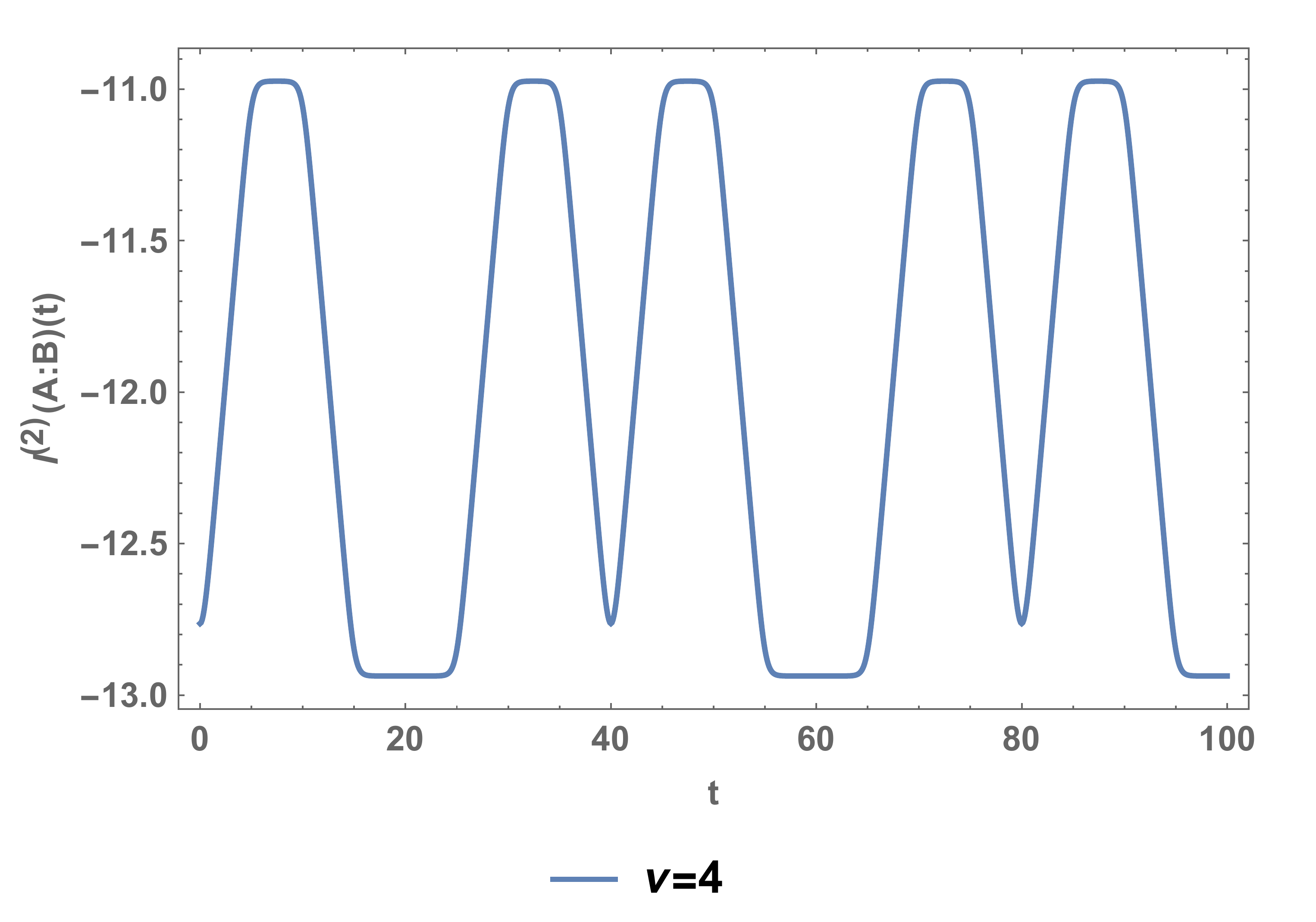}
    \caption{Plots of bipartite operator mutual information for disjoint intervals for the $c=1$ Dirac fermion with $n=2$, $\epsilon=1$ and periodic boundary conditions along the Euclidean time direction. The setup is identical to the one in figure \ref{FiniteSystemBOMI_DisjointIntervals} but with $\nu=4$ instead.}
\end{figure}

\newpage
\section{Free Bosonic Theories versus Quasi-particle Picture}\label{app:boson}
\subsection{Free Bosonic Theory}
The free bosonic theory is defined similar to the fermion case with
\be
\omega_k=\sqrt{m^2+\frac{4}{\delta^2}\sin^2\left(\frac{\pi k}{N}\right)},
\ee
where the oscillators obey the standard bosonic commutation relations and the TFD state is defined as
\be
|\beta\rangle=\bigotimes_{k=1}^N \left(1-e^{-\beta\omega_k}\right)^{\frac{1}{2}}\sum_{n_k=0}^{\infty}e^{-\beta\omega_k n_k/2}e^{-i\omega_k n_k t}|n_k\rangle_1|n_k\rangle_2.
\ee
The numerical method is the same as what we described in section \ref{sec:numericalmethod} where $q$'s and $p$'s in \eqref{eq:covvec} refer to each oscillator in the decoupled (Fourier) basis.

\subsection{Quasi-particle Picture}
\begin{figure}[t!]
\begin{center}
\includegraphics[scale=.27]{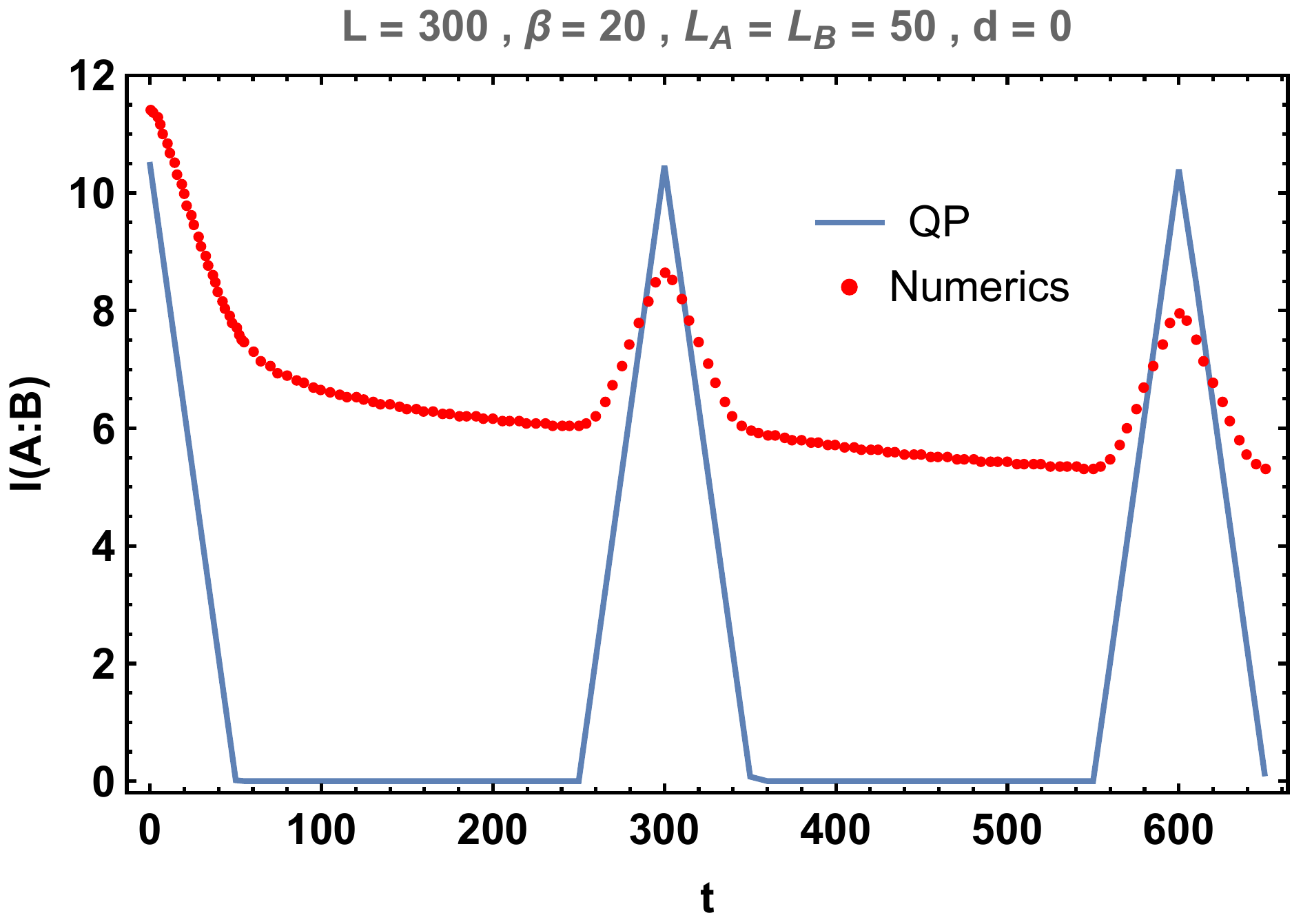}
\includegraphics[scale=.27]{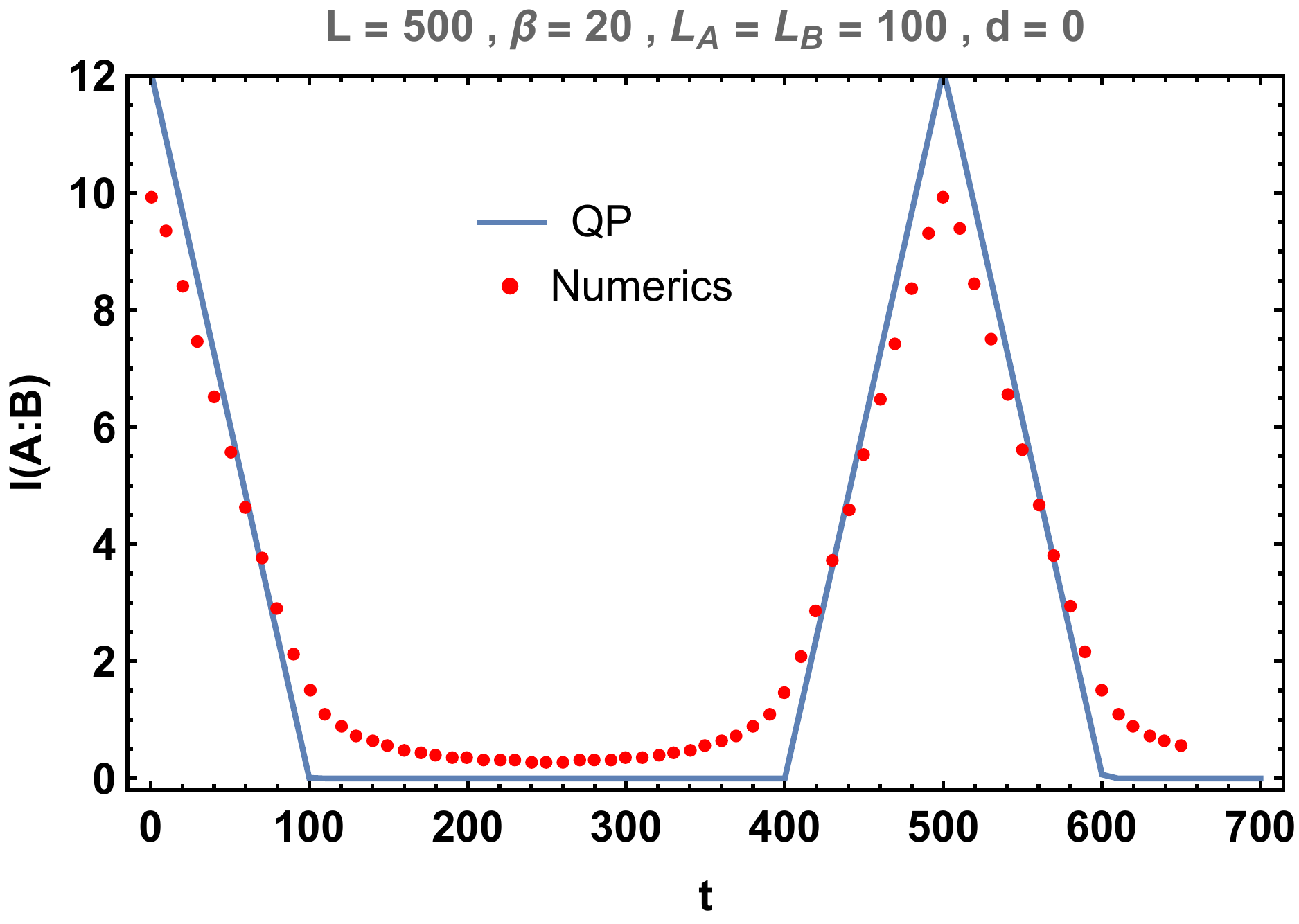}
\includegraphics[scale=.27]{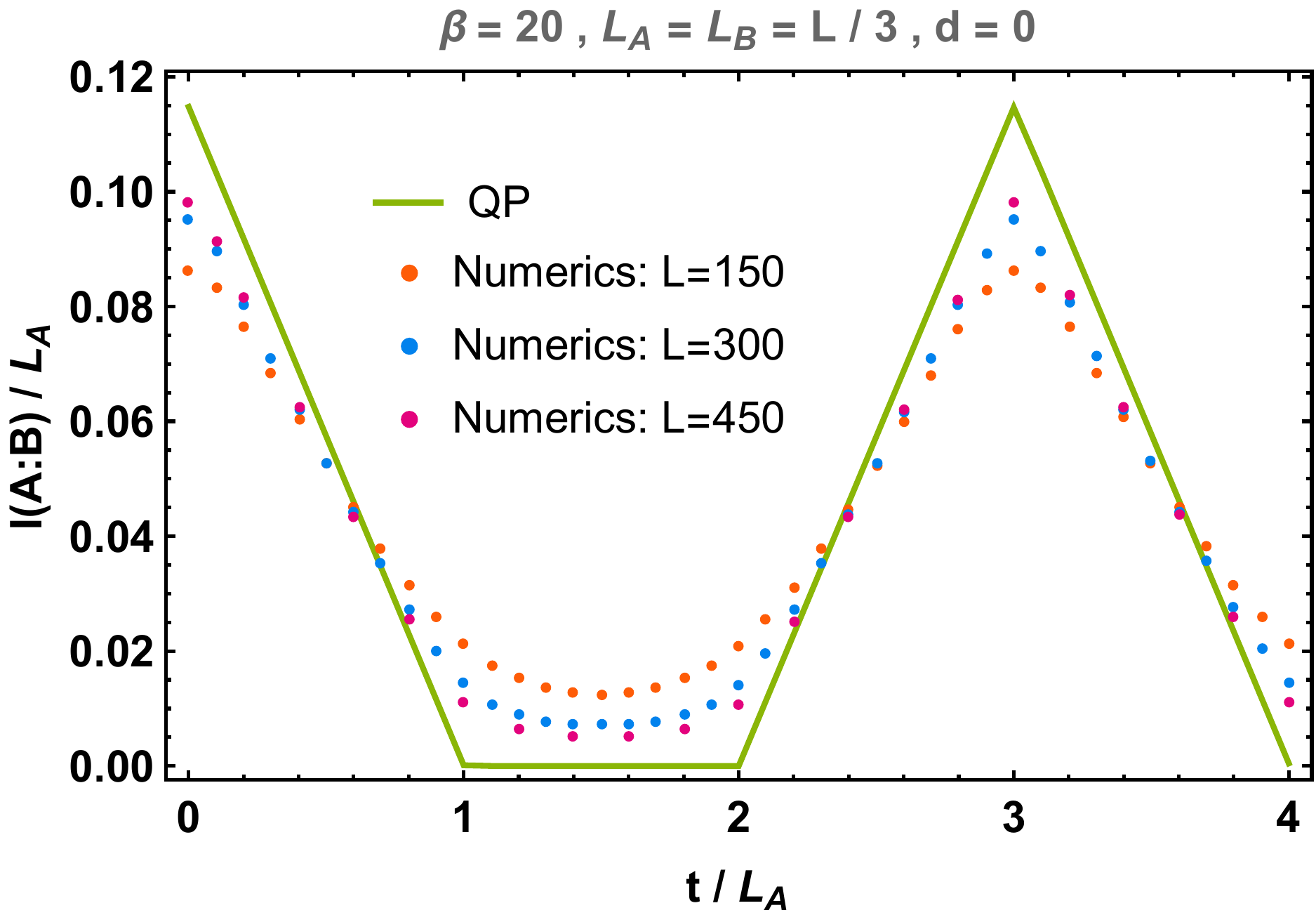}
\end{center}
\caption{Numerical results for free bosonic theory versus quasi-particle picture for periodic and Dirichlet boundary conditions. The left panel corresponds to periodic boundary condition where the zero-mode effect causes significant difference between numerical results and quasi-particle prediction. The middle panel corresponds to Dirichlet boundary condition where the numerical results are much closer to the quasi-particle prediction. The right panel shows that as we approach the scaling limit, the Dirichlet boundary condition numerical results approaches the quasi-particle prediction.  
}
\label{fig:NQPB}
\end{figure}

The analysis of the quasi-particle formula in the main text was quite general for any integrable theory. The main difference between free fermion and boson theory is the role of the zero-mode which makes the case of free boson theory on a compact spatial manifold somehow different from the fermion case. In the fermion case we find perfect agreement with the quasi-particle picture while this is not the case for bosonic theory with periodic boundary condition since the quasi-particle picture does not capture the zero-mode effect. We have presented numerical results in the left panel of Figure \ref{fig:NQPB}. As expected, in the early times ($t\ll L_A$) the zero-mode effect is small and the numerical result is expected to match with the quasi-particle prediction in the scaling limit. As time passes and ($t\sim L_A$), but much before entanglement revival effects come into the game, the amount of entanglement carried by the zero-mode increases, the numerical results deviate more from the quasi-particle prediction. 

Moreover, to avoid the zero-mode effect and confirm our quasi-particle formulae beyond fermionic theories, we have also studied the free scalar theory on a compact spatial direction with Dirichlet boundary condition where there is no zero-mode by construction. The logic of working out the quasi-particle formula for this case is the same as what we discussed for the periodic case, but the formulae of different boundary conditions are slightly different. In the Dirichlet case we find a very good agreement between the quasi-particle prediction and our numerical results (see the middle and right panel of Figure \ref{fig:NQPB}), confirming the consistency of the quasi-particle picture for both fermionic and bosonic theories.

\newpage

\bibliographystyle{ieeetr}
\bibliography{main}

\end{document}